\documentclass[12pt]{article}
\pdfoutput=1
\usepackage{comment}
\usepackage{amsmath}
\usepackage{amssymb}
\usepackage{graphicx}
\usepackage{here}
\usepackage{subcaption}
\usepackage{url}
\usepackage[sort&compress, numbers, merge]{natbib}
\usepackage{braket}
\usepackage{physics}
\usepackage[compat=1.1.0]{tikz-feynhand}
\usepackage{slashed}
\usepackage{mathtools}
\usepackage[bottom]{footmisc}

\usepackage{todonotes}

\numberwithin{equation}{section}

\usepackage[colorlinks=true, linkcolor=blue, citecolor=blue,
urlcolor=black]{hyperref}

\begin{document}
\def\ps{\mathbf{p}}
\def\PS{\mathbf{P}}
\baselineskip 0.6cm
\def\simgt{\mathrel{\lower2.5pt\vbox{\lineskip=0pt\baselineskip=0pt
           \hbox{$>$}\hbox{$\sim$}}}}
\def\simlt{\mathrel{\lower2.5pt\vbox{\lineskip=0pt\baselineskip=0pt
           \hbox{$<$}\hbox{$\sim$}}}}
\def\simprop{\mathrel{\lower3.0pt\vbox{\lineskip=1.0pt\baselineskip=0pt
             \hbox{$\propto$}\hbox{$\sim$}}}}
\def\tr{\mathop{\rm tr}}
\def\SU{\mathop{\rm SU}}
\def\GFtree{G_F^0}

\begin{titlepage}

\begin{flushright}
IPMU23-0048
\end{flushright}

\vskip 1.1cm

\begin{center}

{\Large \bf
Precise Estimate of Charged Higgsino/Wino Decay Rate
}

\vskip 1.2cm
Masahiro Ibe$^{a,b}$,
Yuhei Nakayama$^{a}$ and
Satoshi Shirai$^{b}$
\vskip 0.5cm

{\it

$^a$ {ICRR, The University of Tokyo, Kashiwa, Chiba 277-8582, Japan}

$^b$ {Kavli Institute for the Physics and Mathematics of the Universe
 (WPI), \\The University of Tokyo Institutes for Advanced Study, \\ The
 University of Tokyo, Kashiwa 277-8583, Japan}
}

\vskip 1.0cm

\abstract{
Higgsinos and Winos in the supersymmetric Standard Model are prime candidates for dark matter due to their weakly interacting nature. The mass differences between their charged components (charginos) and neutral components (neutralinos) become degenerate when other superparticles are heavy, resulting in long-lived charginos. In the case of the Winos, the mass difference is approximately 160\,MeV across a wide range of the parameter space. Consequently, the chargino decays into the lightest neutralino, emitting a single charged pion. For Higgsinos, however, mass differences ranging from 
$O(0.1)$\,GeV to $O(1)$\,GeV are possible, leading to a variety of decay channels. 
In this paper, we extend our previous analysis of Wino decay to the chargino with a larger mass difference. 
We emphasize characterizing its decay signatures through leptonic and hadronic modes. 
By utilizing the latest experimental data, we perform a comprehensive study of the decay rate calculations incorporating these hadronic modes to determine the impact on the predicted chargino lifetime. Additionally, we conduct next-to-leading order (NLO) calculations for the leptonic decay modes.
Our NLO results can be applied to the case of more general fermionic electroweak multiplets, e.g., quintuplet dark matter.
}

\end{center}
\end{titlepage}

\setcounter{tocdepth}{2}
{\hypersetup{linkcolor=black}
\tableofcontents
}

\section{Introduction} 

The absence of dark matter (DM) candidates in the Standard Model (SM)
stands as compelling evidence for the existence of new physics. 
The supersymmetric (SUSY) extension of the SM offers a range of theoretical 
and phenomenological benefits, particularly its compatibility with DM properties 
if the $R$-parity is conserved. 
In particular, the SUSY model in which the Higgsino or Wino is identified as the lightest SUSY particle (LSP) and key components of DM is intriguing.

The Higgsino is the superpartner of the Higgs boson in the SUSY Standard Model (SSM),
which is the vector-like SU(2)$_L$ doublet fermion
with hypercharge $1/2$.
As the mass parameter of the Higgsino, 
i.e., the $\mu$-parameter,
directly contributes to the Higgs boson mass parameter, it is a crucial parameter 
for the naturalness of the SSM~\cite{Barbieri:1987fn},
which implies a small $\mu$-parameter or a light Higgsino.

The Wino, identified as a Majorana fermion and an $\mathrm{SU}(2)_{L}$ triplet with zero hypercharge, is the superpartner of the weak gauge boson in the SSM. 
The Wino is likely the LSP and a candidate for DM,
in the anomaly mediation model~\cite{Randall:1998uk, Giudice:1998xp}.
Following the discovery of the Higgs boson, the minimal anomaly mediation, often referred to as ``mini-split SUSY," has gained prominence \cite{Hall:2011jd, Hall:2012zp, Nomura:2014asa, Ibe:2011aa, Ibe:2012hu, Arvanitaki:2012ps, ArkaniHamed:2012gw}.

These Higgsino/Wino-like dark matter appear in various models, not only in the SUSY model.
For instance, the minimal dark matter model~\cite{Cirelli:2005uq,*Cirelli:2007xd,*Cirelli:2009uv,Essig:2007az}, which aims to elucidate the nature of dark matter with the fewest possible modifications to the SM, incorporates electroweak interacting massive 
particles such as the Higgsino/Wino as a DM.

The electroweak interactions involving the Higgsino/Wino-like DM facilitate the exploration and detection of DM. 
Those interactions enable various methods of investigation, including direct and indirect detection,
as well as the collider experiments.

The ``Dirac" Higgsino DM 
is in tension with
the direct detection experiments 
since
the $Z$-boson exchange process
provides a huge spin-independent scattering cross section with the target material \cite{Nagata:2014wma}.
This constraint can be evaded by 
the mixings with the neutral gauginos, i.e., the Bino and the neutral Wino.
The mixings split the Dirac neutral fermion into two Majorana fermions, 
and hence
the DM does not have the spin-independent $Z$-boson exchange process.
Instead,
the DM has the coupling to the Higgs boson,
which induces 
the small spin-independent cross section 
for the direct detection.
Note that the coupling of the DM to the Higgs boson 
also affects the mass splittings 
among the charged Higgsino and the two Majorana Higgsinos.
As the mass splittings correlate with the spin-independent cross section via the Higgs exchange \cite{Fukuda:2019kbp},
the Higgsino-like DM with 
the mass difference 
larger than $O(1)$\,GeV has 
been disfavored by the direct detection experiments \cite{LZ:2022lsv,XENON:2023cxc}.

If the mass difference between the charged and neutral Higgsino, $\mathit{\Delta}m_{\pm}$, 
is small, the chargino's decay length can become macroscopic, making it accessible at the collider experiment. 
In this case the charged Higgsino mainly decays into mesons and its decay length is
\begin{align}
\label{eq:ctau}
c\tau_{{\chi}^\pm \to \chi^0_{1} \pi^{\pm}  } = O({1})\,\mathrm{mm}\times \left(\frac{\mathit{\Delta}m_\pm}{500\,\mathrm{MeV}}\right)^{-3}\, .
\end{align}
When the decay length exceeds $O({1})$\,cm, the charged Higgsino exhibits disappearing tracks \cite{ATLAS:2022rme,CMS:2023mny}.
Even for shorter decay lengths, the pions resulting from the charged Higgsino decay produce soft and displaced tracks, offering a distinct signature from the SM background~\cite{Fukuda:2019kbp}.

The pursuit of long-lived particles holds equal importance in the context of Wino DM. 
For the Wino DM,
the mass difference 
between the charged and the neutral Wino
scarcely depends on the details of the model, as it is predominantly determined by the electroweak loop. 
Consequently, the mass difference is estimated to be approximately $160$~MeV
~\cite{Yamada:2009ve,Ibe:2012sx}, making the disappearing charged track search a critical method at the collider experiments.

In the long-lived chargino searches, the signal acceptance depends exponentially on the lifetime of the charged states. 
Therefore, precise estimate of the charged Higgsino/Wino lifetime is essential for the probe of the Higgsino/Wino DM at collider experiments.
In most experimental analyses, 
the radiative corrections to the 
charged Higgsino/Wino decay rates 
have not been taken into account.
As we will see, however, there are about $10$\% 
ambiguities in the leading order estimate of the decay rate
depending on the choice of the ``tree-level"
parameters. This ambiguity results in an uncertainty of several tens of percent in the acceptance of 
disappearing track signals at the LHC.

In Ref.~\cite{Ibe:2022lkl}, we have 
performed the next-to-leading order (NLO) analysis
of the charged Wino decay into a single pion
aiming at the mass difference $\mathit{\Delta}m_\pm \simeq 160$\,MeV.
In this work, 
we extend the previous estimate of the decay rate
of the chargino for larger mass difference
so as to apply our analysis to more general cases including the Higgsino decay.
The major difference from our previous analysis is in the inclusion of the leptonic mode,
the single Kaon mode, and other multi-meson modes.
With these extensions, we achieve a precision of less than 1\% in calculating the chargino decay rate for $\mathit{\Delta} m_\pm \lesssim 1.5$\,GeV.

In this paper, we perform 
the NLO analysis by taking 
the pure Higgsino as the example.
However, 
we show that 
the NLO corrections to the decay rate,
including electroweak correction, only depend on the difference of the QED charges of the initial and final states 
for the small mass splitting.
Therefore, our current results can be applied 
to the case of more general fermionic electroweak multiplets, e.g., quintuplet dark matter~\cite{Cirelli:2005uq,*Cirelli:2007xd,*Cirelli:2009uv}.

The organization of the paper is as follows.
In Sec.~\ref{sec: Summary of Results}, we 
summarize our results.
In Sec.~\ref{sec: model discripition}, 
we review the model of 
the chargino and neutralino
focusing on the Higgsino-like case.
In Sec.~\ref{sec: EW corrections to FF operator}, 
we define the tree-level four-fermion interaction relevant for the chargino decay
and compute the short-distance electroweak corrections.
In this section, we will
see 
that ambiguities which exist
in ``tree-level'' computation
will be almost resolved
by the short-distance corrections.
In Sec.~\ref{sec: radiative corrections to leptonic mode},
we analytically compute the long-distance corrections to the leptonic decay mode.
In Sec.~\ref{sec: decay into single pion}, we 
update our previous estimation
of NLO corrections to
the single pion mode.
In Sec.~\ref{sec: multimeson},
we estimate other hadronic decay modes.
In particular, we apply
updated data of the hadronic 
tau lepton decay
to estimation of the multi-meson
modes of the chargino.
Sec.~\ref{sec: Conclusions} is devoted to our conclusions.

\section{Summary of Results}
\label{sec: Summary of Results}

\begin{figure}[tb]
\centering
  \includegraphics[width=0.7\linewidth]{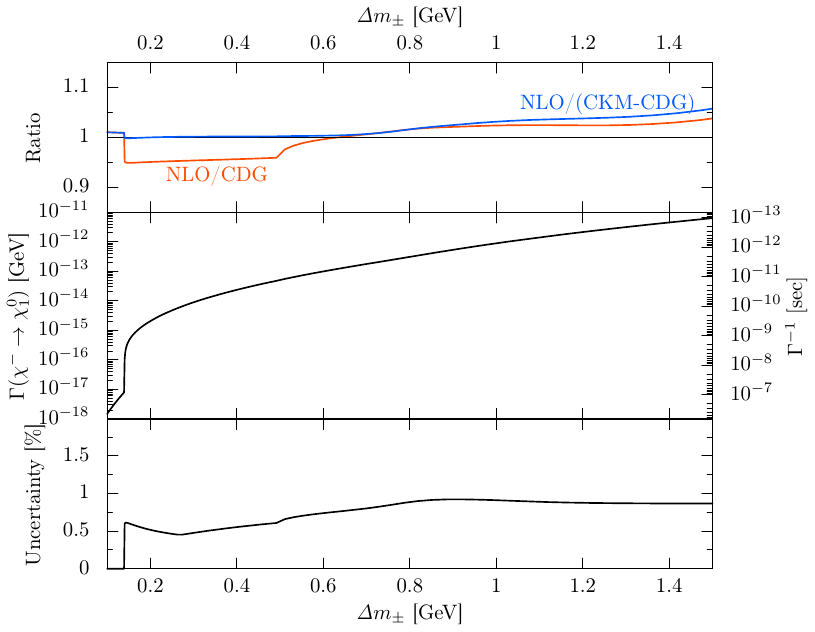}
\caption{The NLO chargino decay rate into one of the neutralinos for a 300\,GeV Higgsino, including the associated uncertainty, is presented in the middle and bottom panel. The red line in the top panel depicts the ratio of our NLO result to the previous CDG estimate. 
In contrast, the blue line indicates the ratio to the CDG estimate corrected for the inclusion of the CKM matrix and the single Kaon mode.
The data
are available at \url{https://member.ipmu.jp/satoshi.shirai/Chargino_Decay/}.
}\label{fig:rate}
\end{figure}

\begin{figure}[tb]
\centering
\includegraphics[width=0.7\linewidth]{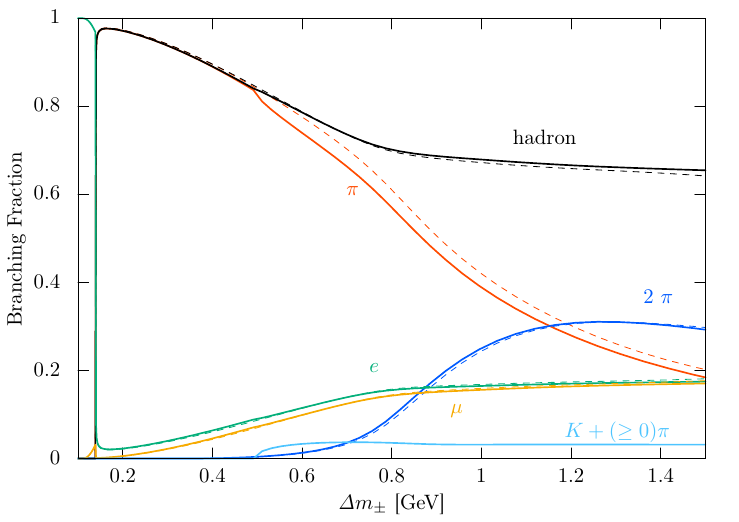}
\caption{The branching fractions 
of each decay mode
for 300\,GeV Higgsino/Wino
together with comparison
with previous estimate
by CDG.
Our NLO results are
shown by the solid lines
and CDG's results are represented by 
the dashed lines.
The black solid line shows
the sum of all the hadronic modes.
The black dashed line represents
the sum of the $2\pi$ and $3\pi$ modes
estimated by CDG.
The Kaon modes, which are not included
in the CDG analysis, have 
a few percent branching fraction.
}\label{fig:bf}
\end{figure}

Before discussing details of our analysis,
we first summarize the key findings of this paper. Fig.~\ref{fig:rate} displays the NLO chargino decay rate into one of the neutralinos for a 300\,GeV Higgsino, including the associated uncertainties. Additionally, we provide a comparison with the previous estimates made by Chen, Drees, and Gunion  (abbreviated as CDG)~\cite{Chen:1996ap}.
The NLO decay rates for other parameter points are available from the URL in Fig.~\ref{fig:rate}.

When the charged Higgsino is capable of decaying into another neutral Higgsino, it is essential to include this decay rate to calculate the chargino lifetime. 
In the scenario of a pure Higgsino, for instance, the rate should be doubled.
Furthermore, we have ascertained that the NLO correction to the decay rate is identical for both the Higgsino and the Wino,
i.e.,
\begin{align}
\left.\frac{\Gamma_\mathrm{NLO}}{\Gamma_\mathrm{tree}}\right|_\mathrm{Higgsino}
=
\left.\frac{\Gamma_\mathrm{NLO}}{\Gamma_\mathrm{tree}}\right|_\mathrm{Wino}\ .
\end{align}
The tree-level decay rate of the charged Wino is four times greater than the Higgsino case, as the charged current interaction of the Wino is twice of the Higgsino (see Eqs.~\eqref{eq: mixing matrix phase}). 
As a result, the decay rate depicted in the figure can be adapted for the charged Wino case by multiplying it by four.

As we will explain in subsection~\ref{sec: ambiguity},
the ``tree-level'' or leading-order (LO)
approximation to the decay rates suffers from more than about 7\% uncertainty.
The NLO computation 
reduces this uncertainty substantially.
The remaining uncertainties come from theoretical estimation 
of the long-distance corrections
and the experimental errors 
of the hadronic spectral functions.
As a result, we provide
the decay width at a precision 
less than 1\% for 
a wide range of mass difference.

The original CDG analysis does not account for the effect of the CKM matrix element. This omission leads to an approximate 5\% deviation in the decay rate compared to our results, as illustrated by the red line in Fig.~\ref{fig:rate}.
We also present the CDG decay rate, now corrected by incorporating the CKM matrix. Additionally, we have extended their analysis on the pion mode to include the single Kaon mode, as represented by the blue line in Fig.~\ref{fig:rate}.
After these corrections, 
our results are about 3\% larger than the CDG estimate for the mass difference larger than 1\,GeV.
These differences come from the multi-meson modes.

Fig.~\ref{fig:bf} illustrates the branching fractions of each decay mode, along with a comparison to the previous estimates by CDG. Our updated results are depicted using solid lines, while CDG's results are represented by dashed lines.
Our new estimate slightly deviates from CDG results for the hadronic modes. 
The deviation mainly comes from 
the difference of 
the hadronic spectral functions.
The Kaon modes, which were not included in the CDG analysis, also affect the branching fractions.

\subsubsection*{Application to Wino Decay Rate}
\label{app: Wino Decay}
\renewcommand{\theequation}{\thesection.\arabic{equation}}

\begin{figure}[tb]
\centering
  \includegraphics[width=0.7\linewidth]{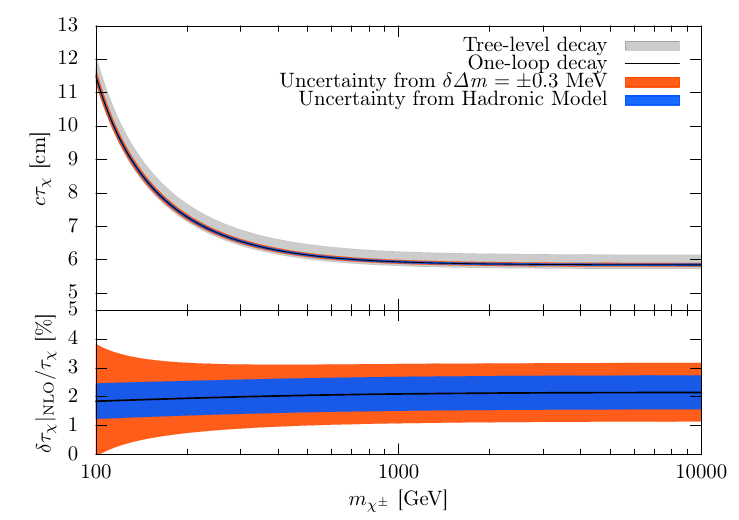}
\caption{
The decay length of the Wino as a function of its mass, denoted as $m_{\chi^\pm}$, is depicted using black solid lines. These lines represent the NLO estimation with the mass difference approximated at the two-loop level, approximately given by $\mathit{\Delta}m_\pm \simeq 160\,\mathrm{MeV}$. The blue bands illustrate the uncertainty of our one-loop estimation due to strong dynamics. In contrast, the gray band indicates the error in the tree-level analysis, which has been significantly reduced in the NLO analysis. Additionally, the red bands represent the uncertainty arising from the three-loop correction to the Wino mass difference, quantified as $\delta\mathit{\Delta}m = \pm 0.3\,\mathrm{MeV}$, as discussed in Ref.~\cite{Ibe:2012sx}. Here, we define $\delta \tau_\chi|_\mathrm{NLO} := \tau_\chi|_\mathrm{NLO} - \tau_\chi|_\mathrm{LO}$.
}\label{fig:ctauWino}
\end{figure}

The present NLO calculation can be applied to the Wino, yielding a result that suggests the Higgsino rate should be multiplied by four. 
In Fig.~\ref{fig:ctauWino}, we show the 
result of the NLO Wino decay length.
The black line
shows the NLO prediction of the Wino decay length for a given chargino mass.
The blue bands around the black solid line show the uncertainty of the NLO decay rate estimation from the strong dynamics. 
The gray band shows the uncertainty of the case of the LO calculation
which is reduced to the blue band by the NLO analysis.

In the figure, we have used the prediction on $\mathit{\Delta}m_\pm$ at the two-loop 
for the pure Wino in Ref.~\cite{Ibe:2012sx} (see also Ref.~\cite{McKay:2017xlc}),
adopting a fitting formula
\begin{align}
    \mathit{\Delta} m_\pm/\mathrm{MeV}|_{\mathrm{tree}} =  
\frac{ 21.8641 + 8.68343 t + 0.0568066 t^2}{1 + 0.0530366 t + 0.000345101t^2} ~~~ \left(t = \frac{m_{\chi^\pm}}{\mathrm {GeV}} \right)\ .
\end{align}
This fitting formula provides a stable fitting for $m_{\chi} > 90$\,GeV.
In the figure, the red band shows  the uncertainty of the lifetime from the higher loop corrections to the Wino mass difference, $\delta\mathit{\Delta}m_{\pm}$ = 0.3 MeV, in Ref.~\cite{Ibe:2012sx}.

Our calculation can be used for more general mass splitting for the Wino.
However, for the Wino, achieving a substantial mass difference between the neutralino and chargino, larger than the size of the radiative correction, proves challenging. 
The mass splitting of the Wino is heavily suppressed by the Higgsino mass,\,\footnote{In the case of $M_1 = M_2$, the neutralinos become what are known as Photino and Zino, which greatly differ from the pure Wino. 
In our analysis, however, we do not consider such extreme situations, and we will address scenarios where $|M_1| - |M_2|$ is separated by about $O(10)$ GeV or more.}
\begin{align}
    \label{eq: tree-level mass difference of Wino}
    \mathit{\Delta}m_\pm
    = \frac{m_Z^4 s_{2\beta}^2s_W^2c_W^2}{(M_1-M_2)\mu^2}
    \simeq\frac{140\,\mathrm{MeV}}{\tan^2\beta}\qty(\frac{300\,\mathrm{GeV}}{M_1-M_2})\qty(\frac{1\,\mathrm{TeV}}{\mu})^2\ .
\end{align}
Definitions of the parameters in this equation are given in Sec.~\ref{sec: model discripition}.
A large mass difference necessitates a smaller Higgsino mass $\mu$ or $M_1 \to M_2$, which in turn would make the Wino significantly deviate from the electroweak gauge interaction eigenstates. 
As a result, our NLO calculation is not applicable in such scenarios.

\subsubsection*{Update from Previous Study}
Let us summarize the main points of the update from Ref.~\cite{Ibe:2022lkl}:
\begin{itemize}
    \item The wave function renormalization factors in the heavy Wino limit used in the short distance correction are changed to those given in Appendix~\ref{app: explicit electroweak}. 
    \item We include the uncertainty of $f_{VW}$ in the analysis of the single pion mode (see subsubsection~\ref{sec: pion error}). 
    \item We include the NLO decay rate into the lepton mode.

    \item We include the Kaon  mode as well as multi-hadron modes
    with short-distance NLO corrections.
\end{itemize}

\section{Model of Higgsino Dark Matter}
\label{sec: model discripition}
In this paper, we perform the NLO analysis of the chargino decay rate by taking the Higgsino as an example.
As emphasised above, however, the following NLO results can be applied to the cases of the more general 
electroweak multiplets.
To provide a detailed demonstration of the calculations, the Higgsino will serve as our illustrative case study. Discussions on applications to other generic electroweak-interacting particles will follow in subsequent sections.

Here, we summarize the
set up of the Higgsino LSP scenario where the lightest neutralino
and the chargino are dominated by 
the Higgsino (vector-like SU$(2)_L$ doublets fermion) contributions.
Such a setup is realized when the 
gauginos and sfermions 
are rather heavy compared with
the Higgsinos.
In this limit, the mass splitting between the neutralinos 
as well as the mass difference between
the neutralino-chargino are small,
which is of $O(1)$\,GeV or less
for the gauginos and the sfermions in the multi-TeV range.
In the following, we review the effective Lagrangian of the Higgsino sector obtained from the minimal SSM (MSSM) setup.
The physical parameters used in our analysis are 
given in Table.~\ref{tab:input}.

\begin{table}[t]
    \caption{Physical parameters used in our analysis. 
    We adopt the absolute values of the CKM matrix elements provided in 
    Ref.~\cite{Charles:2004jd}. The other parameters are taken from Ref.\,\cite{Workman:2022ynf}. We use the weak mixing angle defined in the on-shell scheme in Eq.\,\eqref{eq: on-shell weak mixing angle}.
    }
    \centering
    \begin{tabular}{ccc}
    \hline
    \bf{Quantity} & \bf{Symbol} & \bf{Value}\\
    \hline
    QED fine-structure constant    & $\alpha=e^2/(4\pi)$ & 1/137.035 999 084(21)\\
    Fermi constant & $G_F$ & $1.1663788\times10^{-5}\, \mathrm{GeV}^{-2}$ \\
    $\abs{V_{ud}}$ & & 0.974353(53)\\
$\abs{V_{us}}$ & & 0.22500(23)\\
$W^\pm$-boson mass   &  $m_W$ &80.379(12)\,GeV  \\
$Z$-boson mass & $m_Z$ &91.1876(21)\,GeV \\
$e$ mass & $m_e$ & 
0.51099895000(15)\,MeV\\
$\mu$ mass & $m_\mu$ & 105.6583755(23)\,MeV\\
$\tau$ mass & $m_\tau$ & 1776.86(12)\,MeV\\
$\pi^\pm$ mass & $m_{\pi^\pm}$ & 139.57039(18)\,MeV\\
$\pi^0$ mass & $m_{\pi^0}$ & 134.9768(5)\,MeV\\
$K^\pm$ mass & $m_K$ & 493.677(13)\,MeV\\
Pion decay constant & $\sqrt{2}F_\pi\abs{V_{ud}}$ & 127.13(13)\,MeV \\
Kaon decay constant & $\sqrt{2}F_K\abs{V_{us}}$ & 35.09(5)\,MeV \\
Charged pion lifetime &$\tau_\pi$&
$ 2.6033(5)\times 10^{-8}\,\mathrm{sec}$\\
Charged Kaon lifetime &$\tau_K$&
$ 1.2380(20)\times 10^{-8}\,\mathrm{sec}$\\
$B(\pi^\pm \to \mu^\pm + \nu(+\gamma))$
& &$ 99.98770(4)$\%
\\
$B(K^\pm \to \mu^\pm + \nu(+\gamma))$
& &$ 63.56(11)$\% \\
\hline\\
    \end{tabular}
    \label{tab:input}
\end{table}

\subsection{Mass Spectrum}
In the MSSM, 
the Higgsinos $\tilde{H}_u$
and $\tilde{H}_d$ form 
the SU$(2)_L$ doublets, 
\begin{align}
    \tilde{H}_u = 
    \begin{pmatrix}
        \tilde{H}_u^+ \\
        \tilde{H}_u^0
    \end{pmatrix}\ ;
    \quad
    \tilde{H}_d = 
    \begin{pmatrix}
        \tilde{H}_d^0 \\
        \tilde{H}_d^-
    \end{pmatrix}\ .
\end{align}
with hypercharges 
$1/2$ and $-1/2$, respectively. 
Here those fermions are the left-handed Weyl spinor.%
\footnote{
In this paper, we use the symbol $\Psi$
to represent four-component spinors.}
The mass term for the Higgsinos
is given by
\begin{align}
    \mathcal{L}_\mathrm{Higgsino\,\, mass}
    = -\mu\epsilon^{ij}(\tilde{H}_u)_i(\tilde{H}_d)_j
    +\mathrm{h.c.}\ ,
\end{align}
where $\epsilon^{ij}$ is an antisymmetric
tensor with $\epsilon^{12} = -\epsilon^{21} = +1$.
The covariant derivative is given by
\begin{align}
    D_\mu = \partial_\mu - ig t^aW^a_\mu - ig'Q_YB_\mu\ ,
\end{align}
where  $g$ and $g'$ denote the 
gauge coupling constants of SU(2)$_L\times$U(1)$_Y$ gauge interactions,
 $t^a$ the halves of the Pauli matrices,
$W_\mu^a(a =1,2,3)$ the SU(2)$_L$ gauge bosons, $Q_Y$ the hypercharge, and $B_\mu$ the U(1)$_Y$ gauge boson.

In the MSSM, the gauge-eigenstates of the neutralino $\psi^0$ 
and the chargino $\psi^\pm$ are given 
by
\begin{align}
    \psi^0 &= \qty(\tilde{B}, \tilde{W}^3, \tilde{H}_d^0, \tilde{H}_u^0)^T; \\
    \psi^- &= \qty(\tilde{W}^-,\,\, \tilde{H}_d^-)^T;\,\,\,
    \psi^+ = \qty(\tilde{W}^+, \,\,\tilde{H}_u^+)^T\ ,
\end{align}
where $\tilde{W}^{3,\pm}$
and $\tilde{B}$
are the Winos and the Bino. 
Again, they are represented by 
two-component Weyl spinors.
After the electroweak symmetry breaking, 
the mass terms of the neutralinos and 
the charginos are given by,
\begin{align}
    \mathcal{L}_\text{neutralino mass} 
    &= -\frac{1}{2}\qty(\psi^0)^T M_{\tilde{N}}\psi^0 + \mathrm{h.c.}\ ; \\
    \mathcal{L}_\text{chargino mass} 
    &= -\qty(\psi^-)^T M_{\tilde{C}}\psi^+  + \mathrm{h.c.}\ .
\end{align}
Here $M_{\tilde{N}}$ is a $4\times 4$ complex symmetric matrix,
\begin{align}
    M_{\tilde{N}} = 
    \mqty*(M_1 & 0 & g'\langle H_d^0\rangle/\sqrt{2} & -g'\langle H_u^0\rangle/\sqrt{2} \\
    0 & M_2 & -g\langle H_d^0\rangle/\sqrt{2} & g\langle H_u^0\rangle/\sqrt{2} \\
    g'\langle H_d^0\rangle/\sqrt{2} & -g\langle H_d^0\rangle/\sqrt{2} & 0 & -\mu  \\
    -g'\langle H_u^0\rangle/\sqrt{2} & g\langle H_u^0\rangle/\sqrt{2} & -\mu & 0 )\ ,
\end{align}
and $M_{\tilde{C}}$ is a $2\times 2$ complex matrix,
\begin{align}
    M_{\tilde{C}} = \mqty(M_2 & -g\langle H_u^0\rangle \\
    -g\langle H_d^0\rangle & \mu)\ .
\end{align}
The complex parameters $M_{1,2}$ 
are the Bino and the Wino mass parameters and $\langle H_{u,d}^0\rangle $ are the vacuum expectation values of the neutral components of the up- and down-type Higgs bosons.

We can go to the diagonal basis of the neutralino by Takagi's decomposition, that is, 
there exists a unitary matrix $N_{ia}$ such that
\begin{align}
    (M_{\tilde{N}})_{ab} =
    (N^T)_{ai}\mqty(\dmat{m_{\tilde{N}_1}, m_{\tilde{N}_2}, m_{\tilde{N}_{3}}, m_{\tilde{N}_{4}}})_{ij} N_{jb} \ ,
\end{align}
where the neutralino in the mass basis is given by
\begin{align}
\label{eq: Weyl neutralino mixing}
    \tilde{N}_i = N_{ia}\psi^0_a \ .
\end{align}
Hereafter,  $a,b = 1, \cdots, 4$
denote gauge interaction basis,
and $i,j = 1, \cdots, 4$ denote the mass eigenbasis for the neutralinos.
Similarly, we can find the mass basis of the charginos by the singular decomposition by a pair of unitary matrices $(U, V)$ such that
\begin{align}
    (M_{\tilde{C}})_{ab} = (U^T)_{ai}\mqty(\dmat{m_{\tilde{C}_1},m_{\tilde{C}_2}})_{ij}V_{jb} \ .
\end{align}
The mass basis of the chargino can be defined by
\begin{align}
\label{eq: Weyl chargino mixing}
    \tilde{C}^+_i = V_{ia}\psi^+_a\ , \quad \tilde{C}^-_i = U_{ia}\psi^-_a\ ,
\end{align}
with $a,b = 1, 2$ and 
$i,j = 1, 2$
being the gauge interaction 
and the mass eigenbasis, respectively.
Here, $\tilde{N}_{3,4}$ and $\tilde{C}^\pm_2$ 
consist of  the Higgsino-like states.
When $M_{1,2}$ and $\mu$ are real-valued,
the mass eigenvalues of them
are given by
\begin{align}
    m_{\tilde{N}_3} &= 
    \mu - \frac{(1+s_{2\beta})m_Z^2}{2}
    \qty(\frac{c_W^2}{M_2 - \mu}+\frac{s_W^2}{M_1 - \mu}
    )
    +O({m_Z^4})\ ;\\
    m_{\tilde{N}_4} &= 
    -\mu - \frac{(1-s_{2\beta})m_Z^2}{2}
    \qty(\frac{c_W^2}{M_2 + \mu}+
    \frac{s_W^2}{M_1 + \mu}
    )
    + O({m_Z^4}) \ ;\\
    m_{\tilde{C}_2} &= \mu -
    \frac{c_W^2 m_Z^2(s_{2\beta}M_2+\mu)}{M_2^2 - \mu^2}+ O({m_Z^4})\ .
\end{align}
Here,
$s_W$ and 
$c_W$ stand for 
$\sin \theta_W$ and $\cos \theta_W$
with $\theta_W$ being the weak mixing angle, respectively.
The $Z$-boson mass is denoted by $m_Z$.
The ratio of the Higgsino mass is defined by
$\tan\beta = \langle H_u^0\rangle/\langle H_d^0\rangle > 0$ 
with $s_{2\beta} = \sin2\beta$.
Note that we take the mass eigenvalues $m_{\tilde{N}_i,\tilde{C}_i}$
are complex valued in general
and the mass ordering
is not determined.

In the following, we consider a heavy gaugino scenario, specifically $|M_{1,2}|\gg|\mu|$, where the Higgsino-like neutralinos $\chi^0_{1,2}$ are ordered based on the tree-level mass parameters:
\begin{align}
    \label{eq: neutralino mass 1}
    m_{\chi^0_1} &:= \min\left(|m_{\tilde{N}_3}|, |m_{\tilde{N}_4}|\right)\ ; \\
    \label{eq: neutralino mass 2}
     m_{\chi^0_2} &:= \max\left(|m_{\tilde{N}_3}|, |m_{\tilde{N}_4}|\right)\ ,
\end{align}
and the Higgsino-like chargino $\chi^-$ is composed of $\tilde{C}_2^\pm$ with a tree-level mass defined by
\begin{align}
    \label{eq: chargino mass}
    m_{\chi^\pm} := \abs{m_{\tilde{C}_2}}\ .
\end{align}

When $M_{1,2}$ and $\mu$ are real-valued, 
the mass difference between $\chi_{1,2}^{0}$ is given by
\begin{align}
     \mathit{\Delta}m_0^\mathrm{tree}:=
    m_{\chi^0_2}-m_{\chi^0_1}
    \simeq m_Z^2\qty|\frac{c_W^2}{M_2}
    +\frac{s_W^2}{M_1}|\ ,
\end{align}
for $|M_{1,2}|\gg |\mu|$.
The mass differences 
between the lightest $\chi_{1,2}^0$ 
and $\chi^\pm$
are given by
\begin{align}
    \mathit{\Delta} m_{\pm,1}^\text{tree}
    &= m_{\chi^\pm}-m_{\chi_1^0}  \simeq \frac{\mathit{\Delta} m_0^\text{tree}}{2}    -
    \frac{1}{2}s_{2\beta}\,m_Z^2\left(\frac{c_W^2}{M_2} - \frac{s_W^2}{M_1}\right)\ ,  \\
    \mathit{\Delta} m_{\pm,2}^\text{tree}
    &= m_{\chi^\pm}-m_{\chi_2^0} \simeq -\frac{\mathit{\Delta} m_0^\text{tree}}{2}    -
    \frac{1}{2}s_{2\beta}\,m_Z^2\left(\frac{c_W^2}{M_2} - \frac{s_W^2}{M_1}\right)\ .
\end{align}
Thus, for $|M_{1,2}|$ in the TeV range and $\tan\beta\gg 1$, 
the neutralinos and the chargino 
are quasi-degenerate with mass differences of $O({1})$\,GeV or smaller.

The radiative corrections 
also induce the 
chargino-neutralino mass
splitting.
As we assume the heavy gaugino/sfermions,
the mass splitting from the 
radiative corrections 
is dominated by the 
electroweak contributions,
which result in
\begin{align}
    \mathit{\Delta} m_\pm^\text{rad} &\simeq \frac{\alpha_{\overline{\mathrm{MS}}}(m_Z)\abs{\mu}}{2\pi}  \int^1_0  {\rm d}t(1+t)\log
\left[1+\frac{m_Z^2(1-t)}{\abs{\mu}^2 t^2}\right]\ , 
\end{align}
where $\alpha_{\overline{\mathrm{MS}}}(m_Z)^{-1}=127.93$ is the $\overline{\mathrm{MS}}$ fine structure constant in the SM evaluated at the $Z$-boson mass.
In the limit of $|\mu| \to \infty$, 
\begin{align}
    \mathit{\Delta} m^\text{rad}_\pm \simeq  353\,\text{MeV} \ .
\end{align}
Here, we have included the two-loop corrections to the mass splitting,
which is about $-3$\,MeV in Ref.~\cite{Yamada:2009ve}.
Note that when considering a finite Higgsino mass, the radiative mass splitting shows a mild mass dependence.
 This effect can influence the mass splitting by $O(10)$\% for a Higgsino whose mass is on the order of the weak scale~\cite{Nagata:2014wma}.
 
As a result, the mass differences between the neutralinos and the chargino are given by,
\begin{align}
     \mathit{\Delta} m_{\pm,i} &= 
      \mathit{\Delta} m_{\pm,i}^\text{tree} + 
       \mathit{\Delta} m_{\pm}^\text{rad}\ ,
\end{align}
for $i= 1,2$, respectively.
As we are interested in the Higgsino-like dark matter scenario, 
we assume the spectrum
\begin{align}
    \label{eq: mass spectrum}
    m_{\chi^0_1} < m_{\chi^\pm}\ ,
\end{align}
where the ordering between $\chi^\pm$ and $\chi_2^0$ is not specified.

\subsection{Gauge Interaction}
\label{sec: Gauge Interaction}
The charged Higgsino decay mediated 
by the Higgs and the sfermions 
are highly suppressed,
and hence, the decay process
is dominated by the $W$-boson
exchange.
In the two-component fashion, 
the gauge interactions with the $W$-boson, $W_\mu^+= (W_\mu^1 - i W_\mu^2)/\sqrt{2}$, 
are written by 
\begin{align}
    \mathcal{L}_\mathrm{MSSM}\supset
    gW^+_\mu\qty[\tilde{W}^{3\dagger}\bar{\sigma}^\mu \tilde{W}^-
    -\tilde{W}^{+\dagger}\bar{\sigma}^\mu\tilde{W}^3
    +\frac{1}{\sqrt{2}}\qty(\tilde{H}_d^{0\dagger}\bar{\sigma}^\mu\tilde{H}_d^-+\tilde{H}_u^{+\dagger}\bar{\sigma}^\mu\tilde{H}_u^0)]
    +\mathrm{h.c.}
\end{align}
We define four-component spinors for the Wino and Higgsino as
\begin{align}
\begin{split}
    \Psi_{\tilde{W}}^- &= \mqty(\tilde{W}^- \\ \tilde{W}^{+\dagger}); \,\,\,
    \Psi_{\tilde{H}}^- = \mqty(\tilde{H}^-_d \\ -\tilde{H}_u^{+\dagger}); \\
    \Psi_{\tilde{W}}^0 &= \mqty(\tilde{W}^3 \\ \tilde{W}^{3\dagger}); \,\,\,
    \Psi_{\tilde{H}_d}^0 = \mqty(\tilde{H}_d^0 \\ \tilde{H}_d^{0\dagger}); \,\,\,
    \Psi_{\tilde{H}_u}^0 = \mqty(\tilde{H}_u^0 \\ \tilde{H}_u^{0\dagger}).
\end{split}
\end{align}
Note that $\Psi^0_{\tilde{W}}, \Psi^0_{\tilde{H}_d}$ and $\Psi^0_{\tilde{H}_u}$ are Majorana fields.
In terms of these four-component fields,
the gauge interactions are rewritten as
\begin{align}
    \mathcal{L}_\mathrm{MSSM}\supset
    gW^-_\mu\qty[\bar{\Psi}^-_{\tilde{W}}\gamma^\mu\Psi^0_{\tilde{W}}
    +\frac{1}{\sqrt{2}}\qty(\bar{\Psi}^-_{\tilde{H}}\gamma^\mu P_L\Psi^0_{\tilde{H}_d}+\bar{\Psi}^-_{\tilde{H}}\gamma^\mu P_R\Psi^0_{\tilde{H}_u})]
    +\mathrm{h.c.}
\end{align}
To express the mixings \eqref{eq: Weyl neutralino mixing} and \eqref{eq: Weyl chargino mixing} in the four-component spinors, 
we introduce
\begin{align}
    \qty(\Psi_{\tilde{C}})_i = \mqty(\tilde{C}_i^- \\ \tilde{C}_i^{+\dagger})\ ; \,\,\, 
    \qty(\Psi_{\tilde{N}})_j = \mqty(\tilde{N}_j \\ \tilde{N}^{\dagger}_j)\ , 
\end{align}
whose mass terms are given by
\begin{align}
    \mathcal{L}_\mathrm{MSSM}
     \supset
     -m_{\tilde{C}_i}\qty(\bar{\Psi}_{\tilde{C}})_i\qty(\Psi_{\tilde{C}})_i
     -\frac{1}{2}m_{\tilde{N}_j}\qty(\bar{\Psi}_{\tilde{N}})_j\qty(\Psi_{\tilde{N}})_j\ ,
\end{align}
with $i=1,2$  and $j=1,\cdots 4$.
We can readily see that
\begin{align}
\begin{split}
    \Psi^-_{\tilde{W}} &= U^*_{i1}P_L\qty(\Psi_{\tilde{C}})_i + V_{i1}P_R\qty(\Psi_{\tilde{C}})_i\ ; \\
    \Psi^-_{\tilde{H}} &= U^*_{i2}P_L\qty(\Psi_{\tilde{C}})_i - V_{i2}P_R\qty(\Psi_{\tilde{C}})_i\ ; \\
    \Psi^0_{\tilde{W}} &= N^*_{j2}P_L\qty(\Psi_{\tilde{N}})_j + N_{j2}P_R\qty(\Psi_{\tilde{N}})_j\ ; \\
    \Psi^0_{\tilde{H}_d} &= N^*_{j3}P_L\qty(\Psi_{\tilde{N}})_j + N_{j3}P_R\qty(\Psi_{\tilde{N}})_j\ ; \\
    \Psi^0_{\tilde{H}_u} &= N^*_{j4}P_L\qty(\Psi_{\tilde{N}})_j + N_{j4}P_R\qty(\Psi_{\tilde{N}})_j\ .
\end{split}
\end{align}
Therefore, we obtain
\begin{align}
    \label{eq: neutralino-chargino-W}
    \mathcal{L}_\mathrm{MSSM}\supset
    gW^+_\mu\qty(\bar{\Psi}^0_{\tilde{N}})_j\gamma^\mu\qty[(O^W_L)_{ji}P_L+(O^W_R)_{ji}P_R](\Psi^-_{\tilde{C}})_i\ ,
\end{align}
where
\begin{align}
    (O^W_L)_{ji} = N_{j2}U^*_{i1} + \frac{1}{\sqrt{2}}N_{j3}U^*_{i2}\ ; \quad
    (O^W_R)_{ji} = N^*_{j2}V_{i1} - \frac{1}{\sqrt{2}}N^*_{j4}V_{i2}\ .
\end{align}

In the limit of $\abs{M_{1,2}} \gg \abs{\mu} \gg m_Z$,
these mixing matrices are approximated as
\begin{align}
(O^W_L)_{32} &\simeq
    \frac{1}{2}
    -\frac{(c_\beta^2-s_\beta^2)}{8}
    \frac{m_Z^2(M_1c_W^2+M_2s_W^2)}{M_1M_2\mu}\ ;
    \,\,\,
    (O^W_R)_{32} \simeq
    \frac{1}{2}
    +\frac{(c_\beta^2-s_\beta^2)}{8}
    \frac{m_Z^2(M_1c_W^2+M_2s_W^2)}{M_1M_2\mu}\ ; \\
    (O^W_L)_{42} &\simeq
    \frac{1}{2}
    +\frac{(c_\beta^2-s_\beta^2)}{8}
    \frac{m_Z^2(M_1c_W^2+M_2s_W^2)}{M_1M_2\mu}\ ;
    \,\,\,
    (O^W_R)_{42} \simeq
    -\frac{1}{2}
    +\frac{(c_\beta^2-s_\beta^2)}{8}
    \frac{m_Z^2(M_1c_W^2+M_2s_W^2)}{M_1M_2\mu}\ ,
\end{align}
where we have assumed that $M_{1,2}$ and $\mu$ are real.
Moreover, we find
\begin{align}
\label{eq:OLW}
    (O^W_L)_{32}
    &\simeq \frac{1}{2}
    \qty(1+\frac{\mathit{\Delta}m_{0}^\mathrm{tree}}{4\mu})\ ; \quad
    (O^W_R)_{32} \simeq \frac{1}{2}\qty(1-\frac{\mathit{\Delta}m_{0}^\mathrm{tree}}{4\mu})\ , \\
\label{eq:ORW}
    (O^W_L)_{42}
    &\simeq \frac{1}{2}
    \qty(1-\frac{\mathit{\Delta}m_{0}^\mathrm{tree}}{4\mu})\ ; \quad
    (O^W_R)_{42} \simeq -\frac{1}{2}\qty(1+\frac{\mathit{\Delta}m_{0}^\mathrm{tree}}{4\mu})\ ,
\end{align}
up to $O(\mathit{\Delta}m_0^\mathrm{tree}/M_{1,2})$ contribution.
These couplings are related to $\mathit{\Delta}m_{\pm,1}^{\mathrm{tree}}$ via 
$\mathit{\Delta}m_{\pm,1}^{\mathrm{tree}} =
\mathit{\Delta}m_{0}^{\mathrm{tree}}/2$ for $\tan\beta \gg 1$.
In this paper, we consider the case of the quasi-degenerate Higgsino, 
where $\abs{\mu} \gtrsim 100$\,GeV
and $\mathit{\Delta}m_{\pm, 1}^\mathrm{tree} \lesssim O(1)$\,GeV.
In such a scenario, 
the $W$-boson interaction of the Higgsino-like states 
are close to those 
of the pure-Higgsino case.
A similar observation can be made for the $Z$-boson interactions.
Therefore, the gauge interactions of the Higgsino-like states can be well approximated as those of the pure Higgsinos.

Before closing this section, let us comment on the mass mixing effect
to the decay rates.
As we will see later, the decay rates depend on  
those mixing matrices through the following combinations,
\begin{align}
    \abs{O_L^W}^2 + \abs{O_R^W}^2\ , \quad
   O_L^W O_R^{W*} + O_L^{W*} O_R^W\ .
\end{align}
In both combinations, the effect of $O(\mathit{\Delta}m_0^\mathrm{tree}/\mu)$ terms cancels, as we can see
Eqs.~\eqref{eq:OLW} and \eqref{eq:ORW}. The remaining effect comes from
terms of $O(\mathit{\Delta}m_0^\mathrm{tree}/M_{1,2})$. We have checked
that this contribution to the decay rates 
is less than $O(0.1)\%$ when the gaugino masses
$M_{1,2}$ is greater than around $3$\,TeV.

\section{Four-Fermion Interaction and Radiative Corrections}
\label{sec: EW corrections to FF operator}

The decay process of the chargino
is described by the effective four-fermion 
operators.
To estimate its decay rate precisely, 
we should compute radiative corrections to
the decay processes. Those corrections are composed of the 
electroweak, QED and QCD corrections.
The electroweak interactions give rise to
the short-distance corrections, whose computational
framework we will show in this section.
The perturbative QCD effects of ${O}(\alpha^0\alpha_s)$ are absent from the short-distance corrections.
The long-distance QED and QCD effects
are discussed in Sec.~\ref{sec: radiative corrections to leptonic mode} and Sec.~\ref{sec: decay into single pion}.

In our analysis, 
we adopt the on-shell scheme of the electroweak theory~\cite{Bohm:1986rj}.
The weak mixing angle $\theta_W$
is defined by 
\begin{align}
    \label{eq: on-shell weak mixing angle}
   c_W=\cos\theta_W = \frac{m_W}{m_Z}\ ; \quad
    s_W=\sin\theta_W = \sqrt{1-\frac{m_W^2}{m_Z^2}}
\end{align}
to all order of perturbation, 
where $m_W$ and $m_Z$ are the physical masses of the $W$- and $Z$-bosons, 
respectively.

\subsection{Tree-Level Four-Fermion Interaction}
The gauge interactions of the chargino and neutralino exhibit 
a complex structure attributed to mass mixing, 
as detailed in Sec.~\ref{sec: model discripition}. 
Nevertheless, in the pure-Wino/Higgsino scenario, the chargino decay can be examined 
within a simplified model at tree-level as outlined below.

We would like to discuss the weak decay of the 
chargino $\chi^-$ into a neutralino $\chi^0$ 
with emission of the SM particles, $\chi^-\to\chi^0+\mathrm{SM}$.
We define the Dirac fermion $\Psi_{\chi^-}$ for the chargino 
and the Majorana fermion $\Psi_{\chi^0}$ for the neutralino. Their mass terms 
are written in the form of
\begin{align}
    \label{eq: mass terms}
    \mathcal{L}_\mathrm{mass}
     = -m_{\chi^\pm}\overline{\Psi}_{\chi^-}\Psi_{\chi^-}
     -\frac{1}{2}m_{\chi^0}\overline{\Psi}_{\chi^0}\Psi_{\chi^0}\ .
\end{align}
We have rearranged
the complex phases of the chargino/neutralino
fields so that the mass parameters $m_{\chi^\pm}$ and $m_{\chi^0}$ are real and positive.
In the Higgsino DM scenario,
we have the two independent neutralinos, $\chi_1^0$ and $\chi^0_2$.
In this case,
the chargino mass is given by Eq.~\eqref{eq: chargino mass},
whilst the neutralino mass is either Eqs.~\eqref{eq: neutralino mass 1}
or \eqref{eq: neutralino mass 2}. 

The decay of the chargino proceeds through the exchange of the $W$-boson, as we consider the scenario where all sparticles, except for the chargino and the neutralinos, are decoupled.
In this case, the decay process would be characterized mainly by the mass difference, 
\begin{align}
    \label{eq: abstract mass difference}
    \mathit{\Delta}m_\pm = m_{\chi^\pm}-m_{\chi^0}\ .
\end{align}
For $\mathit{\Delta}m_\pm \lesssim O(1)$ GeV,  
the $W$-boson exchange can be well approximated by the current-current 
operator,
 \begin{align}
    \label{eq: current-current interaction with SM}
    \mathcal{L}_\mathrm{CC} 
        &= -2\sqrt{2}G_FJ^{-\mu}_\chi J^+_{\mathrm{SM}\mu}
        +\mathrm{h.c.}\ ,
\end{align}
where
$G_F$ denotes the Fermi constant determined by the muon decay~\cite{Workman:2022ynf},
$J^\mu_\chi$ is the weak current
of the chargino and the neutralino,
and $J_\mathrm{SM}^\mu$ is 
the weak current composed of 
the SM quarks and the SM leptons.
The small correction from the $W$-boson propagator
is suppressed by $\qty(\mathit{\Delta}m_\pm/m_{W})^2$
and can be safely neglected in the following discussion.
By recalling that the mass mixing effects appear in the weak interaction
as in Eq.~\eqref{eq: neutralino-chargino-W},
the weak current of the chargino and neutralino can be written as
\begin{align}
    \label{eq: Majorana Higgsino current}
    J^{-\mu}_\chi
     = 
    \sqrt{2}\,\overline{\Psi}_{\chi^0}
    \gamma^\mu\qty(O^W_LP_L+O^W_RP_R)\Psi_{\chi^-}\ ,
\end{align}
where $O_{L,R}^W$ are complex constants. 
We can rotate away the complex phases of $O_{L,R}^W$
which are relevant for the chargino decay.
We follow
the convention on the phases so that\,\footnote{Strictly speaking,
for the Higgsino-like neutralino $\tilde{N}_4$ in Sec.~\ref{sec: model discripition},
$O_L^W = O_R^W = i/2$. The difference from Eq.~\eqref{eq: mixing matrix phase}
does not affect the following discussion.}
\begin{align}
    \label{eq: mixing matrix phase}
    \begin{split}
        &O_L^W = +1\ ; \quad O_R^W = +1\quad\text{(pure-Wino case)}\ ; \\
        &O_L^W = +\frac{1}{2}\ ; \quad O_R^W = +\frac{1}{2}
        \quad\text{(pure-Higgsino case)}\ .
    \end{split}
\end{align}
The weak charged current of 
the $u,d,s$-quarks  and the leptons
$\ell, \nu_\ell$ is given by
\begin{align}
    J_\mathrm{SM}^{+\mu} 
    =
    V^*_{ud}\overline{\Psi}_d\gamma^\mu P_L \Psi_u
    +V^*_{us}\overline{\Psi}_s\gamma^\mu P_L \Psi_u
    +\overline{\Psi}_\ell\gamma^\mu P_L 
    \Psi_{\nu_\ell}\ ,
\end{align}
where $V_{uD}\,(D=d,s)$ is the CKM matrix element and the SM particles are 
denoted by four-component spinors.

\begin{figure}[t]
    \centering
    \subcaptionbox{\label{fig: leptonic}}{\includegraphics[width=0.23\textwidth]{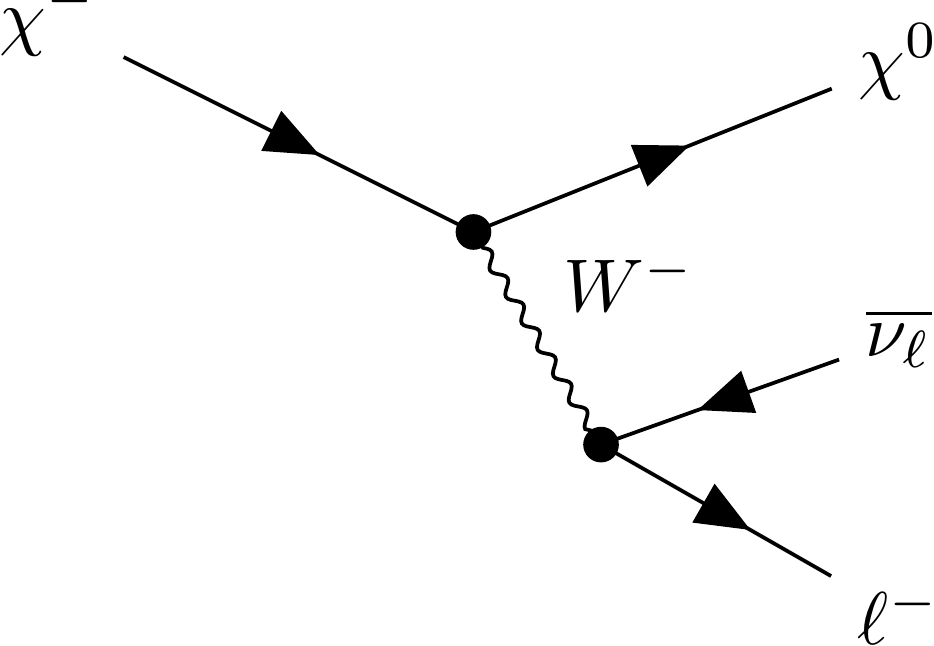}}
    \subcaptionbox{\label{fig: singlepi}}{\includegraphics[width=0.23\textwidth]{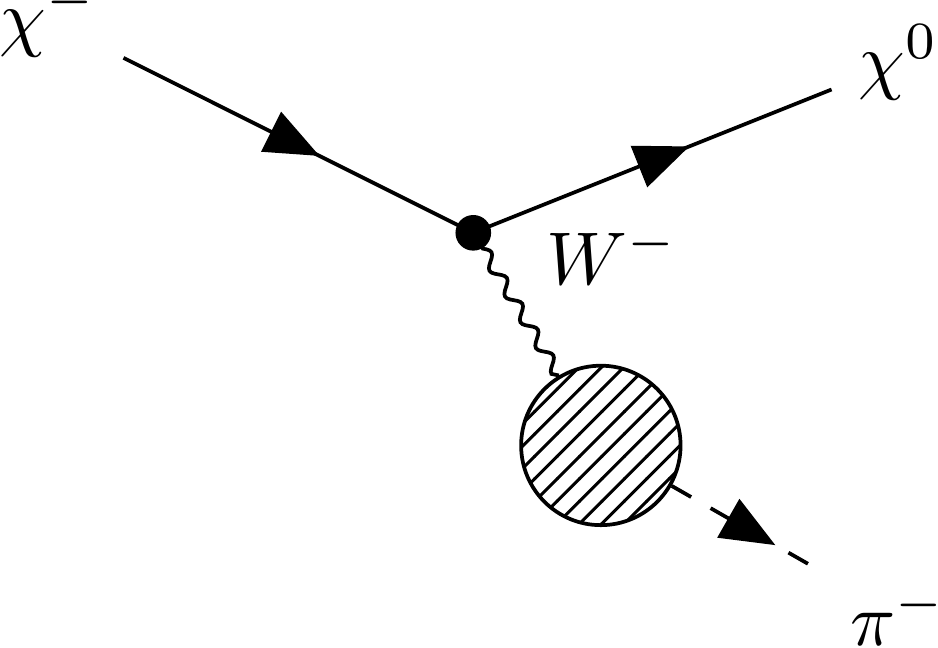}}
    \subcaptionbox{\label{fig: twopi}}{\includegraphics[width=0.23\textwidth]{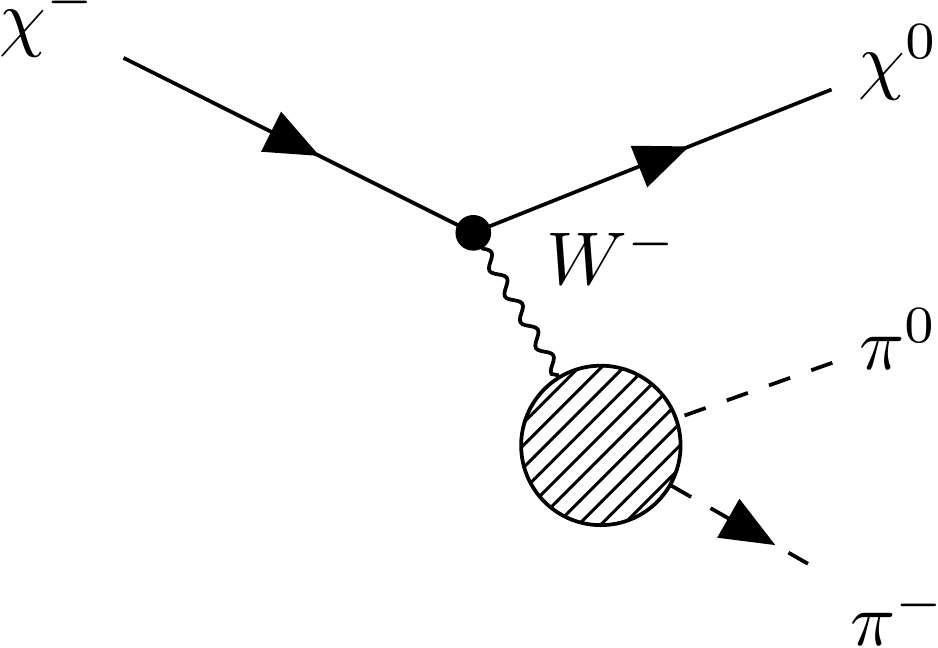}}
    \subcaptionbox{\label{fig: threepi}}{\includegraphics[width=0.23\textwidth]{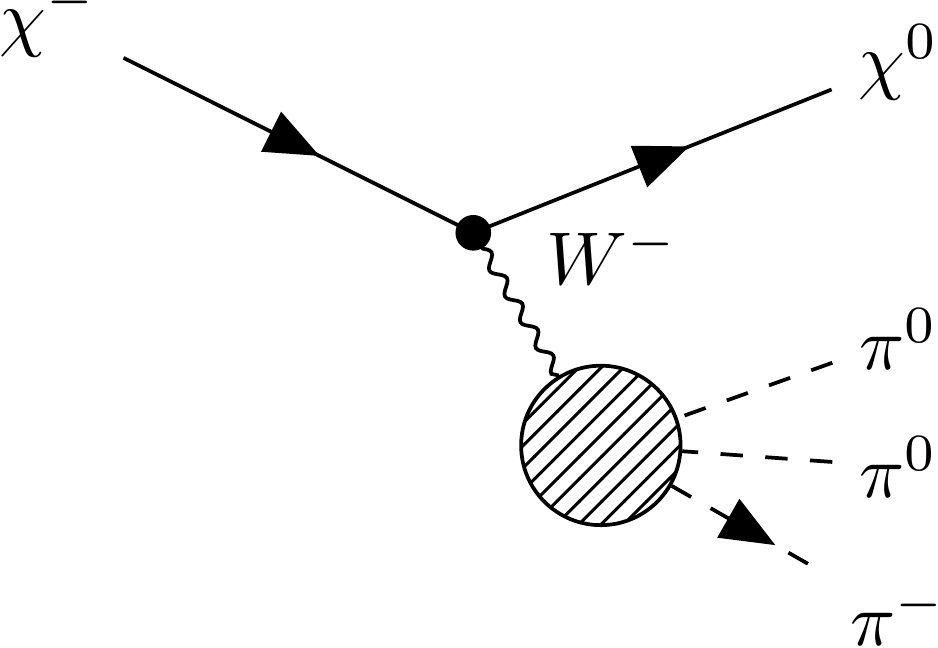}}
    \caption{Example diagrams of chargino decay}
    \label{fig: Higgsino decays}
\end{figure}

Since we consider cases that $\mathit{\Delta}m_\pm\lesssim O(1)$ GeV,
we have to treat the quarks in the final state
as hadrons (See Fig.~\ref{fig: singlepi}-\ref{fig: threepi}\,\footnote{In this paper, the Feynman diagrams are drawn by using \texttt{TikZ-FeynHand}\,
\cite{Ellis:2016jkw,Dohse:2018vqo}.}).
Then, the decay channels for the chargino-neutralino mass difference less than
$O(1)$\,GeV are
\begin{itemize}
    \item Leptonic modes ($\chi^-\to 
    \chi^0 +   e/\mu +\bar{\nu}$,
    Sec.~\ref{sec: radiative corrections to leptonic mode})
    \item Single pion mode ($\chi^- \to 
    \chi^0 + \pi^-$, Sec.~\ref{sec: decay into single pion})
    \item Single Kaon and multi-meson modes
    ($\chi^-\to \chi^0 
    + K^-$ and $\chi^-\to \chi^0 
    + (\ge 2)P, \, (P = \pi, K)$, Sec.~\ref{sec: multimeson})
\end{itemize}
For a detailed discussion of each decay mode, see the sections indicated in the parentheses.
In the following, we calculate the universal short-distance corrections, which is a common contribution for all the modes.

\subsection{Electroweak Correction
to Four-Fermion Interaction}
We calculate the short-distance correction to the four-fermion interaction~\eqref{eq: current-current interaction with SM}. 
We are interested in corrections of $\mathcal{O}\qty(\alpha)$
and neglect the contributions suppressed by 
$\qty(\mathit{\Delta}m_{\pm}/m_W)^2$.
In this approximation, we can adopt the pure-Higgsino/Wino theory 
to estimate the electroweak corrections.

In the following, we will provide a detailed description 
of the pure-Higgsino case.
The pure-Higgsino theory is defined by the following Lagrangian,
\begin{align}
\label{eq: pure-Higgsino Lagrangian}
    \mathcal{L}_\mathrm{tree}^\mathrm{pure} =\,
    &\overline{\Psi}_0 i \gamma^\mu \partial_{\mu} \Psi_0 - m_{\chi}\overline{\Psi}_0\Psi_0
    + \bar{\Psi}_- i \gamma^\mu \partial_{\mu} \Psi_- - m_{\chi}\overline{\Psi}_-\Psi_- \notag \\
    &+\frac{e}{c_W s_W}\qty(\frac{1}{2}\overline{\Psi}_0\gamma^\mu \Psi_0
    -\frac{1}{2}\overline{\Psi}_-\gamma^\mu \Psi_- +s_W^2\overline{\Psi}_-\gamma^\mu \Psi_-)Z_\mu
    -e \overline{\Psi}_-\gamma^\mu\Psi_- A_\mu \notag \\
    &+\frac{e}{\sqrt{2}s_W}\overline{\Psi}_- \gamma^\mu\Psi_0 W^-_\mu
    +\frac{e}{\sqrt{2}s_W}\overline{\Psi}_0 \gamma^\mu\Psi_- W^+_\mu\ . 
\end{align}
Here, $m_\chi$ is the SU$(2)$-symmetric mass parameter,
which is taken to be equal to the physical charged Higgsino
mass $m_{\chi^\pm}$ by tuning the $\mathrm{SU}(2)$
symmetric mass counterterm.
The charged and neutral pure-Higgsinos are represented by the 
Dirac fields, $\Psi_-$ and $\Psi_0$,
respectively.
These two fields 
form the SU(2)$_L$ doublet with 
the U(1)$_Y$ hypercharge $-1/2$.
We choose the phases of $\Psi_-$ and $\Psi_0$
so that $m_\chi$ is real and positive.
We can see the equivalence to the gauge interactions
written in the basis
described in subsection~\ref{sec: Gauge Interaction}
by the replacement
\begin{align}
    \label{eq: matching pure-Higgsino with MSSM}
    m_\chi = \mu\ ; \quad
    \Psi_- = i\qty(\Psi_{\tilde{C}})_2\ ;\quad
    \Psi_0 = \frac{i}{\sqrt{2}}\qty[\qty(\Psi_{\tilde{N}}^0)_3-\gamma_5\qty(\Psi_{\tilde{N}}^0)_4]\ .
\end{align}

The four-fermion operator relevant for the charged Higgsino decay
is given by\,\footnote{We have dropped the CKM matrix element, $V_{ud}$ or $V_{us}$,
just for notational convenience although we should multiply
the decay amplitude with either of 
these matrix elements
when $f$ and $f'$ are free quarks.}
\begin{align}
    \label{eq: pure four-fermion}
   \mathcal{L}_\mathrm{FF}^{\mathrm{pure},0}
   = -2\sqrt{2}\GFtree
   \qty(\overline{\Psi}_0\gamma^\mu \Psi_-)
        \qty(\overline{\Psi}_{f}\gamma_\mu P_L\Psi_{f'})
        +\mathrm{h.c.}\ ,
\end{align}
where $f, f'$ represent the SU$(2)_L$ SM doublet fermions
and $\GFtree$ is defined by
\begin{align}
    \label{eq: GF0}
    \GFtree
    &= \frac{e^2}{4\sqrt{2}s_W^2m_W^2}\ .
\end{align}
 In the rest of this section, physical quantities with the superscript $0$ should be understood as those  
determined by $\GFtree$, not by 
the Fermi constant $G_F$.
The $W$-boson mass $m_W$ is related to 
the Fermi constant as
\begin{align}
    \label{eq: GFmuon}
    \GFtree  = G_F\times (1-\mathit{\Delta}r)\ ,
\end{align}
where $\mathit{\Delta}r = 0.0354055$.
At one-loop level, $\mathit{\Delta}r$
is given by~\cite{Sirlin:1980nh}\,\footnote{This relation holds even if the $W$-boson self-energy includes
almost pure-Higgsino/Wino loop.}
\begin{align}
\label{eq:SirlinFacToFermi}
    \mathit{\Delta}r|_{\text{1-loop}} 
    &= \frac{\alpha}{4\pi}
    \qty(\displaystyle{\frac{\hat{\Sigma}_{WW}^\mathrm{1PI}(0)}{m_W^2}}
   +\frac{6}{s_W^2}+\frac{7-4 s_W^2}{s_W^4} \log c_W)\ .
\end{align}
Note that $\mathit{\Delta}r|_{\text{1-loop}}$
dominates $\mathit{\Delta}r$ by including
higher-order QCD effects in the self-energy of $W$-boson 
$\qty(\alpha/(4\pi))\times\hat{\Sigma}_{WW}^\mathrm{1PI}(0)$~\cite{Freitas:2001zs}.

By following Ref.~\cite{Braaten:1990ef},
we compute the electroweak radiative correction to 
the tree-level amplitude,
\begin{align}
    \label{eq: tree-level amplitude of pure Higgsino}
    \mathcal{M}_\mathrm{tree}^0
     = -2\sqrt{2}\GFtree\bar{u}_0(p_2)\gamma^\mu u_-(p_1)
    \bar{u}_f(p_3)\gamma_\mu P_Lv_{f'}(p_4)\ ,
\end{align}
where the momentum assignment is determined as 
$\chi^-(p_1)\to\chi^0(p_2)f(p_3)\overline{f'}(p_4)$.
To derive
the matching conditions to the low-energy effective
theories, we keep the QED charge difference
$\overline{Q}=Q_f-\qty(-Q_{f'})$ arbitrary~\cite{Descotes-Genon:2005wrq}.
The SM fermion masses in the final state will be neglected in the short-distance corrections since they are not relevant in matching to low-energy effective theories. 
In addition, the finite momenta of $p_{3,4}$
correspond to contributions of operators
with dimension higher than six and suppressed by
additional power of $m_W^2$.
In the following, therefore, we approximate the momenta as $p_1\simeq p_2$ and $p_{3,4}\simeq 0$ in the computation
of short-distance corrections to the four-fermion operator~\eqref{eq: current-current interaction with SM}.

The radiative corrections in electroweak theory are composed of wave-function renormalizations,
box contributions, vertex corrections, and
the self-energy of the virtual $W$-boson.
We describe computational detail in Appendix~\ref{app: explicit electroweak}.
In our analysis, we calculate the radiative corrections with the help of   
$\mathtt{FeynArts}$~\cite{Hahn:2000kx}, 
$\mathtt{FeynCalc}$~\cite{Mertig:1990an,*Shtabovenko:2016sxi,*Shtabovenko:2020gxv} and 
$\mathtt{Package\mbox{--}X}$~\cite{Patel:2016fam}
through
$\mathtt{FeynHelper}$~\cite{Shtabovenko:2016whf}.
Putting all the contributions together, 
we can write down the electroweak corrections in the form of
\begin{align}
    \mathcal{M}_\mathrm{Virtual(EW)}^0 
    &= \mathcal{M}_\mathrm{tree}^0
    \times \frac{\alpha}{4\pi}\qty[3\log\frac{m_W}{m_\gamma}
    + \frac{\hat{\Sigma}_{WW}^\mathrm{1PI}(0)}{m_W^2}
    + F^\mathrm{Virtual(EW)}_V(r_W)]\label{eq: vector weak corrections}
    \notag \\
    &\hspace{20pt}
    +\mathcal{M}_A^0\times \frac{\alpha}{4\pi}F^\mathrm{Virtual(EW)}_A(r_W)\ ,
\end{align}
where $m_\gamma$ is the photon mass to regulate IR divergences, $r_W = m_W/m_{\chi^\pm}$,
$\mathcal{M}_A^0$ is the axial current 
contribution given by
\begin{align}
    \label{eq: axial amplitude}
    \mathcal{M}_A^0
     = -2\sqrt{2}\GFtree\bar{u}_0(p_2)\gamma^\mu \gamma_5 u_-(p_1)
    \bar{u}_f(p_3)\gamma_\mu P_Lv_{f'}(p_4)\ ,
\end{align}
and
$F_{V,A}^\mathrm{Virtual(EW)}(r_W)$ represent
non-logarithmic electroweak corrections.

For later use,
we provide the analytical expression of $F_V^{\mathrm{Virtual(EW)}}(r_W)$ here:
\begin{align}
    \label{eq: vector-like form factor}
   &F_V^{\mathrm{Virtual(EW)}}(r_W) 
    = 
   \frac{3(2-r_W^2)}{2s_W^2}
   +\frac{2-s_W^2}{2s_W^2c_W^2}\times r_W^4\log r_W\cr
   \cr
   &\hspace{20pt}
   +\frac{2(4-5s_W^2-2s_W^4+3s_W^6)+(c_W^2-s_W^2)r_W^4}{2s_W^4c_W^2}\log c_W\cr
   &\hspace{20pt}
   +\frac{8-4s_W^2+(1+s_W^2)r_W^2(2-r_W^2)}{2s_W^4} 
   \times
   \frac{r_W}{\sqrt{4-r_W^2}}\cot^{-1}\qty(\frac{r_W}{\sqrt{4-r_W^2}})\cr
   &\hspace{20pt}
   +\frac{-4(2-s_W^2)c_W^4-2(1-3s_W^2+2s_W^4)r_W^2+(c_W^2-s_W^2)r_W^4}{2s_W^4c_W^2}\cr
   &\hspace{40pt}
   \times\frac{r_W}{\sqrt{4c_W^2-r_W^2}}
   \cot^{-1}\qty(\frac{r_W}{\sqrt{4c_W^2-r_W^2}})\ .
\end{align}
Note that, in the degenerate limit $\mathit{\Delta}m_\pm\to0$, the finite part of the one-loop electroweak corrections to the four-fermion interaction~\eqref{eq: pure four-fermion}
is independent of the $\mathrm{SU}(2)_L\times\mathrm{U}(1)_Y$ representation
of the chargino/neutralino multiplet (see Appendices~\ref{sec: wave-function and vertices} and \ref{sec: box contributions}).

Consequently, the short-distance correction factor described in Eq.\,\eqref{eq: vector-like form factor} 
is universal to weak decays of colorless fermions
with $|\mathit{\Delta}Q_\mathrm{QED}| = 1$ 
and with small mass splitting.
Therefore, our current results can be utilized for the Wino and quintuplet minimal dark matter~\cite{Cirelli:2005uq,*Cirelli:2007xd,*Cirelli:2009uv} as well as for more generic electroweak-interacting dark matter \cite{Nagata:2014aoa}, particularly in scenarios involving the mass degenerate limit.

The self-energy of the $W$-boson is removed
by using the relation~\eqref{eq: GFmuon}.
In other words, 
the perturbative expansion 
can be rearranged as
\begin{align}
    \label{eq: equivalence of GF and GF0}
    \mathcal{M}_\mathrm{1-loop} 
    &= \mathcal{M}_\mathrm{tree}^0
    +\mathcal{M}_\mathrm{Virtual(EW)}^0 \notag \\
    &= \mathcal{M}_\mathrm{tree}
    +\qty(\mathcal{M}_\mathrm{tree}^0-\mathcal{M}_\mathrm{tree}+\mathcal{M}_\mathrm{Virtual(EW)}^0)\ ,
\end{align}
where $\mathcal{M}_{\mathrm{tree}}$ 
denotes the tree-level amplitude obtained by
replacing $\GFtree$ with $G_F$
in Eq.~\eqref{eq: tree-level amplitude of pure Higgsino}. In this case, the electroweak correction is given by
\begin{align}
    \mathcal{M}_\mathrm{Virtual(EW)}
    :=\,\, &\mathcal{M}_\mathrm{tree}^0-
    \mathcal{M}_\mathrm{tree}+\mathcal{M}_\mathrm{Virtual(EW)}^0 \cr
    \label{eq: EW virtual corrections}
    =\,\, &\mathcal{M}_\mathrm{tree}
    \times \frac{\alpha}{4\pi}\qty[
    3\log\frac{m_W}{m_\gamma}
    +F^\mathrm{Virtual(EW)}_V(r_W)
    -\qty(\frac{6}{s_W^2}+\frac{7-4 s_W^2}{s_W^4}\log c_W)]
    \cr
    &+\mathcal{M}_A\times \frac{\alpha}{4\pi}F^\mathrm{Virtual(EW)}_A(r_W)\ ,
\end{align}
where $\mathcal{M}_A$ also 
represents the amplitude~\eqref{eq: axial amplitude} with $\GFtree$ replaced by $G_F$.

Finally, we would like to stress that
these contributions become constant in the heavy chargino limit $m_{\chi^\pm}\to\infty$
in the radiative correction~\eqref{eq: EW virtual corrections}. This is physically sensible behaviour
since the recoil energy received by the heavy neutralino in the final state
is very small and the total decay rate should be insensitive to the chargino mass
in the limit of $m_{\chi^\pm}\to\infty$. 
The situation is similar to the limiting behaviour argued by Appelquist-Carrazone's
decoupling theorem~\cite{Appelquist:1974tg},
but different from the theorem in that in our case the heavy particles are
external ones. In our analysis, we assume the decoupling of external particles to all order,
which will be used as a sanity check to our computation.

\subsection{Ambiguities in Definition of ``Tree-Level''}
\label{sec: ambiguity}
\begin{figure}[tb]
\centering
  \includegraphics[width=0.7\linewidth]{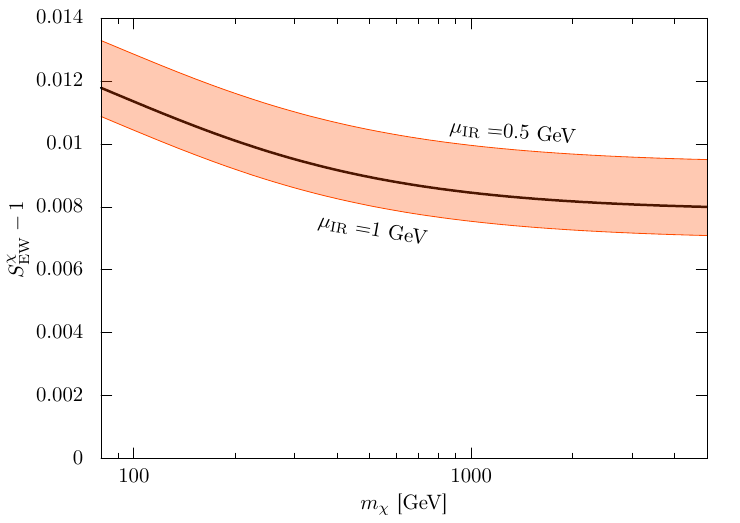}
\caption{The NLO electroweak and short-distance QED correction factor $S_\mathrm{EW}^{\chi}$.
This factor is universal to weak decays of colorless fermions
with $|\mathit{\Delta}Q_\mathrm{QED}| = 1$ 
and with small mass splitting.
Therefore,
$S_\mathrm{EW}^\chi$ is common to both the Wino and the Higgsino.
The orange band
shows the dependence of $S_\mathrm{EW}^\chi$ on 
the infrared cut-off $\mu_\mathrm{IR}$.
The solid line represents the case of $\mu_\mathrm{IR}=m_\rho$. 
}\label{fig:SEW}
\end{figure}
As stressed in Introduction, 
there is an ambiguity in definition
of the ``tree-level'' decay rate.
Namely, we can adopt the four-fermion operator~\eqref{eq: current-current interaction with SM} using the Fermi constant
$G_F$ instead of the operator~\eqref{eq: pure four-fermion} using $G_F^0$ defined by Eq.~\eqref{eq: GF0}.
The difference in the choice
of the tree-level coupling
introduces about 7\% error
in the decay rate:
\begin{align}
    \label{eq: GF uncertainty}
    \frac{G_F^2}{\qty(G_F^0)^2}-1
    \simeq 0.07\ .
\end{align}
We cannot determine which coupling to be used
and hence the tree-level estimation has the uncertainty provided by Eq.~\eqref{eq: GF uncertainty}
as long as we stick to the leading-order computation.

As we stated below Eq.~\eqref{eq:SirlinFacToFermi},
the difference 
between $G_F^0$ and $G_F$ is dominated
by $\left.\mathit{\Delta}r\right|_\mathrm{1-loop}$
with higher-order QCD effects included in the $W$-boson
self energy $\qty(\alpha/(4\pi))\times\hat{\Sigma}_{WW}^\mathrm{1PI}(0)$.
Hence, by using the NLO amplitude, Eq.~\eqref{eq: vector weak corrections} or Eq.~\eqref{eq: EW virtual corrections}, the result is almost free from the ambiguity~\eqref{eq: GF uncertainty}.

We should also note that the constant $G_F^0$ can be defined 
in terms of the running 
weak gauge coupling constant renormalized at high energy scale.
For example,  
the weak gauge coupling at the scale
larger than 1\,TeV
results in the deviation
of the ``tree-level" decay 
rate by more than $5$\%.
Our NLO analysis resolves 
this uncertainty too
since we obtained 
the renormalization scale 
independent virtual correction
at the one-loop level
(see Eq.\,\eqref{eq: EW virtual corrections}).

A caveat is that
in Eqs.~\eqref{eq: vector weak corrections} and \eqref{eq: EW virtual corrections}, the IR cutoff $m_\gamma$ still remains
since so far we only computed the virtual corrections.
These dependencies on the photon mass vanish if we include real emission processes.
In this paper, we present the complete calculation
of long-distance corrections 
for the leptonic mode and single pion mode 
in Sec.~\ref{sec: radiative corrections to leptonic mode}
and Sec.~\ref{sec: decay into single pion}, respectively.

For later convenience, 
we introduce the multiplicative
factor $S_\mathrm{EW}^\chi$ defined by,
\begin{align}
    \label{eq: SEW}
   S^\chi_\mathrm{EW}(\mu_\mathrm{IR}) = \,\,
   &1+\frac{\alpha}{4\pi}\qty[6\log\frac{m_W}{\mu_\mathrm{IR}}
   +2F^\mathrm{Virtual(EW)}_V\qty(\frac{m_W}{m_{\chi^\pm}})
   -2\qty(\frac{6}{s_W^2}+\frac{7-4 s_W^2}{s_W^4} \log c_W)
   ]\ .
\end{align}
This factor
encapsulates the electroweak and short-distance QED corrections above $\mu_\mathrm{IR}$
to the decay rates computed in terms of 
the Fermi constant $G_F$ rather $G_F^0$ up to contributions of $O(\alpha^2, \alpha\alpha_s)$.
Here,
the logarithmic factor in $S_\mathrm{EW}^\chi$
arises from the renormalization group (RG) effect\,\footnote{The analysis of the RG effect on the Fermi interaction is discussed
in Appendix~\ref{sec: RG analysis}.}
and $\mu_\mathrm{IR}$ is some infrared cutoff.
In Fig.~\ref{fig:SEW}, we show the dependence of $S_\mathrm{EW}^{\chi}-1$
on the chargino mass.
The dependence of 
$S_\mathrm{EW}^{\chi}-1$
on $\mu_\mathrm{IR}$ (from $0.5\,\mathrm{GeV}$ to $1\,\mathrm{GeV}$) 
is depicted as the orange band.
In the limit of $m_{\chi^\pm}\to\infty$, 
we obtain a numerical estimation with $\mu_\mathrm{IR}$ around the $\rho$ meson mass, $m_\rho = 0.77$\,GeV,
\begin{align}
    \label{eq: numerical SEW}
    S_\mathrm{EW}^{\chi}-1
    \simeq 
    \frac{\alpha}{4\pi}\cdot
    6\log\qty(\frac{0.77\,\mathrm{GeV}}{\mu_\mathrm{IR}})
   + 
    7.88\times{10}^{-3}\ .
\end{align}

For the single Kaon and multi-meson modes,
we will approximate electroweak corrections 
by multiplying the decay rates
by $S_\mathrm{EW}(m_\rho)$ without performing
full one-loop computation,
since the branching fractions of these modes
are small.

\section{Decay into Leptons}
\label{sec: radiative corrections to leptonic mode}

The chargino can decay into leptons 
if kinematically allowed.
In this section, we review 
the tree-level decay rates and perform
analytical computation of 
one-loop corrections to the leptonic decay.

\subsection{Tree-Level Decay Rate}
\label{sec: tree-level lepton}
The chargino decays into a neutralino $\chi^0_1$ (or $\chi^0_2$ if possible) 
and leptons through the operators,
\begin{align}
    \label{eq: HiggsinoEffIntWithLepton1}
    \mathcal{L}^\mathrm{lepton}_\mathrm{CC} 
    = -4G_F\,\overline{\Psi}_\ell\gamma^\mu P_L\Psi_{\nu_\ell}\times
    \overline{\Psi}_{\chi^0}\gamma^\mu\qty(O^W_LP_L+O^W_RP_R)\Psi_{\chi^-}
    +\mathrm{h.c.}\ ,
\end{align}
where $\Psi_{\nu_\ell}$ and $\Psi_\ell$ are four-component spinors 
to indicate the neutrino and charged lepton.
At tree-level, we obtain the decay rate,
\begin{align}
    \label{eq: chargino leptonic decay rate}
    \Gamma_\mathrm{tree}(\chi^-\to\chi^0\ell^-\overline{\nu}_\ell)
    = 
   \frac{G_F^2}{\qty(2\pi)^3}
    \int_{m_\ell^2}^{(\mathit{\Delta}m_\pm)^2} 
    ds\,\frac{s^3}{m_{\chi^\pm}^3}
   \lambda_{\chi^\pm\chi^0}^{1/2}(s)
   \times
    K^{\mathrm{lep}}(s)\ ,
\end{align}
where 
$m_\ell$ is the charged lepton mass,
$\lambda_{AB}(s)$ is the kinematical function defined by
\begin{align}
    \label{eq: lepton kallenlambda}
    \lambda_{AB}(s) &=
    \lambda(1,m_A^2/s, m_B^2/s)\ ;\cr
    \lambda(a, b, c) &= a^2+b^2+c^2-2ab-2bc-2ca\ ,
\end{align}
and
\begin{align}
    \label{eq: lepton kernel}
    K^{\mathrm{lep}}(s)
    &= 
    \qty(1-\frac{m_\ell^2}{s})^2
    \qty[\qty(\abs{O^W_L}^2+\abs{O^W_R}^2)K_1^{\mathrm{lep}}(s)+
    \qty(O^W_LO^{W*}_R+O^W_RO^{W*}_L)K_2^{\mathrm{lep}}(s)]\ ;\\
    K_1^{\mathrm{lep}}(s) &= 
    \frac{1}{6}\qty{\qty(\frac{m_{\chi^\pm}^2+m_{\chi^0}^2}{s}-1)\qty(1-\frac{m_\ell^2}{s})
    +\qty[\qty(\frac{m_{\chi^\pm}^2-m_{\chi^0}^2}{s})^2-1]\qty(1+\frac{2m_\ell^2}{s})}\ ;\\
    K_2^{\mathrm{lep}}(s) &= -\frac{m_{\chi^\pm}m_{\chi^0}}{s}\ .
\end{align}
If we consider 
the Higgsino DM scenario,
that is, if we take the limit of $\abs{M_{1,2}}\gg\abs{\mu}\, (\gtrsim m_Z)$, 
then the decay rate is approximated as
\begin{align}
    \label{eq: expanded lepton decay rate}
    \Gamma_\mathrm{tree}(\chi^-\to\chi^0\ell^-\overline{\nu}_\ell)
    = \Gamma^{\chi\to\ell}_{\mathrm{tree}, m_\ell=0}
    \times f\qty(\frac{m_\ell}{\mathit{\Delta}m_\pm})
    \qty[1-\frac{3 \mathit{\Delta}m_\pm}{2 m_{\chi^\pm}}
    +O\qty(\qty(\frac{\mathit{\Delta}m_\pm}{m_{\chi^\pm}})^2)]\ ,
\end{align}
where
$\Gamma_{\mathrm{tree},m_\ell=0}^{\chi\to\ell}=G_F^2\qty(\mathit{\Delta}m_\pm)^5/(30\pi^3)$
and the function
\begin{align}
    f(x) = \frac{1}{2}\qty[(2-9x^2-8x^4)\sqrt{1-x^2}+15x^4\tanh^{-1}\qty(\sqrt{1-x^2})]
\end{align}
reflects the nonzero lepton mass.
As we saw at the end of subsection~\ref{sec: Gauge Interaction},
the mass mixing effects to the decay rate is small, 
and hence we have neglected them in Eq.~\eqref{eq: expanded lepton decay rate}.
The $O\qty((\mathit{\Delta}m_\pm/m_{\chi^\pm})^2)$
contribution in Eq.~\eqref{eq: expanded lepton decay rate},
which is negligible in the present setup, 
comes from the expansion of the integrand~\eqref{eq: lepton kernel}.
We can obtain the decay rate into leptons 
in the pure-Wino case,
$O_{L/R}^W\simeq1$,
by multiplying Eq.~\eqref{eq: expanded lepton decay rate}
by four.

As stressed in subsection~\ref{sec: ambiguity},
there is an ambiguity in definition
of the ``tree-level'' decay rate coming from 
the difference between $G_F$ and $G_F^0$,
$G_F^2/(G_F^0)^2-1\simeq 0.07$.
Although this ambiguity has been 
mostly resolved 
by the short-distance NLO corrections
computed in Sec.~\ref{sec: EW corrections to FF operator},
the dependence on the photon mass $m_\gamma$ introduced in Eq.~\eqref{eq: vector weak corrections} remains.
In the following, we perform the NLO
computation of the
long-distance corrections 
and will obtain a result free from 
the photon mass dependence.

\subsection{Computation of Radiative Corrections}
We compute NLO corrections to the lepton mode
according to the following strategy. 
First of all, we define the effective 
Lagrangian described 
by the four-fermion operators,
whose counterterms are tuned 
so that the four-fermion theory
reproduces the on-shell amplitude
in the electroweak theory.
Next, we compute virtual-photon corrections 
and real-photon emissions based on 
the four-fermion theory defined above
and obtain a UV and IR-finite result.

For definiteness, we assume the chargino and neutralino are Higgsino-like in this section.
We can obtain
the NLO leptonic decay rate of the Wino
just by multiplying the result of the Higgsino case
by four, because the radiative corrections are common
to both cases up to $O(\alpha\mathit{\Delta}m_\pm/m_{\chi^\pm})$
contributions.

\subsubsection{Effective Fermi Theory}
\label{sec: four-fermi for leptonic mode}
We define the effective four-fermion theory 
of the Higgsinos,
\begin{align}
   \label{eq: Higgsino-lepton FF reordered by Fierz}
   \mathcal{L}_\mathrm{FF}^\mathrm{lepton}
        =\, &-2G_F\qty(\overline{\Psi}_\ell\gamma^\mu P_L\Psi_{\chi^-})
        \qty(\overline{\Psi}_{\chi^0}\gamma_\mu P_L\Psi_{\nu_\ell})
        +4G_F\qty(\overline{\Psi}_{\ell} P_R\Psi_{\chi^-})
        \qty(\overline{\Psi}_{\chi^0} P_L\Psi_{\nu_\ell})
        +\mathrm{h.c.}\ ,
\end{align}
which is obtained by applying Fierz reorderings to 
Eq.~\eqref{eq: HiggsinoEffIntWithLepton1} with $O_L^W = O_R^W =1/2$.
The operator basis~\eqref{eq: Higgsino-lepton FF reordered by Fierz} is convenient for calculation of
radiative corrections since the neutral fermion line
is not affected by virtual photon exchanges.
We add counterterms to the above four-fermion interaction
as\,\footnote{The fields are defined to be bare ones. 
See the matching conditions~\eqref{eq: lepton matching 1}
and \eqref{eq: lepton matching 2}.}
\begin{align}
   \label{eq: Higgsino-lepton FF counterterms}
   \delta\mathcal{L}_\mathrm{FF}^\mathrm{lepton}
        =\, &-2G_F\delta_V^L\qty(\overline{\Psi}_{\ell}\gamma^\mu P_L\Psi_{\chi^-})
        \qty(\overline{\Psi}_{\chi^0}
        \gamma_\mu P_L{\nu_\ell})
        +4G_F\delta_Y^R\qty(\overline{\Psi}_{\ell} P_R\Psi_{\chi^-})
        \qty(\overline{\Psi}_{\chi^0} P_L\Psi_{\nu_\ell})
        +\mathrm{h.c.}
\end{align}
The electroweak corrections
computed in Sec.~\ref{sec: EW corrections to FF operator}
will be incorporated via 
the finite parts of these counterterms.
That is, 
$\delta_{V}^L$ and $\delta_Y^R$ are determined
so that the effective theory 
\eqref{eq: Higgsino-lepton FF reordered by Fierz}
reproduces the amplitude in the electroweak theory 
\eqref{eq: EW virtual corrections}
when we neglect the lepton mass.

Here is a minor note on matching with the electroweak theory 
in the pure-Higgsino case.
In the pure-Higgsino scenario, we have the two neutralinos
$\chi_1^0$ and $\chi_2^0$ with different masses.
In principle, 
the coefficients of the counterterms $\delta$'s 
should be prepared for each neutralino.
Under the situation where the mass mixing effects are sufficiently small,
we may use the $\delta$'s common to both $\chi^0_1$ and $\chi^0_2$ because
the electroweak and QED interaction do not distinguish the two neutralinos.
In addition, the two Majorana field, $\Psi_{\chi^0_1}$ and $\Psi_{\chi^0_2}$,
can be collected into the single Dirac field of the pure-neutral Higgsino,
which is denoted by $\Psi_0$ in Sec.~\ref{sec: EW corrections to FF operator}.
Therefore, the two copies of the Fermi theory
defined by Eqs.~\eqref{eq: Higgsino-lepton FF reordered by Fierz} and \eqref{eq: Higgsino-lepton FF counterterms}
are equivalent to the Fermi theory of the pure-Higgsinos described by
Dirac spinors,
\begin{align}
    \label{eq: Dirac Fermi Theory for lepton}
    \mathcal{L}&\supset
    \mathcal{L}_\mathrm{FF, lepton}^\mathrm{Dirac}
    +\delta\mathcal{L}_\mathrm{FF, lepton}^\mathrm{Dirac}
    +\mathrm{h.c.}\ ; \\
    \mathcal{L}_{\mathrm{FF, lepton}}^\mathrm{Dirac}
    &= -2\sqrt{2}G_F\qty(\overline{\Psi}_{\ell}\gamma^\mu P_L\Psi_{-})
        \qty(\overline{\Psi}_{0}\gamma_\mu P_L\Psi_{\nu_\ell})
        +4\sqrt{2}G_F\qty(\overline{\Psi}_{\ell} P_R\Psi_{-})
        \qty(\overline{\Psi}_{0} P_L\Psi_{\nu_\ell})\ ; \\
    \delta\mathcal{L}_{\mathrm{FF, lepton}}^\mathrm{Dirac}
    &= -2\sqrt{2}G_F\delta_V^L\qty(\overline{\Psi}_{\ell}\gamma^\mu P_L\Psi_{-})
        \qty(\overline{\Psi}_{0}
        \gamma_\mu P_L\Psi_{\nu_\ell})
        +4\sqrt{2}G_F\delta_Y^R\qty(\overline{\Psi}_{\ell} P_R\Psi_{-})
        \qty(\overline{\Psi}_{0} P_L\Psi_{\nu_\ell})\ ,
\end{align}
which can be directly compared with the theory defined by Eq.~\eqref{eq: pure-Higgsino Lagrangian}.

By explicit computation,
we find that these counterterms
should be chosen as
\begin{align}
    \label{eq: lepton matching 1}
    \delta_V^L =
    &\frac{\alpha}{4\pi}
    \qty(
    \frac{3}{2}\log\frac{m_W^2}{m_{\chi^{\pm}}^2}-\frac{11}{4}) \cr
    &+
    \frac{\alpha}{4\pi}\qty[F^\mathrm{Virtual(EW)}_V\qty(\frac{m_W}{m_{\chi^\pm}}) 
    - \qty(\frac{6}{s_W^2}+\frac{7-4 s_W^2}{s_W^4}\log c_W)
    -\left. F^\mathrm{Virtual(EW)}_A\qty(\frac{m_W}{m_{\chi^\pm}})\right|_{\overline{Q}=-1}]
    \ ;
    \\
    \label{eq: lepton matching 2}
    \delta_Y^R =\,\, &-\frac{\alpha}{4\pi}
    \qty(3
    \log\frac{\Lambda_\mathrm{FF}^2}{m_{\chi^\pm}^2} + \frac{3}{2}\log\frac{m_W^2}{m_{\chi^\pm}^2}
    -\frac{1}{4})\cr
    &+\frac{\alpha}{4\pi}
    \qty[
    F^\mathrm{Virtual(EW)}_V\qty(\frac{m_W}{m_{\chi^\pm}}) 
    - \qty(\frac{6}{s_W^2}+\frac{7-4 s_W^2}{s_W^4}\log c_W) 
    + \left. F^\mathrm{Virtual(EW)}_A\qty(\frac{m_W}{m_{\chi^\pm}})\right|_{\overline{Q}=-1}
    ]
    \ ,
\end{align}
to reproduce the electroweak theory (see Eq.~\eqref{eq: EW virtual corrections}).
Here, $\Lambda_\mathrm{FF}$ is the Pauli-Villars regulator mass introduced to regularize the photon loop.

\subsubsection{Virtual Corrections}
\begin{figure}[t]
    \centering
    \includegraphics[width=0.75\textwidth]{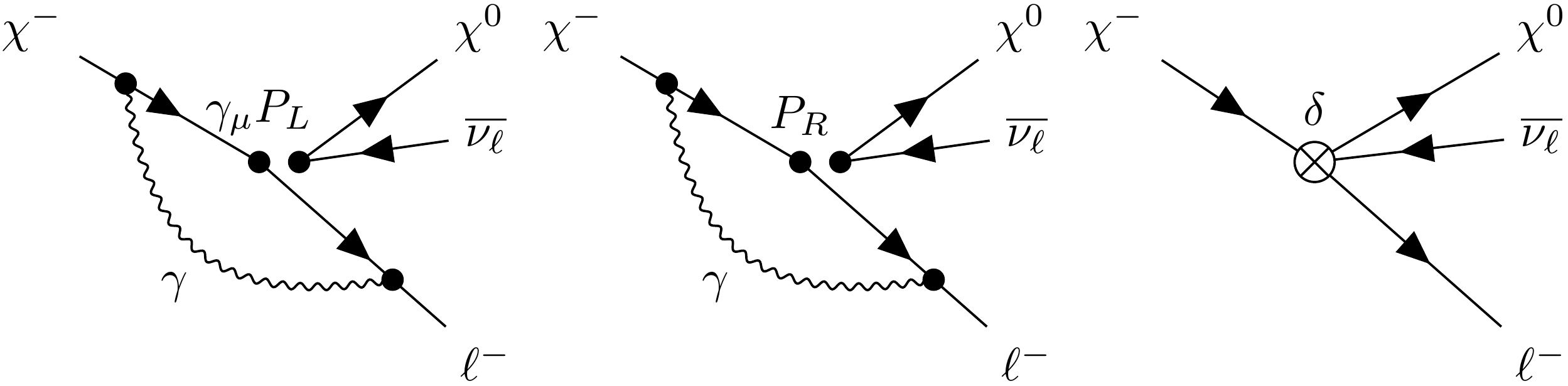}
    \caption{Feynman diagrams of 1PI virtual corrections
    and counterterm contribution to the leptonic mode.}
    \label{fig: leptonicvirtual}
\end{figure}
We perform computation of 
the virtual photon corrections 
including the finite lepton mass.
These corrections consist of wave-function renormalizations,
1PI corrections, and counterterm contributions
(for the last two contributions, see Fig.~\ref{fig: leptonicvirtual}).
The resultant one-loop corrections can be 
broken up into the three pieces:
\begin{align}
    \delta\Gamma^{\chi\to\ell}_\mathrm{virtual} 
    = \delta\Gamma^{\chi\to\ell}_\mathrm{SD(EW)}
    + \delta\Gamma^{\chi\to\ell}_\mathrm{virtual,IR} 
    + \delta\Gamma^{\chi\to\ell}_\mathrm{LD(QED)}\ .
\end{align}

The first term represents the electroweak non-logarithmic
corrections provided by the counterterms,
\begin{align}
    \delta\Gamma^{\chi\to\ell}_\mathrm{SD(EW)}
    &= \Gamma^{\chi\to\ell}_\mathrm{tree}\times
    \frac{\alpha}{4\pi}
    \cdot 2\qty[F^\mathrm{Virtual(EW)}_V\qty(\frac{m_W}{m_{\chi^\pm}}) -\qty(\frac{6}{s_W^2}+\frac{7-4 s_W^2}{s_W^4}\log c_W)]\ .
\end{align}
Here, $\Gamma^{\chi\to\ell}_\mathrm{tree}$
is the tree-level rate\,\footnote{This decay rate coincides with  
Eq.~\eqref{eq: expanded lepton decay rate} in the limit of
$\mathit{\Delta}m_\pm\ll m_{\chi^\pm}$.}
in the limit of 
$\mathit{\Delta}m_\pm\ll m_{\chi^\pm}$:
\begin{align}
    \Gamma^{\chi\to\ell}_\mathrm{tree}
    = \Gamma_{\mathrm{tree},m_\ell=0}^{\chi\to\ell}
    \times
    \int_x^1 dw\,\rho_\mathrm{tree}(w)\ ,
\end{align}
where 
$w$ is the charged lepton energy normalized by $\mathit{\Delta}m_\pm$, 
$x = m_\ell/\mathit{\Delta}m_\pm$, 
and
\begin{align}
    \label{eq: tree-level spectrum}
    \rho_\mathrm{tree}(w)
   &=30w^2(1-w)^2\sqrt{1-\frac{x^2}{w^2}}\ .
\end{align}

The second term
$\delta\Gamma^{\chi\to\ell}_\mathrm{virtual, IR}$
is the infrared divergent part, 
\begin{align}
    \label{eq: IRvirtual}
    \delta\Gamma^{\chi\to\ell}_\mathrm{virtual,IR}
    &= 4\log\frac{m_\ell^2}{m_\gamma^2}
    \times\frac{\alpha}{4\pi}
    \Gamma_{\mathrm{tree},m_\ell=0}^{\chi\to\ell}
    \int_x^1 dw\,
    \rho_\mathrm{tree}(w)
    \qty[1-
    \frac{w}{\sqrt{w^2-x^2}}\cosh^{-1}\qty(\frac{w}{x})]\ ,
\end{align}
which will be canceled with real photon corrections.

The third term 
$\delta\Gamma^{\chi\to\ell}_\mathrm{LD(QED)}$
represents 
the long-distance corrections from 
the virtual photon exchange.
In the limit of 
$\mathit{\Delta}m_\pm\ll m_{\chi^\pm}$, 
it is given by
\begin{align}
    \delta\Gamma^{\chi\to\ell}_\mathrm{LD(QED)}
    = \Gamma^{\chi\to\ell}_{\mathrm{tree},m_\ell=0}
    \times
    \frac{\alpha}{4\pi}\int_x^1 dw\,\rho_\mathrm{LD(QED)}(w)\ ,
\end{align}
where
\begin{align}
\label{eq: LDQED}
    \rho_\mathrm{LD(QED)}(w)
    =\, &\qty(-\frac{5}{2}+ 3\log\frac{m_W^2}{m_\ell^2})
    \rho_\mathrm{tree}(w)\notag \\
    &+ 120w^2(1-w)^2
    \qty[\cosh^{-1}\qty(\frac{w}{x})
    -\qty(\cosh^{-1}\qty(\frac{w}{x}))^2
    -\mathrm{Li}_2\qty(\frac{2\sqrt{w^2-x^2}}{w+\sqrt{w^2-x^2}})
    ]\notag \\
    &-120x^2(1-w)^2\cosh^{-1}\qty(\frac{w}{x})\ .
\end{align}
Note that the axial contribution 
from the interference between the axial and the tree-level amplitudes in Eqs.\,\eqref{eq: tree-level amplitude of pure Higgsino}
and \eqref{eq: axial amplitude}
is suppressed by $\alpha\cdot m_\ell/m_{\chi^\pm}$.

We find no divergent terms
in the limit of $m_{\chi^\pm}\to\infty$
as in the case of the electroweak correction.
However, 
the decay spectrum $\rho_\mathrm{LD(QED)}(w)$ has
logarithmic singularities with respect to 
the charged lepton mass,
\begin{align}
    \label{eq: mass singularity in virtual corrections}
    \rho_\mathrm{LD(QED)}(w)
    \xrightarrow[m_\ell\ll\mathit{\Delta}m_\pm]{}\, 
    &-6\log x\times\rho_\mathrm{tree}(w) \cr
    &- 120\qty[\log x+\qty(\log x)^2]\times w^2(1-w)^2
    \cr
    &+ 240\log x\times w^2(1-w)^2\log(w+\sqrt{w^2-x^2})\cr
    &=: \rho_\mathrm{LD(QED)}^\mathrm{singular}(w)\ ,
\end{align}
which correspond to divergences arising when the charged lepton and photon become collinear.
According to 
the Kinoshita-Lee-Nauenberg (KLN) theorem~\cite{Kinoshita:1962ur, Lee:1964is},
if we include real photon emissions,
such singularities in decay spectra will disappear after integrating
them over the phase space.
We will see that the cancellation of the mass singularities will actually occur
in the next subsubsection.

\subsubsection{Real Photon Emission}
\begin{figure}[t]
    \centering
    \includegraphics[width=0.5\textwidth]{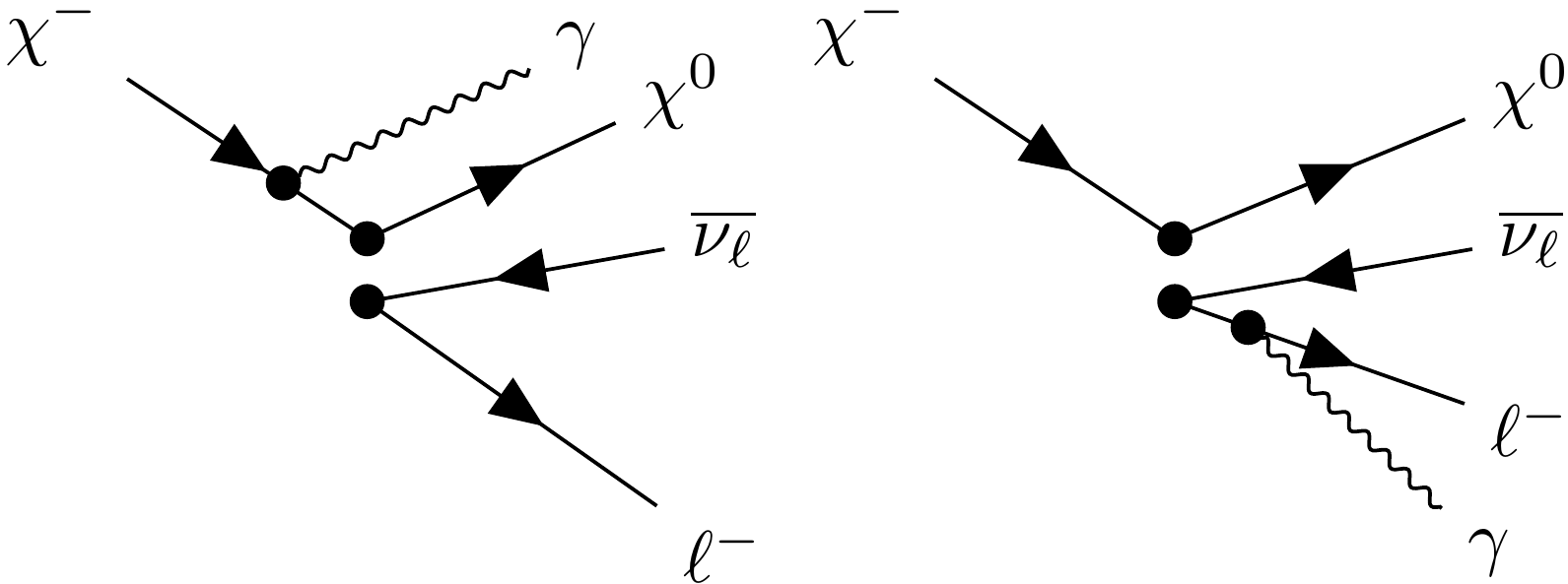}
    \caption{Feynman diagrams of real photon emissions
    in the leptonic mode.}
    \label{fig: leptonicreal}
\end{figure}
For completeness, we calculate the real photon emission processes
shown in Fig.~\ref{fig: leptonicreal}.
The resultant decay rate is composed of the infrared divergent part 
and finite part,
\begin{align}
    \Gamma^{\chi\to\ell}_\mathrm{emit} 
    = \Gamma^{\chi\to\ell}_\mathrm{emit,IR}
    +\Gamma^{\chi\to\ell}_\mathrm{emit,finite}\ .
\end{align}
The infrared divergent part is given by
\begin{align}
    \label{eq: IRreal}
    \Gamma^{\chi\to\ell}_\mathrm{emit, IR}
    = -4\log\frac{m_\ell^2}{m_\gamma^2}
    \times\frac{\alpha}{4\pi}
    \Gamma_{\mathrm{tree},m_\ell=0}^{\chi\to\ell}
    \int_x^1 dw\,
    \rho_\mathrm{tree}(w)
    \qty[1-
    \frac{w}{\sqrt{w^2-x^2}}\cosh^{-1}\qty(\frac{w}{x})]\ ,
\end{align}
which exactly cancels the IR divergence in the virtual correction~\eqref{eq: IRvirtual}.
In the limit of 
$\mathit{\Delta}m_\pm\ll m_{\chi^\pm}$, 
the part free from IR divergences is given by
\begin{align}
    \Gamma^{\chi\to\ell}_\mathrm{emit, finite}
    = \Gamma^{\chi\to\ell}_{\mathrm{tree},m_\ell=0}
    \times
    \frac{\alpha}{4\pi}\int_x^1 dw\,\rho_\mathrm{emit}(w)\ ,
\end{align}
where
\begin{align}
    \label{eq: real emission spectrum}
    \rho_\mathrm{emit}(w)
    &=
    240w^2(1-w)^2
    \qty[-\log x + \log(1-w)]
    \qty[-\sqrt{1-\frac{x^2}{w^2}}+\cosh^{-1}\qty(\frac{w}{x})]\cr
    &+120w^2(1-w)^2\sqrt{1-\frac{x^2}{w^2}}\,(1-2\log 2)\cr
    &+240(\log 2) w^2(1-w)^2\cosh^{-1}\qty(\frac{w}{x})\cr
    &+120w^2(1-w)^2\cosh^{-1}\qty(\frac{w}{x})
    \qty[1+\log x-\log(2w)]\cr
    &+120w^2(1-w)^2\qty[
    \mathrm{Li}_2\qty(-\sqrt{1-\frac{x^2}{w^2}})-\mathrm{Li}_2\qty(\sqrt{1-\frac{x^2}{w^2}})
    ]\cr
    &+60w^2(1-w)^2\qty[
    \mathrm{Li}_2\qty(\frac{w+\sqrt{w^2-x^2}}{2w})
    -\mathrm{Li}_2\qty(\frac{w-\sqrt{w^2-x^2}}{2w})
    ]\cr
    &+360w^2(1-w)^2
    \qty[\sqrt{1-\frac{x^2}{w^2}}-\cosh^{-1}\qty(\frac{w}{x})]\cr
    &+10(1-w)^3\qty[-8\sqrt{w^2-x^2}
     +(1+7w)\cosh^{-1}\qty(\frac{w}{x})]\ .
\end{align}
This estimation includes hard photon emissions.
As in the case of the virtual corrections, 
the spectrum does not depend on 
the chargino/neutralino mass
in the limit of $m_{\chi^\pm}\to\infty$.

The decay spectrum~\eqref{eq: real emission spectrum} exhibits 
the mass singularity in the limit of $m_\ell\ll\mathit{\Delta}m_\pm$:
\begin{align}
   \rho_\mathrm{emit}(w)
   \xrightarrow[m_\ell\ll\mathit{\Delta}m_\pm]{}\,
   &120\log x\times w^2(1-w)^2\cr
    &\hspace{10pt}\times
    \qty[2\sqrt{1-\frac{x^2}{w^2}}-\log(w+\sqrt{w^2-x^2})-2\log(1-w)-1+\log\frac{w}{2}]\cr
    &+120(\log x)^2w^2(1-w)^2\cr
    &+360\log x\times w^2(1-w)^2-10\log x\times (1-w)^3(1+7w)\cr
    &=: \rho_\mathrm{emit}^\mathrm{singular}(w)\ .
\end{align}
However,  after combining the limiting behaviour of the virtual correction~\eqref{eq: mass singularity in virtual corrections} and integration over the phase space,
we find
\begin{align}
    \int_x^1 dw\,
    \qty[\rho_\mathrm{LD(QED)}^\mathrm{singular}(w)+\rho_\mathrm{emit}^\mathrm{singular}(w)]
    = -10x + O(x^2)\ .
\end{align}
That is, the mass singularities completely vanish 
and the result is regular at 
$x = m_\ell/\mathit{\Delta}m_\pm = 0$, 
as argued by the KLN theorem.

\subsubsection{NLO Decay Rate}
By combining the virtual and real photon corrections,
we obtain the inclusive decay rate at one-loop level
in the limit of $\mathit{\Delta}m_\pm\ll m_{\chi^\pm}$:
\begin{align}
\label{eq: lepton NLO}
    \Gamma_\mathrm{NLO}\qty(\chi^-\to\chi^0\ell^-\overline{\nu}_\ell(\gamma))
    =
    \Gamma_\mathrm{tree}\qty(\chi^-\to\chi^0\ell^-\overline{\nu}_\ell)
    +\delta\Gamma^{\chi\to\ell}_\mathrm{one\mathchar`-loop}\ ,
\end{align}
where
\begin{align}
    \delta\Gamma^{\chi\to\ell}_\mathrm{one\mathchar`-loop}
    = \Gamma^{\chi\to\ell}_{\mathrm{tree},m_\ell=0}
    \times
    \frac{\alpha}{4\pi}\int_x^1 dw\,\rho_\mathrm{one\mathchar`-loop}(w)\ ,
\end{align}
and 
\begin{align}
    \rho_\mathrm{one\mathchar`-loop}(w)
    =\,\,
    &2\qty[
    F^\mathrm{Virtual(EW)}_V\qty(\frac{m_W}{m_{\chi^\pm}}) 
    -\qty(\frac{6}{s_W^2}+\frac{7-4 s_W^2}{s_W^4}\log c_W)
    ]\rho_\mathrm{tree}(w)\cr
    &+\rho_\mathrm{LD(QED)}(w)+\rho_\mathrm{emit}(w)\ .
    \label{eq: lepton 1-loop spectrum}
\end{align}
From Eqs.~\eqref{eq: LDQED} and \eqref{eq: lepton 1-loop spectrum},
we can see that the universal short-distance correction
$S^\chi_\mathrm{EW}(\mathit{\Delta}m_\pm)$ is contained in the NLO decay rate~\eqref{eq: lepton NLO}.

\subsection{Numerical Results}
\begin{figure}[t!]
	\centering
  	\subcaptionbox{\label{fig:lepton_rate}}
	{\includegraphics[width=0.47\textwidth]{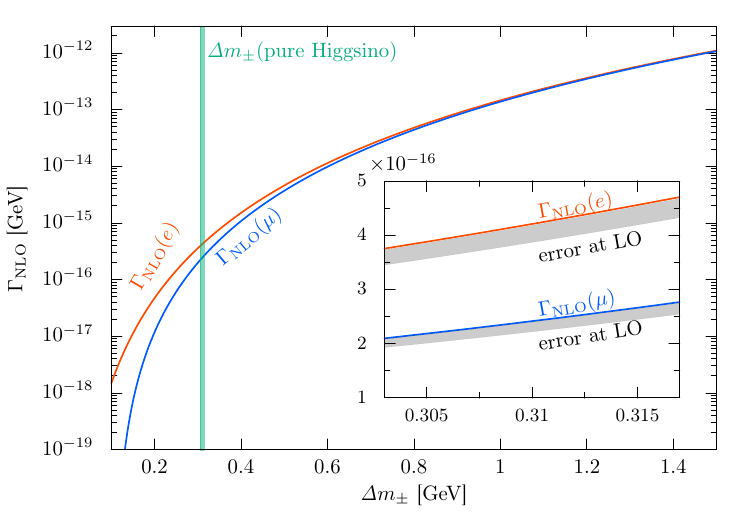}} 
  	\subcaptionbox{\label{fig:lepton_ratio}}
	{\includegraphics[width=0.47\textwidth]{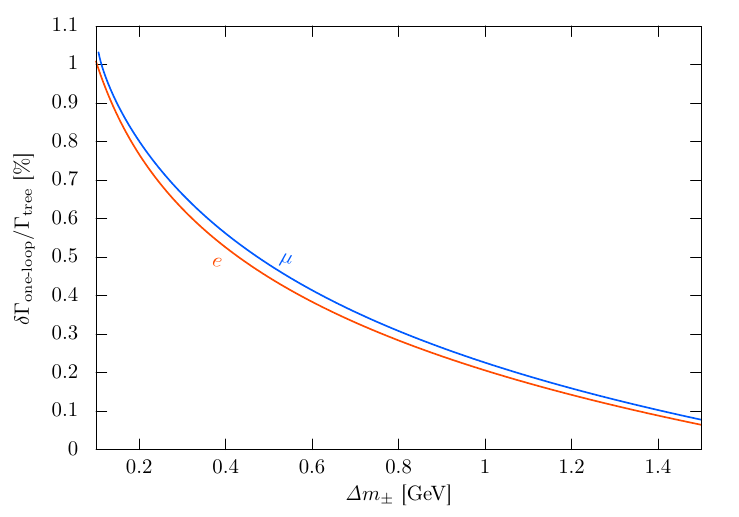}} 
\caption{
(Left) The NLO decay rate into the leptons
as functions of $\mathit{\Delta}m_\pm$ 
in the case of $300$\,GeV Higgsino.
The blue and red line represent
the muon and electron modes, respectively.
The inserted window is the close-up view 
of the intersection between $\Gamma_\mathrm{NLO}$
and the vertical green band, 
which corresponds to the case of
the 300\,GeV pure-Higgsino.
The gray bands represent the uncertainty of about 7\% in the leading order 
decay rate
(see Eq.~\eqref{eq: GF uncertainty}),
which shrink to too small values to be shown 
in the NLO results.
(Right) The ratio of the 
one-loop correction to the tree-level decay rate
as functions of the mass difference for $m_{\chi^\pm} = 300\,\mathrm{GeV}$.
}\label{fig:lepton}
\end{figure}
In Fig.~\ref{fig:lepton_rate}, we show the 
NLO decay rate of the 300\,GeV Higgsino into the leptons as
functions of $\mathit{\Delta}m_\pm$ 
for the muon mode (blue) and the electron mode (red).
Here, only the decay to $\chi^0_1$ 
is considered.
In the pure-Higgsino case, there is another neutralino $\chi^0_2$ whose mass is identical to that of $\chi^0_1$ and  $\Gamma(\chi^- \to \chi^0_1)= \Gamma(\chi^- \to \chi^0_2)$.
Therefore, the total decay rate of the charged Higgsino should be multiplied by two.
The decay rates in Fig.~\ref{fig:lepton_rate} are translated into
the NLO decay rate of the charged Wino into the leptons
just by multiplying the rate by four.
The inserted window is the close-up view 
of the intersection between $\Gamma_\mathrm{NLO}$
and the vertical green band, 
which corresponds to the mass difference of
the 300\,GeV pure-Higgsino~\cite{Thomas:1998wy, Yamada:2009ve, Nagata:2014wma}.
The gray bands represent the uncertainty of about 7\% in the leading order 
decay rate
(see Eq.~\eqref{eq: GF uncertainty}).
The NLO calculation determines the decay rate
to a precision of $O(\alpha/\pi\cdot\mathit{\Delta}m_\pm/m_{\chi^\pm})$.

In Fig.~\ref{fig:lepton_ratio},
we show the ratio of the 
one-loop correction to the tree-level decay rate
as functions of the mass difference 
for the 300\,GeV Higgsino.
The NLO correction provides 
about $0.1$--$1$\% modification to the tree-level rate.

\section{Decay into Single Pion}
\label{sec: decay into single pion}

In this section, we reanalyze
the single pion mode of the chargino decay
and update our previous estimate in 
Ref.~\cite{Ibe:2022lkl} as 
mentioned in Sec.~\ref{sec: Summary of Results}.

\subsection{Tree-Level Decay Rate}
\label{sec: tree-level single meson}
The decay into the single charged pion $\pi^\pm$
is caused by the operators, 
\begin{align}
    \label{eq: decay operator of single pion}
    \mathcal{L}_\mathrm{CC}^\mathrm{pion}
    = -2\sqrt{2}\,G_F V_{ud}^*F_\pi
    (\partial_\mu \pi^-)^*\times
    \overline{\Psi}_{\chi^0}\gamma^\mu\qty(O^W_LP_L+O^W_RP_R)\Psi_{\chi^-}+\mathrm{h.c.}\ ,
\end{align}
where $V_{ud}$ is the CKM matrix element and $F_\pi$ is the pion decay constant (see Table~\ref{tab:input}).
When $\mathit{\Delta}m_\pm$ is larger than the charged pion mass $m_{\pi^\pm}$, 
the chargino $\chi^-$ can decay into a neutralino $\chi^0$
and a single pion $\pi^-$ through this coupling.
The tree-level decay rate is given by
\begin{align}
    \label{eq: tree-level single pion rate}
    \Gamma_\mathrm{tree}\qty(\chi^-\to\chi^0+\pi^-)
    =
    &\frac{F_\pi^2 G_F^2 \abs{V_{ud}}^2}{8\pi m_{\chi^\pm}}
    \sqrt{\lambda\qty(1, m_{\chi^0}^2/m_{\chi^\pm}^2, m_{\pi^\pm}^2/m_{\chi^\pm}^2)}\cr
    &\times\Bigg\{\abs{O^W_L+O^W_R}^2
    \qty(m_{\chi^\pm} - m_{\chi^0})^2
    \qty[\qty(m_{\chi^\pm} + m_{\chi^0})^2-m_{\pi^\pm}^2]\cr
    &\hspace{40pt}
    +\abs{O^W_L-O^W_R}^2
    \qty(m_{\chi^\pm}+m_{\chi^0})^2
    \qty[\qty(m_{\chi^\pm}-m_{\chi^0})^2-m_{\pi^\pm}^2]
    \Bigg\}\ ,
\end{align}
where $\lambda(a, b ,c)$ is given by
\begin{align}
    \label{eq: pion kallenlambda}
    \lambda(a, b, c) = a^2+b^2+c^2-2ab-2bc-2ca\ .
\end{align}
For the almost pure-Higgsino case that $\abs{M_{1,2}}\gg\abs{\mu}\gtrsim m_Z$, 
the above rate can be approximated as
\begin{align}
\label{eq: single pion mode at tree-level}
    \Gamma_\mathrm{tree}\qty(\chi^-\to\chi^0+\pi^-)
    =\,
    &\frac{1}{\pi} F_\pi^2 G_F^2\abs{V_{ud}}^2 
    \qty(\mathit{\Delta}m_\pm)^3
    \sqrt{1-\qty(\frac{m_{\pi^\pm}}{\mathit{\Delta}m_\pm})^2
    }\cr
    &\times\qty[1-\frac{\mathit{\Delta}m_\pm}{m_{\chi^\pm}}
    +\frac{\qty(\mathit{\Delta}m_\pm)^2 - m_{\pi^\pm}^2}{4m_{\chi^\pm}^2}]^{3/2}\ ,
\end{align}
where we have neglected the deviation from the 
pure-Higgsino assumption (see subsection~\ref{sec: Gauge Interaction}).

\subsection{``Tree-Level'' and ``Leading-Order''
Decay Rates}
\label{sec: tree ambiguity in pion}
As we have mentioned in Sec.~\ref{sec: ambiguity},
we face ambiguities in choice of 
the coupling constants of Lagrangian
as long as we stick to the tree-level computation.
Since the single pion mode is the most important
decay mode for searches of the almost pure-Wino/Higgsino
at collider experiments, 
we elucidate the ambiguities of the tree-level prediction
of the decay rate to the single pion.

In subsection~\ref{sec: tree-level single meson}, we used 
the tree-level Lagrangian with the coefficient given by the Fermi constant $G_F$.
We can also consider
the tree-level decay rate with $G_F$ replaced by
$G_F^0$ given by Eq.\,\eqref{eq: GF0},
\begin{align}
    \label{eq: Gamma0 single pion}
    \Gamma^0_\mathrm{tree}\qty(\chi^-\to\chi^0+\pi^-)
    :=
    \left.\Gamma_\mathrm{tree}\qty(\chi^-\to\chi^0+\pi^-)\right|_{G_F\to G_F^0}\ .
\end{align}
Alternatively, we may consider 
the leading-order 
decay rate defined as
the tree-level decay rate normalized  
by the $\pi_{\mu 2}$ decay rate, i.e., \cite{Ibe:2012sx}
\begin{align}
    \label{eq: pion decay at leading by BF}
    \Gamma_\mathrm{LO}\qty(\chi^-\to\chi^0+\pi^-)
    := B\qty(\pi^-\to\mu^-+\overline{\nu_\mu}(+\gamma))
        \times\Gamma^\mathrm{tot}_\pi
        \times\frac{\Gamma_\mathrm{tree}\qty(\chi^-\to\chi^0+\pi^-)}
        {\Gamma_\mathrm{tree}\qty(\pi^-\to\mu^-+\overline{\nu_\mu})}\ ,
\end{align}
where
\begin{align}
    \label{eq: pion decay at tree-level}
    \Gamma_\mathrm{tree}\qty(\pi^-\to\mu^-+\overline{\nu_\mu})
     = \frac{1}{4\pi}G_F^2\abs{V_{ud}}^2
     F_\pi^2m_{\pi^\pm} m_\mu^2\qty(1-\frac{m_\mu^2}{m_{\pi^\pm}^2})^2\ ,
\end{align}
with 
$\Gamma^\mathrm{tot}_\pi = \tau_\pi^{-1}$ 
and 
$B(\pi^-\to\mu^-+\overline{\nu_\mu}(+\gamma))$
being  
the charged pion decay width 
and the branching 
fraction of the $\pi_{\mu2}$ mode
given in Table~\ref{tab:input}.
Note that the dependence on $G_F\abs{V_{ud}}F_\pi$
is canceled in 
the ratio of the tree-level decay rates, 
\begin{align}
    \label{eq: tree-level branching of single pion mode}
    \frac{\Gamma_\mathrm{tree}\qty(\chi^-\to\chi^0+\pi^-)}
    {\Gamma_\mathrm{tree}\qty(\pi^-\to\mu^-+\overline{\nu_\mu})}
    \simeq\,
    &\frac{4\qty(\mathit{\Delta}m_\pm)^3}{m_{\pi^\pm} m_\mu^2} 
    \sqrt{1-\qty(\frac{m_{\pi^\pm}}{\mathit{\Delta}m_\pm})^2}
    \qty[1-\frac{\mathit{\Delta}m_\pm}{m_{\chi^\pm}}
    +\frac{\qty(\mathit{\Delta}m_\pm)^2 - m_{\pi^\pm}^2}{4m_{\chi^\pm}^2}]^{3/2}
    \qty(1-\frac{m_\mu^2}{m_{\pi^\pm}^2})^{-2} \ .
\end{align}

When focusing on the ``tree-level'' decay rate, it is important to note that the three decay rates $\Gamma_\mathrm{tree}$, $\Gamma_\mathrm{tree}^0$,
and $\Gamma_\mathrm{LO}$ are 
all considered to be a correct zeroth-order of a perturbative calculation.
Numerically, however, we find that those tree-level decay rates differ by about 10\%, 
\begin{align}
   \frac{\Gamma_\mathrm{tree}\qty(\chi^-\to\chi^0+\pi^-)}
    {\Gamma_\mathrm{LO}\qty(\chi^-\to\chi^0+\pi^-)} \simeq 0.98 \ ;
    \qquad      
    \frac{\Gamma^0_\mathrm{tree}\qty(\chi^-\to\chi^0+\pi^-)}
    {\Gamma_\mathrm{LO}\qty(\chi^-\to\chi^0+\pi^-)} \simeq 0.91 \ .
\end{align}
This uncertainty profoundly affects searches for meta-stable charginos at collider experiments, where the detection efficiency strongly depends on the chargino lifetime.

\subsection{Prelude to Radiative Corrections: Two-step Matching Procedure}
\label{sec: two-step matching}
\begin{figure}[t]
\centering{{\includegraphics[width=0.99\textwidth]{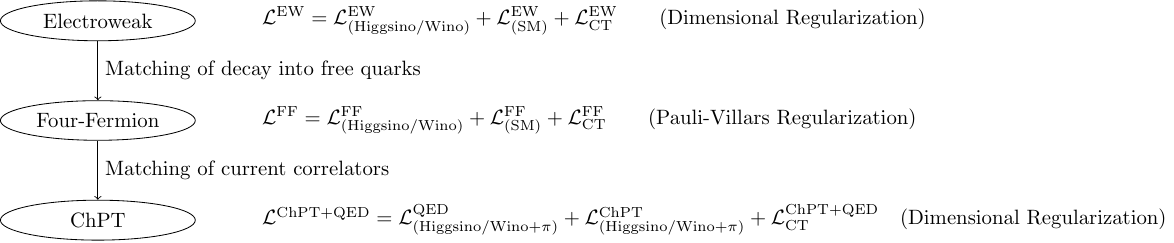}} }
 \caption{
Flowchart of the matching procedure.
 }
\label{fig:flowchart}
\end{figure}
The decay into a single pion is described by the non-renormalizable 
interaction~\eqref{eq: decay operator of single pion}.
In the presence of the QED interactions, 
the derivative in this operator should be replaced with 
the covariant one, $D_\mu = \partial_\mu - ieA_\mu$.
As a result, we obtain the interaction term, 
\begin{align}
    \mathcal{L}_\mathrm{CC}^\mathrm{meson}= 
    -2\sqrt{2}\,G_F V_{ud}^*F_\pi\qty(D_\mu \pi^-)^*\times
    \overline{\Psi}_{\chi^0}\gamma^\mu
    \qty(O_L^WP_L+O_R^WP_R)\Psi_{\chi^-}+\mathrm{h.c.}
\end{align}
This theory can no longer be renormalized
multiplicatively since the QED interaction explicitly violates the shift symmetry of the pion field.

We encounter a similar situation in the $\pi_{\ell 2}$ decay,
which Knecht et al.~\cite{Knecht:1999ag}  addressed within the framework of the chiral perturbation theory (ChPT).
In that literature, the authors enumerate $O(e^2p^2)$ 
low-energy constants (LECs)
to the pion-lepton interactions. By using those LECs, they provide
a UV-finite decay rate 
of the charged pion at one-loop level.
Note that the finite part of the LECs 
cannot be determined unambiguously
unless
ChPT is matched with some UV-complete and 
renormalizable theory.
Descotes-Genon and Moussallam~\cite{Descotes-Genon:2005wrq} 
(hereafter referred to as D\&M) 
match ChPT with the electroweak theory and 
fix the finite part 
of the LECs relevant for the $\pi_{\ell 2}$ decay.

In our previous analysis~\cite{Ibe:2022lkl}, we introduced the LECs $Y_i$ 
for the Wino-pion interaction following Ref.~\cite{Knecht:1999ag}
and determined their finite parts by
D\&M's matching procedure (See Fig.~\ref{fig:flowchart}).
First, we tune the counterterms of the intermediate four-fermion theory, $\mathcal{L}^\mathrm{FF}_\mathrm{CT}$,
to reproduce the amplitude
of the decay into the free quarks calculated 
by the electroweak theory including the Wino,
$\mathcal{L}^\mathrm{EW}$.
In the second step, we determine the LECs $Y_i$
in $\mathcal{L}_\mathrm{CT}^\mathrm{ChPT+QED}$
by matching ChPT with the four-fermion theory 
through the so-called current correlators.
We found that the LECs $Y_i$ obtained by D\&M's matching method
successfully provide the UV-finite Wino decay rate into a single pion.

As we will see in the following subsections,
the computational framework used in the Wino case
can be extended straightforwardly to 
more generic electroweak fermions including the Higgsino. 
In this case,
we can safely neglect $O(\alpha\mathit{\Delta}m_\pm/m_{\chi^\pm})$ contributions
in the computation of the radiative corrections.
In addition, we can omit the mass mixing effect discussed in Sec.~\ref{sec: model discripition}
because 
it is suppressed by $\alpha$ and
the Wino or the Higgsino mass parameter.
Therefore, the radiative corrections can be computed based on 
the pure-Higgsino/Wino Lagrangian.
In the following analysis,
we compute the radiative corrections to the decay rate of the charged Higgsino into a single pion
based on the computational method used in Ref.~\cite{Ibe:2022lkl}.

\subsection{Matching Four-Fermion Theory with Electroweak Theory}
\label{sec: matching FF with EW}
Following the analysis conducted by D\&M, 
we introduce the intermediate step
to define the four-fermion theory 
which reproduces the decay rate
into free quarks under the
electroweak interactions.
The four-fermion theory obtained here
will be matched with ChPT.

Let us consider the four-fermion theory 
with counterterms,
\begin{align}
    \label{eq: Majorana Fermi Theory}
    \mathcal{L}&\supset
    \mathcal{L}_\mathrm{FF, quark}
    +\delta\mathcal{L}_\mathrm{FF, quark}
     + \mathrm{h.c.}\ ; \\
    \mathcal{L}_{\mathrm{FF, quark}}
    &= -2G_F V_{ud}^* (\overline{\Psi}_{d}\gamma^\mu P_L \Psi_u)
    (\overline{\Psi}_{\chi^0}\gamma^\mu \Psi_{\chi^-})\ ; \\
    \label{eq: CT for Majorana Fermi Theory}
    \delta\mathcal{L}_{\mathrm{FF, quark}}
    &= -2G_F V_{ud}^* (\overline{\Psi}_{d}\gamma^\mu P_L \Psi_u)
    (\overline{\Psi}_{\chi^0}\gamma^\mu \Psi_{\chi^-})
     \notag \\
    &\hspace{20pt}\times e^2 \qty[f_{\chi\chi}Q_\chi^2 + f_{d\bar{u}}(Q_d+Q_{\bar{u}})^2 
    + f_{\chi d}Q_\chi Q_d - f_{\chi \bar{u}}Q_\chi Q_{\bar{u}}]\ ,
\end{align}
where $Q_d$ and $Q_{\bar{u}}$ are the charges of the $d$-quark and $u$-antiquark, respectively. The coefficients $f_{\chi\chi}, f_{d\bar{u}}, f_{\chi d}$, and $f_{\chi\bar{u}}$ should be determined by matching with the results of Sec.~\ref{sec: EW corrections to FF operator}.
It is important to leave the difference of the charges,
$\overline{Q}=Q_d-Q_{\bar{u}}$, arbitrary
to derive matching conditions.

As we noted in subsubsection~\ref{sec: four-fermi for leptonic mode}, 
if we consider the case of the Higgsino-like DM, then
we have the two Majorana neutralinos
$\chi_1^0$ and $\chi_2^0$ with different masses
and 
the counterterms should be prepared for each neutralino.
However, since we address the situation
where the mass mixings are sufficiently small,
the two copies of the four-fermion theory~\eqref{eq: Majorana Fermi Theory} 
are equivalent to the four-fermion theory of the pure-Higgsinos,
\begin{align}
    \label{eq: Dirac Fermi Theory}
    \mathcal{L}&\supset
    \mathcal{L}_\mathrm{FF, quark}^\mathrm{Dirac}
    +\delta\mathcal{L}_\mathrm{FF, quark}^\mathrm{Dirac}
    +\mathrm{h.c.}\ ; \\
    \mathcal{L}_{\mathrm{FF, quark}}^\mathrm{Dirac}
    &= -2\sqrt{2}G_F V_{ud}^* (\overline{\Psi}_{d}\gamma^\mu P_L \Psi_u)
    (\overline{\Psi}_0\gamma^\mu \Psi_-)\ ; \\
    \delta\mathcal{L}_{\mathrm{FF, quark}}^\mathrm{Dirac}
    &= -2\sqrt{2}G_F V_{ud}^* (\overline{\Psi}_{d}\gamma^\mu P_L \Psi_u)
    (\overline{\Psi}_0\gamma^\mu \Psi_-)
     \notag \\
    &\hspace{20pt}\times e^2 \qty[f_{\chi\chi}Q_\chi^2 + f_{d\bar{u}}(Q_d+Q_{\bar{u}})^2 
    + f_{\chi d}Q_\chi Q_d - f_{\chi \bar{u}}Q_\chi Q_{\bar{u}}]\ ,
\end{align}
where $\Psi_-$ and $\Psi_0$
are Dirac spinors for the charged and neutral 
Higgsinos, respectively.
The four-fermion theory~\eqref{eq: Dirac Fermi Theory} can be directly compared with the theory defined by Eq.~\eqref{eq: pure-Higgsino Lagrangian}.

The decay amplitude of the chargino into free quarks can be computed at one-loop level
based on the four-fermion theory~\eqref{eq: Dirac Fermi Theory}. 
The QED correction generates the amplitude including the axial current of the chargino/neutralino,
$\mathcal{M}_A^\mathrm{quark}$.
As we pointed out in Ref.~\cite{Ibe:2022lkl}, however, the axial amplitude provides
only two-loop-level contribution because the interference between
$\mathcal{M}_\mathrm{tree}^\mathrm{quark}$ and $\mathcal{M}_A^\mathrm{quark}$ vanishes
after integration over the phase space. 
Hence, we take into account only the one-loop level amplitude proportional to the tree-level one,
\begin{align}
    \label{eq:virtual FF}
    \frac{\mathcal{M}_{\mathrm{Virtual(FF)}}^\mathrm{quark}}{{\mathcal{M}}_{\mathrm{tree}}^{\mathrm{quark}}}
    &=
    \frac{\alpha}{4\pi}\frac{3}{2}
    \qty(
    \log\frac{\Lambda_\mathrm{FF}^2}{m_\gamma^2}
    - 1
    )+
    e^2\left(f_{\chi\chi} + 
    f_{d\bar{u}} +  \frac{1-\overline{Q}}{2}f_{\chi d}
    - \frac{1+\overline{Q}}{2}f_{\chi \bar{u}}\right)\ ,
\end{align}
where $\Lambda_\mathrm{FF}$ is the Pauli-Villars regulator mass.
We have set $Q_{\chi}=-1$ and $Q_d + Q_{\bar{u}}=-1$ 
while leaving 
$Q_d-Q_{\bar{u}}=\overline{Q}$ arbitrary.
Since we use the bare fermion fields in this analysis, the virtual correction includes both the 1PI vertex corrections and the corrections to the two-point functions of the fermions.
Note that unlike our previous analysis~\cite{Ibe:2022lkl},
we use the Pauli-Villars regularization to avoid the ambiguity in 
four-dimensional anti-symmetric tensor, $\epsilon_{\mu\nu\rho\sigma}$, in generic space-time dimension, which appears in the reduction of the UV divergent vertex correction.

By comparing the amplitude in the four-fermion theory~\eqref{eq:virtual FF} with 
the vector-like part of the amplitude in the electroweak theory~\eqref{eq: EW virtual corrections}, 
we get two independent matching conditions
since the amplitude in the four-fermion theory~\eqref{eq:virtual FF} depends on the charge difference
$\overline{Q}$:\,\footnote{In this matching procedure,
we also introduce $\overline{Q}$ in the electroweak theory
although the vector-like part of the resultant amplitude~\eqref{eq: EW virtual corrections} does
not depend on $\overline{Q}$.}
\begin{align}
    \label{eq:fwd+fwbaru}
    -e^2\frac{f_{\chi d}+f_{\chi\bar{u}}}{2}
     &=0\ ; \\
   \label{eq:FFEWmatching}  
     e^2(f_{\chi\chi} + f_{d\bar{u}} + f_{\chi d}/2-f_{\chi \bar{u}}/2) 
     &= \frac{\alpha}{4\pi}
     \left[
    -\frac{3}{2}\qty(
    \log\frac{\Lambda_{\mathrm{FF}}^2}{m_W^2}-1)\right.\cr
    &\left.\hspace{20pt}
    + F_V^{\mathrm{Virtual(EW)}}\qty(\frac{m_W}{m_{\chi^\pm}}) 
     - \qty(\frac{6}{s_W^2}+\frac{7-4 s_W^2}{s_W^4}\log c_W)
    \right]\ .
\end{align}
These counterterms incorporate
the electroweak corrections into the four-fermion theory.

\subsection{Matching ChPT with Four-Fermion Theory}
\label{sec: matching chpt with FF}
We match the four-fermion theory obtained in the previous step with the extended ChPT including the charginos.
We implement the procedure following D\&M's method.
This step is essential as it yields integral representations of the LECs, 
which encapsulate vital information about the strong dynamics.

\subsubsection{Contributions from Low-Energy Constants to Branching Fraction}
\label{sec: LEC contribution to Branching}
The correction of the decay rate depends on three kinds 
of the LECs: $K_i$ for virtual photon 
introduced by Urech~\cite{Urech:1994hd}; $X_i$ for 
virtual leptons enumerated by Knecht et al.~\cite{Knecht:1999ag}; $Y_i$ for virtual charginos obtained in parallel with $X_i$~\cite{Ibe:2022lkl}.
The radiative correction of the chargino/pion's decay rates
depends on those LECs as 
\begin{align}
    \label{eq:structure dependent Wino}
    \left.\frac{\delta\Gamma_{\chi\to\pi}}{\Gamma_{\chi\to\pi}} \right|_\mathrm{LEC}
    &=
    e^2 
    \Bigg[\frac{8}{3}K_1 + \frac{20}{9}K_5+4 K_{12}\cr
    &\hspace{3cm}
    -\hat{Y}_6 -\frac{4}{3} 
    (Y_1+\hat{Y}_1)
    -4 \left(Y_2+\hat{Y}_2
    - \frac{m_{{\chi^\pm}}}{\mathit{\Delta} m_\pm} Y_3\right)
    \Bigg]\ ; \\
    \label{eq:structure dependent pion}
    \left.\frac{\delta\Gamma_\pi}{\Gamma_{\pi}}\right|_\mathrm{LEC}
    &=
    e^2 
    \Bigg[\frac{8}{3}K_1 + \frac{20}{9}K_5+4 K_{12}\cr &
    \hspace{3cm}-\hat{X}_6 -\frac{4}{3} (X_1+\hat{X}_1)
    -4 (X_2+ \hat{X}_2)
    +4  X_3
    \Bigg]\ .
\end{align}
For the definitions of these LECs, consult subsection\,5.1 of Ref.~\cite{Ibe:2022lkl}. 
Here, we show LECs needed to cancel UV divergences
arising from virtual photon exchange
without taking the ratio between
the chargino and pion decay rates.\,\footnote{Although we included $K_{2,6}$ contributions in our previous analysis~\cite{Ibe:2022lkl},
they do not affect the final result
and have been dropped here.}

We would like to stress that we do not have to determine these LECs individually.
In addition, as we can see from Eqs.~\eqref{eq:structure dependent Wino} and \eqref{eq:structure dependent pion},
the radiative corrections to the 
branching ratio~\eqref{eq: tree-level branching of single pion mode}
does not depend on $K_i$ at all and are free from uncertainties from these LECs.
In our analysis, therefore,
the NLO decay rate is given in the form, 
\begin{align}
    \label{eq: form of NLO single pion}
    \Gamma_\mathrm{NLO}\qty(\chi^-\to\chi^0+\pi^-(+\gamma))
    = \Gamma_\mathrm{LO}\qty(\chi^-\to\chi^0+\pi^-)\times
        \qty(1+
        \frac{\delta\Gamma_{\chi\to\pi}}{\Gamma_{\chi\to\pi}} - \frac{\delta\Gamma_\pi}{\Gamma_\pi})\ ,
\end{align}
where
\begin{align}
    \Gamma_{\chi\to\pi}
    :=\Gamma_\mathrm{tree}\qty(\chi^-\to\chi^0+\pi^-)\ ;\quad
    \Gamma_\pi
    :=\Gamma_\mathrm{tree}\qty(\pi^-\to\mu^-+\overline{\nu_\mu})\ ,
\end{align}
and $\delta\Gamma_{\chi\to\pi}$ and $\delta\Gamma_\pi$ stand for the radiative corrections
to these tree-level decay rates.

\subsubsection{Current Correlator with Minimal Resonance}
To estimate numerically the above LECs' contributions, we
have to introduce a model to describe hadronization effects. 
From the discussion in the previous section,
we find that the combinations of LECs relevant for our discussion are given by
$X_2+\hat{X}_2-X_3$,
$X_1+\hat{X}_1$,
$Y_2+\hat{Y}_2-\qty(m_{\chi^\pm}/\mathit{\Delta}m_\pm)Y_3$,
and $Y_1+\hat{Y}_1$.
By using D\&M's method, 
we find that 
$X_2+\hat{X}_2-X_3$ and
$X_1+\hat{X}_1$
are related to the current correlation functions
in the momentum space,
\begin{align}
    \label{eq: VA with initial pion}
 i f^{abc}\Gamma_{VA}^{\mu\nu}(
 \ell,r):=\int d^4x e^{i\ell x} 
    \langle 0|T J_{V}^{b\mu}(x) J_{A}^{c\nu}(0) | \pi^a(r)\rangle
\end{align}
and
\begin{align}
\label{eq: VV with initial pion}
    d^{abc}\Gamma_{VV}^{\mu\nu}(
 \ell,r)&:=\int d^4x e^{i\ell x} 
    \langle 0|T J_{V}^{b\mu}(x) J_{V}^{c\nu}(0) | \pi^a(r)\rangle\ ,
\end{align}
respectively.
Here, $r$ is the pion momentum.
The vector- and the axial-vector currents
of the light quarks $\Psi_q=\qty(\Psi_u, \Psi_d,\Psi_s)^T$
are 
given by
\begin{align}
    J_V^{a\mu} = \overline{\Psi}_q\gamma^\mu t^a\Psi_q\ ;\quad
    J_A^{a\mu} = \overline{\Psi}_q\gamma^\mu\gamma^5 t^a\Psi_q\ ,
\end{align}
where $t^a=\lambda^a/2$ with $\lambda^a$ being 
the Gell-Mann matrix, and
$f^{abc}$ and $d^{abc}$ are the structure constants given by,
\begin{align}
    f^{abc} = -2i
    \tr(t^a[t^b,t^c])\ , \quad d^{abc} =2\tr(t^a\{t^b,t^c\})\ .
\end{align}
Similarly,
the combinations relevant for the chargino decay,
$Y_2+\hat{Y}_2-\qty(m_{\chi^\pm}/\mathit{\Delta}m_\pm)Y_3$
and $Y_1+\hat{Y}_1$, can be 
connected to 
the current correlation functions
\begin{align}
    \label{eq: VA with final pion}
 i f^{abc}\overline{\Gamma}_{VA}^{\mu\nu}(
 \ell,r):=\int d^4x e^{-i\ell x} 
    \langle \pi^a(r)|T J_{V}^{b\mu}(x) J_{A}^{c\nu}(0) | 0\rangle
\end{align}
and
\begin{align}
\label{eq: VV with final pion}
    d^{abc}\overline{\Gamma}_{VV}^{\mu\nu}(
 \ell,r)&:=\int d^4x e^{-i\ell x} 
    \langle \pi^a(r) |T J_{V}^{b\mu}(x) J_{V}^{c\nu}(0) | 0\rangle\ ,
\end{align}
respectively.
These correlators can be related to Eqs.~\eqref{eq: VA with initial pion}
and \eqref{eq: VV with initial pion} via the crossing symmetry,
\begin{align}
    \overline{\Gamma}_{VA/VV}^{\mu\nu}(\ell, r) = 
    \Gamma_{VA/VV}^{\mu\nu}(-\ell, -r)\ .
\end{align}

As in the Wino case, we adopt the minimal consistent resonance model (MRM)~\cite{Weinberg:1967kj,Moussallam:1994xp,Moussallam:1997xx,Knecht:2001xc}
to estimate these current correlators.
In this model, a current correlation function is represented 
by rational functions with a finite number of 
mesons. 
In particular, for the correlation functions~\eqref{eq: VA with initial pion} and \eqref{eq: VV with initial pion}, 
by including the resonances corresponding to $\rho(770)$ 
and $a_1(1260)$, 
we can give interpolation functions
such that they satisfy the asymptotic constraints put by the operator product expansion
(OPE).
Hence, the current correlators~\eqref{eq: VA with initial pion} and \eqref{eq: VV with initial pion}
are characterized by two mass parameters $M_V$ and $M_A$.
The explicit form of $\Gamma_{VA}^{\mu\nu}$ in the MRM~\cite{Moussallam:1997xx,Knecht:2001xc} is given by
\begin{align}
    \Gamma_{VA}^{\mu\nu}(k,p) = 
    F_0\Bigg[
    \frac{(k^\mu+2 q^\mu)q^\nu}{q^2} - g^{\mu\nu} + F(k^2,q^2) P^{\mu\nu}
    + G(k^2,q^2) Q^{\mu\nu}
    \Bigg] \ ,
\end{align}
where $q= p-k$ with
\begin{align}
    P^{\mu\nu} = q^{\mu} k^{\nu} 
    - (k\cdot q)g^{\mu\nu}\ , 
    \quad
    Q^{\mu\nu} =k^2 q^\mu q^\nu
    +q^2 k^\mu k^\nu - (k\cdot q)
    k^\mu q^\nu - k^2 q^2 g^{\mu\nu} \ ,
\end{align}
and 
\begin{align}
    F(k^2,q^2) = \frac{k^2-q^2+2(M_A^2-M_V^2)}{2(k^2-M_V^2)(q^2 - M_A^2)} 
    \ ,  \quad
     G(k^2,q^2) = \frac{-q^2+2M_A^2}{(k^2-M_V^2)(q^2 - M_A^2)q^2} \ .
\end{align}
The MRM expression of $\Gamma_{VV}^{\mu\nu}$~\cite{Moussallam:1994xp,Knecht:2001xc} 
is provided by
\begin{align}
    \Gamma_{VV}^{\mu\nu}(k, p) = i F_0 \varepsilon^{\mu\nu\rho\sigma}k_{\rho}
    p_\sigma \Gamma_{VV}(k, p)\ ,
\end{align}
with 
\begin{align}
    \Gamma_{VV}(k, p) = 
    \frac{2k^2-2k\cdot p - c_V}{2(k^2 - M_V^2)((p-k)^2 - M_V^2)}\ .
\end{align}
The value of $c_V$ is given by
\begin{align}
\label{eq:cV}
    c_V = \frac{N_c M_V^4}{4\pi^2 F_0^2}\ ,
\end{align}
which is determined by the coupling of the Wess-Zumino-Witten term.

In our analysis, we consider $M_V$ and $M_A$ as model parameters,
and they will be determined under the constraint $M_A = \sqrt{2}M_V$
by fitting with the result of lattice calculations~\cite{Boyle:2009xi}.
Typically, $M_V$ takes a value around the $\rho$ meson mass $m_\rho\simeq 0.77$\,GeV.

\subsubsection{Matching Conditions}
By applying D\&M's method, we can work out
the matching conditions for the combinations of the LECs,
$Y_2+\hat{Y}_2-\qty(m_{\chi^\pm}/\mathit{\Delta}m_\pm)Y_3$
and $Y_1+\hat{Y}_1$.
See Sec.~5 of Ref.~\cite{Ibe:2022lkl} for a detailed explanation
of how to determine these LEC combinations.
Here we list the resultant matching conditions.

The combination of LECs 
$Y_1+\hat{Y}_1$ does not provide any contribution to the decay rate
thanks to the matching condition~\eqref{eq:fwd+fwbaru}:
\begin{align}
\label{eq:Y1condition}
   e^2(Y_1 + \hat{Y}_1) = \frac{1}{4}e^2 (f_{\chi d}+f_{\chi \bar{u}}) = 0\ .
\end{align}
On the other hand,
the combination $Y_2+\hat{Y}_2-\qty(m_{\chi^\pm}/\mathit{\Delta}m_\pm)Y_3$
is given by
\begin{align}
\label{eq:Y2 and Y3 condition}
   -&e^2 \left(Y_2+\hat{Y}_2 -\frac{m_{{\chi^\pm}}}{\mathit{\Delta} m_\pm} Y_3\right) -
   \frac{\alpha}{8\pi}
   \frac{m_{{\chi^\pm}}}{\mathit{\Delta}m_\pm} 
   \left(
   \frac{3}{\bar{\epsilon}_\mathrm{ChPT}} + 
   3 \log\frac{\mu_\mathrm{ChPT}^2}{m_{{\chi^\pm}}^2}+4
   \right)
   \cr
    &
   +\frac{\alpha}{16\pi}
  \left(
   \frac{5}{\bar{\epsilon}_\mathrm{ChPT}} + 
    5\log\frac{\mu_\mathrm{ChPT}^2}{m_{{\chi^\pm}}^2}
   + \log\frac{m_{{\chi^\pm}}^2}{m_\gamma^2}
   +\frac{9}{2}
   -\frac{5}{3}\pi^2 
   -4\log\frac{m_\gamma^2} {m_{\chi^\pm}^2}
   -
   \log^2\frac{m_\gamma^2}{4 \qty(\mathit{\Delta}m_\pm)^2}
   \right) 
   \cr
   =\,
  &\frac{e^2}{4}( f_{\chi d} - f_{\chi \bar{u}})
  +\frac{\alpha}{16\pi}f_{VW}(\mathit{\Delta}m_\pm,M_A,M_V)\cr
  &+\frac{\alpha}{16\pi}\qty[
  5\log\frac{\Lambda_\mathrm{FF}^2}{m_{\chi^\pm}^2}
     -\frac{9}{2}
    + \log\frac{m_{{\chi^\pm}}^2}{m_\gamma^2}
    + \log\frac{m_{{\chi^\pm}}^2}{M_V^2} - 4
    \log\frac{m_\gamma^2}{M_V^2}
- \log^2\frac{m_\gamma^2}{4 (\mathit{\Delta}m_\pm)^2}] \ ,
\end{align}
where $f_{VW}$ is very complicated function of $\mathit{\Delta}m_\pm$ and $M_{A, V}$.
Here, we have used the dimensional regularization with $d=4-2\epsilon_\mathrm{ChPT}$ on the ChPT side,
and
have obtained expression with 
the UV pole
$\qty(\bar{\epsilon}_\mathrm{ChPT})^{-1} = \epsilon_\mathrm{ChPT}^{-1}-\gamma_E+\log 4\pi$
and the arbitrary mass parameter $\mu_\mathrm{ChPT}$.
On the four-fermion theory side, 
we have regularized the loop integral
by the Pauli-Villars mass parameter for the same reason
explained in Sec.~\ref{sec: matching FF with EW},
unlike our previous analysis~\cite{Ibe:2022lkl}.

Note that since in ChPT we prepare 
only
LECs corresponding to operators without
derivatives on the chargino or neutralino,
we treat
the terms higher than 
$O\qty((\mathit{\Delta}m_\pm/M_{A,V})^0)$ 
in the function $f_{VW}$
as uncertainty of our analysis.
For $\mathit{\Delta}m_\pm\ll M_{V,A}$,
the function $f_{VW}$
is approximated as
\begin{align}
    \label{eq: fVW expansion}
    f_{VW}\qty(\mathit{\Delta}m_\pm, M_A, M_V) 
    = &-\frac{4}{\mathit{\Delta}m_\pm}\frac{\pi M_AM_V}{M_A+M_V}\cr
    &-\frac{5\pi^2}{3}+\frac{6M_A^2-9M_V^2}{M_A^2-M_V^2} 
    + 3\frac{M_V^4}{\qty(M_A^2-M_V^2)^2}\log\frac{M_A^2}{M_V^2}
    +O\qty(\frac{\mathit{\Delta}m_\pm}{M_{A,V}})\ .
\end{align}
We simply 
use the expression~\eqref{eq: fVW expansion}
even in the region $\mathit{\Delta}m_\pm \gtrsim M_{A,V}$ 
by neglecting 
$O\qty(\mathit{\Delta}m_\pm/M_{A,V})$ 
contributions to $f_{VW}$.
We will discuss uncertainty of the decay rate caused by the approximation of $f_{VW}$
at the end of subsection~\ref{sec: pion NLO}.

The LEC $\hat{Y}_6$ is determined by matching of the wave-function renormalization of the chargino, 
\begin{align}
    \label{eq:Y6condition}
    e^2\hat{Y}_6 + \frac{\alpha}{4\pi}\frac{1}{\bar{\epsilon}_{\mathrm{ChPT}}} 
    = -2e^2 f_{\chi\chi}+\frac{\alpha}{4\pi}\qty(\log\frac{\Lambda_\mathrm{FF}^2}{\mu_\mathrm{ChPT}^2}+\frac{1}{2})  \ .
\end{align}
Combining the above conditions, the contributions of $Y_i$ can be 
expressed in terms of the counterterms of the four-fermion theory as
\begin{align}
    \left.\frac{\delta\Gamma_{\chi\to\pi}}{\Gamma_{\chi\to\pi}}\right|_\mathrm{LEC}
    &\supset e^2\qty[-\hat{Y}_6 -\frac{4}{3} 
    (Y_1+\hat{Y}_1)
    -4 \qty(Y_2+\hat{Y}_2
    - \frac{m_{{\chi^\pm}}}{\mathit{\Delta} m_\pm} Y_3)]\cr
    &=
    e^2\qty(2f_{\chi\chi}+f_{\chi d}-f_{\chi\bar{u}}) 
    + \frac{\alpha}{4\pi}\cdot 4\log\frac{\Lambda_\mathrm{FF}^2}{m_W^2}\cr
    &\hspace{20pt}
    -\frac{\alpha}{4\pi}\cdot 
    \qty(4-6\frac{m_{\chi^\pm}}{\mathit{\Delta}m_\pm})\qty(\frac{1}{\bar{\epsilon}_\mathrm{ChPT}}+\log\frac{\mu_\mathrm{ChPT}^2}{m_{\chi^\pm}^2})
    +\mathrm{(finite\,\,terms)}\ .
\end{align}

In order to match ChPT with the electroweak theory by using the matching condition~\eqref{eq:FFEWmatching}, we need the matching condition for the LEC $K_{12}$,\,\footnote{In Ref.\,\cite{Ibe:2022lkl}, we erroneously interpreted the subtraction scheme for the $K$-terms and $X$-terms in Ref.\,\cite{Descotes-Genon:2005wrq} as the minimal subtraction, resulting in the incorrect constant terms 
in Eqs.\,(7.13), (7.14) and (7.16) 
in Ref.\,\cite{Ibe:2022lkl}.
The effects on the numerical results are, however, 
negligible compared to the other uncertainties associated with the hadron models. See also Appendix~\ref{app: not MSbar}.
}
\begin{align}
\label{eq:K12}
 &e^2K_{12}  
 +\frac{\alpha}{4\pi}\frac{1}{8}
 \qty(\frac{1}{\bar{\epsilon}_\mathrm{ChPT}}+\log\frac{\mu_\mathrm{ChPT}^2}{m_W^2})\cr
 =\,  &\frac{1}{2}e^2f_{d\bar{u}}\cr
 &+ 
 \frac{\alpha}{4\pi}
 \frac{1}{8}\left(-2\log\frac{\Lambda_\mathrm{FF}^2}{m_W^2}
 -3 \log \frac{m_W^2}{M_V^2}
 + \frac{3(M_A^2 + M_V^2)M_V^2}{(M_A^2-M_V^2)^2}
 \log\frac{M_A^2}{M_V^2} - \frac{6 M_A^2}{M_A^2 - M_V^2}
 - \frac{1}{2}\right)\ .
\end{align}
By using this condition, the LEC contribution to the decay rate can be expressed as
\begin{align}
    \label{eq: matching result in chargino decay rate}
    \left.\frac{\delta\Gamma_{\chi\to\pi}}{\Gamma_{\chi\to\pi}}\right|_\mathrm{LEC}
    &\supset e^2\qty[4K_{12}-\hat{Y}_6 -\frac{4}{3} 
    (Y_1+\hat{Y}_1)
    -4 \qty(Y_2+\hat{Y}_2
    - \frac{m_{{\chi^\pm}}}{\mathit{\Delta} m_\pm} Y_3)]\cr
    &=
    e^2\qty(2f_{\chi\chi}+f_{\chi d}-f_{\chi\bar{u}}+2f_{d\bar{u}}) 
    + \frac{\alpha}{4\pi}\cdot 3\log\frac{\Lambda_\mathrm{FF}^2}{m_W^2}\cr
    &\hspace{20pt}-\frac{\alpha}{4\pi}\cdot 
    \qty(\frac{9}{2}-6\frac{m_{\chi^\pm}}{\mathit{\Delta}m_\pm})\qty(\frac{1}{\bar{\epsilon}_\mathrm{ChPT}}+\log\frac{\mu_\mathrm{ChPT}^2}{m_{\chi^\pm}^2})
    +\mathrm{(finite\,\,terms)}\ .
\end{align}
From Eq.~\eqref{eq:FFEWmatching}, we see that the above combination is
free from the Pauli-Villars regulator mass. The remaining contributions from 
$K_1$ and $K_5$
in Eqs.~\eqref{eq:structure dependent Wino}
and \eqref{eq:structure dependent pion} cancel each other in the correction to
the branching ratio $\Gamma_{\chi\to\pi}/\Gamma_\pi$.

\subsection{NLO Decay Rate}
\label{sec: pion NLO}
\subsubsection{Corrections to Chargino Decay Rate}
To estimate the NLO rate of the chargino decay into the single pion,
we include long-distance virtual photon exchanges.
They consist of wave-function renormalizations
and 1PI virtual photon corrections.
Note that we do not need to 
include virtual meson loops since
they contribute to the chargino decay 
in the same way as the pion decay
and will be canceled in the branching ratio~\eqref{eq: form of NLO single pion}.
In our analysis, therefore, the virtual corrections
are composed of virtual photon loops (see Fig.~14 in Ref.~\cite{Ibe:2022lkl}).
The UV divergences will be subtracted by 
the LECs' contributions determined by Eqs.~\eqref{eq: matching result in chargino decay rate} and~\eqref{eq:FFEWmatching}.

In addition, 
we include the real emission processes (see Fig.~15 in Ref.~\cite{Ibe:2022lkl}),
in order to see the cancellation of infrared divergences
and collinear singularities between real photons
and the on-shell pion.
The resultant radiative correction to the decay
rate is given by
\begin{align}
\label{eq:dGammachi}
 \frac{\delta\Gamma_{\chi\to\pi}}
   {\Gamma_{\chi\to\pi}}
     =\,&  \frac{\alpha}{4\pi}
    \qty[3\log\frac{m_{{\chi^\pm}}^2} {\mathit{\Delta} m_\pm^2}
    + 4f_{{\chi}}\left(\frac{m_{\pi^\pm}}{\mathit{\Delta} m_\pm}\right)]
    \cr
    &+\frac{\alpha}{4\pi}
    \qty[3 \left(\frac{1}{\bar{\epsilon}_{\mathrm{ChPT}}}+\log\frac{\mu_\mathrm{ChPT}^2}{m_{\pi^\pm}^2}\right)
    - \frac{6m_{{\chi^\pm}}}{\mathit{\Delta} m_\pm}\left(\frac{1}{\bar{\epsilon}_\mathrm{ChPT}}+\log\frac{\mu_\mathrm{ChPT}^2}{m_{{\chi^\pm}}^2}
    + \frac{4}{3}\right)]
    \cr
    & + 
    e^2 \qty(\frac{8}{3}K_1 + \frac{20}{9}K_5)\cr
    & + 
    e^2 
    \qty[4K_{12}-\hat{Y}_6 -\frac{4}{3} (Y_1+\hat{Y}_1)
    -4 \left(Y_2+ \hat{Y}_2 -
     \frac{m_{{\chi^\pm}}}{\mathit{\Delta} m_\pm} Y_3\right)]\ ,
\end{align}
where $f_\chi$ is a function arising from the calculation of the real emission process,
\begin{align}
    \label{eq:fchi}
    f_{{\chi}}(z) =\, &
    5 +\frac{5}{2}\log z -2 \log(4(1-z^2)) \cr
    &+\frac{1}{\sqrt{1-z^2}}
    \Bigg\{-\frac{2\pi^2}{3}-\frac{1}{2}\log^2 2 +(1-2\log 2) \log z +6\log^2 z \cr
    &\hspace{30pt}-\frac{1}{2} (\log 2 + 2\log z) \log(1-z^2)-\frac{1}{8} \log^2\left(1-z^2\right) \cr
    &\hspace{30pt}+\frac{1}{2}\qty[-2 + 6\log 2 +3 \log(1-z^2)-20\log z]\log(1-\sqrt{1-z^2}) \cr
    &\hspace{30pt}+\frac{7}{2}\log^2(1-\sqrt{1-z^2})-\mathrm{Li}_2\qty(\frac{1}{2}-\frac{1}{2\sqrt{1-z^2}})
    +3\mathrm{Li}_2\qty(\frac{1-\sqrt{1-z^2}}{1+\sqrt{1-z^2}})\Bigg\}\ .
\end{align}
The last line in Eq.~\eqref{eq:dGammachi} is determined by matching 
conditions~\eqref{eq: matching result in chargino decay rate} and~\eqref{eq:FFEWmatching}.
The LEC contribution in the third line of Eq.~\eqref{eq:dGammachi} subtracts
the UV divergences from virtual photon loops as
\begin{align}
    e^2 
    \qty(\frac{8}{3}K_1+\frac{20}{9}K_5)
    =
    \frac{\alpha}{4\pi}\frac{3}{2}
    \qty(\frac{1}{\bar{\epsilon}_\mathrm{ChPT}}+1)
    +e^2 
    \qty[\frac{8}{3} K_1^r(\mu_\mathrm{ChPT})
    + \frac{20}{9} K_5^r(\mu_\mathrm{ChPT})]\ ,
\end{align}
where $K^r_{1,5}(\mu_\mathrm{ChPT})$ are the finite parts.
By combining the UV pole with the one appearing
in Eq.~\eqref{eq: matching result in chargino decay rate},
we can see that the radiative correction~\eqref{eq:dGammachi} is UV-finite.
Note that
since
the same combination from $K_{1,5}$
appears in the correction to the $\pi_{\ell2}$ decay rate,
we do not need explicit expressions for $K_{1,5}^r(\mu_\mathrm{ChPT})$.

By substitute the explicit expressions of the finite parts of the LECs, 
we obtain the NLO correction to the decay rate of the chargino, 
\begin{align}
\label{eq:dGamchi}
   \frac{\delta\Gamma_{\chi\to\pi}}
   {\Gamma_{\chi\to\pi}}
   =\, &\frac{\alpha}{4\pi}
   \left\{-\frac{4\pi M_AM_V}{\mathit{\Delta}m_\pm (M_A+M_V)}
   + \frac{1}{4} g_\chi\qty(\frac{M_V}{M_A},\frac{\mathit{\Delta}m_\pm}{M_A})+
    4f_\chi\qty(\frac{m_{\pi^\pm}}{\mathit{\Delta}m_\pm})\right.\cr
    &\hspace{20pt}\left.
     -\frac{3}{2}+2F_V^{\mathrm{Virtual(EW)}}(r_W) 
    - 2\qty(\frac{6}{s_W^2}+\frac{7-4 s_W^2}{s_W^4}\log c_W) \right.\cr
    &\hspace{20pt}\left.
    +(4\pi)^2\qty[\frac{8}{3}K_1^r(\mu_\mathrm{ChPT})
    +\frac{20}{9}K_5^r(\mu_\mathrm{ChPT})] \right.\cr 
    &\hspace{20pt}\left. 
    + \frac{3}{2}\log\frac{\mu_\mathrm{ChPT}^2}{M_V^2} 
    + 3\log \frac{m_W^2}{\mu_\mathrm{ChPT}^2}
    + \log\qty(\frac{(\mathit{\Delta}m_\pm)^2 M_V^4}{m_{\pi^\pm}^6}) 
    + 8\log 2
    \right\}\ ,
\end{align}
where we have defined the function
\begin{align}
    g_{{\chi}}(\zeta,\eta) =\, & 
    -\frac{3\qty[1+6 \zeta ^2-7 \zeta ^4+4 \qty(\zeta ^2+3 \zeta ^4) \log\zeta]}{\qty(1-\zeta^2)^2}
    -16 \log(\frac{4 \eta ^2}{\zeta ^2})\ .
\end{align}

We can see that the radiative correction~\eqref{eq:dGamchi}
has no infrared divergences.
Moreover, 
Eq.~\eqref{eq:dGamchi} remains
finite in the limit of $m_{\pi^\pm}\to 0$ (Note that
the $K$-terms do not introduce dependence on the pion mass
since they are determined in the chiral limit).
That is, our result is free from collinear
divergences and respects the KLN theorem as in the case of the leptonic decay.

\subsubsection{Ratio Between Chargino and Pion Decay Rates}
Similarly, we can compute the radiative corrections to $\pi_{\ell2}$ decay rate.
See subsection 7.2 of Ref.~\cite{Ibe:2022lkl} for computational detail.
The total radiative correction to the branching fraction is given by
\begin{align}
    \label{eq:analyticNLO}
     &\frac{\delta\Gamma_{{\chi\to\pi}}}{\Gamma_{{\chi\to\pi}}}
     - \frac{\delta\Gamma_{\pi}}{\Gamma_{\pi}}\cr
     =\, &\frac{\alpha}{4\pi}\left\{
     6\log\frac{m_W}{\mathit{\Delta}m_\pm}
     -8\log\frac{m_Z}{\mathit{\Delta}m_\pm}
     \right.\cr
    &\left.+2F_V^{\mathrm{Virtual(EW)}}\qty(\frac{m_W}{m_{\chi^\pm}}) 
    - 2\qty(\frac{6}{s_W^2}+\frac{7-4s_W^2}{s_W^4}\log c_W)\right.\cr
     &\left.-\frac{4\pi M_AM_V}{\mathit{\Delta}m_\pm (M_A+M_V)}
   + \frac{1}{4}\qty[g_\chi\qty(\frac{M_V}{M_A},\frac{\mathit{\Delta}m_\pm}{M_A})
   -g_\pi\qty(\frac{M_V}{M_A})]
   +12\log\frac{M_V}{\mathit{\Delta}m_\pm}
   \right.\cr
    &\left.+4f_\chi\qty(\frac{m_{\pi^\pm}}{\mathit{\Delta}m_\pm})
    -
    \qty[12\log \frac{m_\mu}{\mathit{\Delta}m_\pm}+9-8\log2-\frac{4}{3}\pi^2 + 2f_{\pi}\left(\frac{m_\mu}{m_{\pi^\pm}}\right)]
     \right\}\ .
\end{align}
Here, $F_V^{\mathrm{Virtual(EW)}}(m_W/m_{\chi^\pm})$ 
represents the non-logarithmic electroweak correction, 
which is applicable
to other $\mathrm{SU}(2)\times \mathrm{U}(1)_Y$ 
representations.
The other functions $f_\pi$ and $g_\pi$ stem from the radiative corrections
to $\pi_{\ell2}$ decay rate. They are given by
\begin{align}
        f_\pi(r)=\,&4 \left(\frac{1+r^2}{1-r^2}\log r - 1\right)\log(1-r^2) + 4 \frac{1+r^2}{1-r^2}
    \mathrm{Li}_2(r^2) \cr
& -\frac{r^2(8-5r^2)}{(1-r^2)^2}\log r - \frac{r^2}{1-r^2} \left(\frac{3}{2}+\frac{4}{3}\pi^2\right)\ ;\cr
    g_{\pi}(\zeta) =\,& -19-\frac{36 \zeta ^2 \log \zeta }{1-\zeta ^2}+2\hat{c}_V\ .
\end{align}
The constant $\hat{c}_V$ is related to the parameter $c_V$~\eqref{eq:cV}
by the relation $\hat{c}_V = c_V/M_V^2$.

The RHS of Eq.~\eqref{eq:analyticNLO} 
can be understood line by line as follows.
The first line is the difference in the leading-log contributions
between the chargino and the pion decay.
The second line arises
because the non-logarithmic electroweak contribution 
to the chargino decay
is different from that to the muon decay.
The $S_\mathrm{EW}^{\chi}$ factor~\eqref{eq: SEW}
is included in the first and second lines.
The third line represents the contributions which depend
on the parameters of the hadron model (the MRM in our computation).
The remaining long-distance corrections are
collected in the fourth line.\,\footnote{The mass singularities appearing in the final line are artificial ones
caused by normalizing the $\pi_{\ell2}$ decay rate.
They do not cause 
physical singularities 
in the $\pi_{\ell2}$ decay
since its decay rate 
is proportional to $m_\mu^2 m_{\pi^\pm}$.}

\subsubsection{Estimation of Error from Hadron Model}
\label{sec: pion error}
First, we consider the uncertainties when the
mass difference is sufficiently
smaller than resonance masses,
$\mathit{\Delta}m_\pm \ll M_V$,
and hence ChPT works well.
In this region, we follow the estimation 
provided in our previous analysis (see subsection~7.3 of Ref.~\cite{Ibe:2022lkl}).
Here, we state briefly how to evaluate those errors. 
For those small mass difference, dominant error comes from the numerically dominant term in the correction,
\begin{align}
    \label{eq: MRM leading}
     \frac{\delta\Gamma_{{\chi\to\pi}}}{\Gamma_{{\chi\to\pi}}}- \frac{\delta\Gamma_{\pi}}{\Gamma_{\pi}}
     \supset -\frac{\alpha}{4\pi}
     \frac{4\pi M_AM_V}{\mathit{\Delta}m_\pm (M_A+M_V)}\ .
\end{align}
This term provides the leading hadronic contribution, whose uncertainty we estimate
by varying $M_V$ in the range of $[0.6, 0.8]$ GeV while keeping $M_A = \sqrt{2}M_V$.
This range was determined by comparing two-point current correlator in the MRM with the lattice simulation~\cite{Boyle:2009xi}.
Following D\&M. 
we put $\pm50\%$ error on
the contributions from the MRM, 
\begin{align}
    \label{eq: MRM subleading}
     \frac{\delta\Gamma_{{\chi\to\pi}}}{\Gamma_{{\chi\to\pi}}}- \frac{\delta\Gamma_{\pi}}{\Gamma_{\pi}}
     \supset \frac{\alpha}{4\pi}\qty{
   \frac{1}{4}\qty[g_\chi\qty(\frac{M_V}{M_A},\frac{\mathit{\Delta}m_\pm}{M_A})-g_\pi\qty(\frac{M_V}{M_A})]
   }\ .
\end{align}

Next, let us consider the case
that the mass splitting is 
large, i.e.,
$\mathit{\Delta}m_\pm\gtrsim M_{V}$.
In this case, our estimation
using ChPT is not valid.
Nevertheless, we extrapolate Eq.~\eqref{eq:analyticNLO}
up to the higher mass difference
by neglecting 
$O\qty(\mathit{\Delta}m_\pm/M_{A,V})$ 
contributions from $f_{VW}$.
We take the difference between $f_{VW}$
and its expansion~\eqref{eq: fVW expansion} as an estimation
of the size of theoretical error, 
which ranges from 0.5\% to 1\%.
Alternatively, we can also estimate uncertainty
when the momentum transfer exceeds the valid range 
of ChPT, from the difference 
between
long-distance corrections in the $\pi_{\ell2}$ decay rate
and the tau decay rate into the single pion.%
\footnote{The difference can be estimated by using the experimental value of the tau decay rate as  $\delta \Gamma_{\tau\to\pi}/\Gamma_{\tau\to \pi} - \delta \Gamma_\pi/\Gamma_\pi = 
(\Gamma_\tau^{\mathrm{(exp)}}B(\tau \to \pi)-\Gamma_\mathrm{LO}(\tau\to \pi))/\Gamma_\mathrm{LO}(\tau\to \pi)$.}
These two estimates provide similar size of the uncertainty 
in the chargino decay rate into the single pion.

\subsection{Numerical Results}
\begin{figure}[t!]
	\centering
  	\subcaptionbox{
   \label{fig:single_rate}}
	{\includegraphics[width=0.47\textwidth]{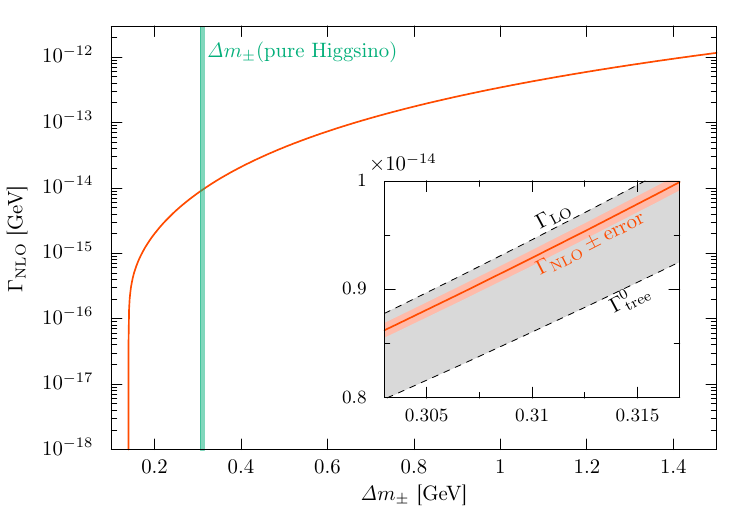}} 
 	\subcaptionbox{
  \label{fig:single_NLO_LO}}
	{\includegraphics[width=0.47\textwidth]{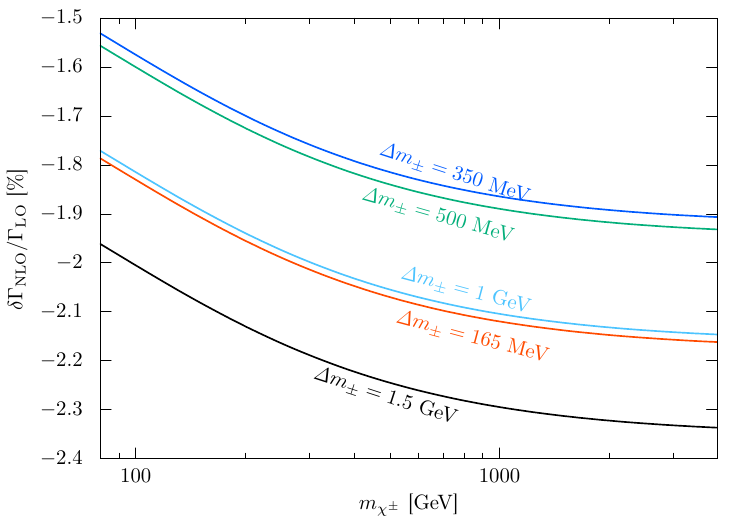}} 
 	\subcaptionbox{\label{fig:single100}}
	{\includegraphics[width=0.47\textwidth]{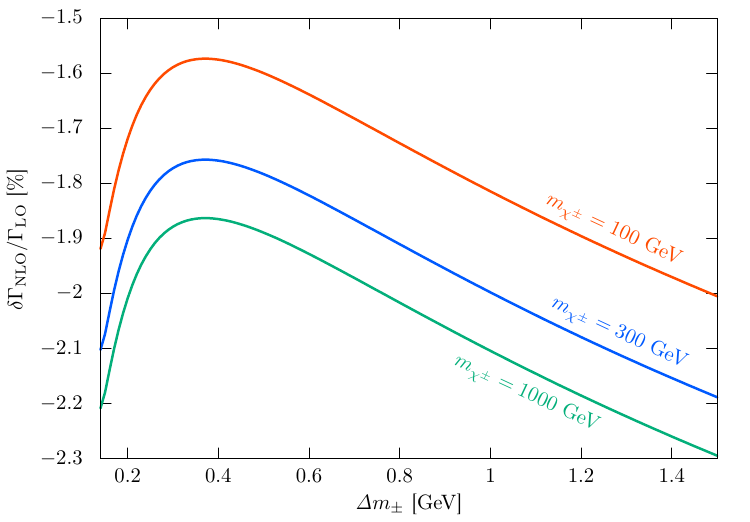}}
 	\subcaptionbox{\label{fig:single_error} }
	{\includegraphics[width=0.47\textwidth]{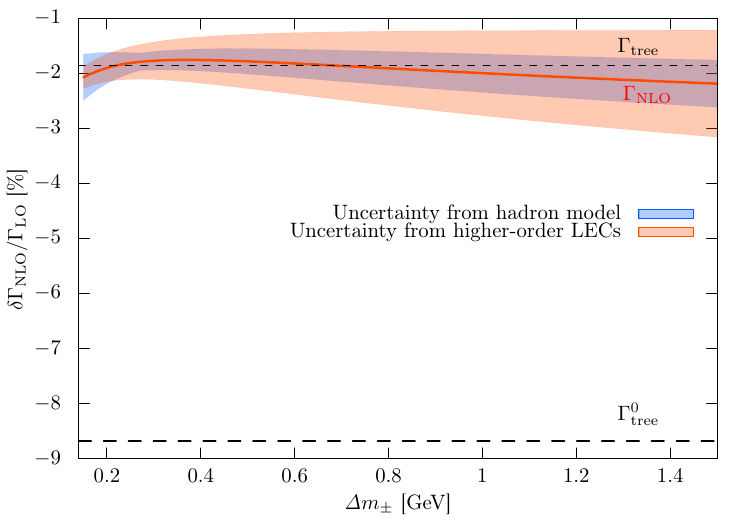}}

\caption{(a) 
The NLO decay rate as a function 
of the mass difference for the 300\,GeV charged Higgsino.
The inserted window is the close-up view 
for the mass difference in the limit 
of the pure Higgsino for $m_{\chi^{\pm}}= 300$\,GeV, $\mathit{\Delta}m_{\pm} \simeq 310$\,MeV.
The gray band represents the uncertainties of the tree-level approximation, which is reduced 
to the very thin orange band at the NLO computation.
(b)
The ratio of the NLO correction
to the leading-order decay rate~\eqref{eq: pion decay at leading by BF}
as a function of the chargino mass.
(c)
The ratio of the NLO correction
to the LO decay rate~\eqref{eq: pion decay at leading by BF}
as a function 
of the mass difference.
(d) 
The deviations of  
$\Gamma_\mathrm{tree}$~\eqref{eq: tree-level single pion rate},
$\Gamma_\mathrm{tree}^0$~\eqref{eq: Gamma0 single pion} and $\Gamma_\mathrm{NLO}$~\eqref{eq: form of NLO single pion} from $\Gamma_\mathrm{LO}$~\eqref{eq: pion decay at leading by BF}
together with the theoretical
uncertainties explained in subsubsection~\ref{sec: pion error}.
The NLO single pion decay rate of the 
Wino can be obtained just by multiplying 
the result in (a)
by four.
}\label{fig:single}
\end{figure}

In Fig.~\ref{fig:single_rate},
we show the NLO decay rate as a function 
of the mass difference for the 300\,GeV 
charged Higgsino.
The inserted window is the close-up view 
of the intersection between $\Gamma_\mathrm{NLO}$
and the vertical green band, 
which corresponds to the mass difference of
the 300\,GeV pure-Higgsino.
The gray band represents uncertainty
in the tree-level approximation explained in
Sec.~\ref{sec: tree ambiguity in pion};
$\Gamma^0_\mathrm{tree}$ and $\Gamma_\mathrm{LO}$
in the figure
are given by Eqs.~\eqref{eq: Gamma0 single pion}
and \eqref{eq: pion decay at leading by BF},
respectively.
In the close-up window, the NLO decay rate is shown as the orange line
accompanied by the very thin light-orange 
band representing the theoretical error.
Fig.~\ref{fig:single_NLO_LO} shows the ratio of the NLO correction
to the leading-order decay rate~\eqref{eq: pion decay at leading by BF}
as a function of the chargino mass for multiple mass differences.
In Fig.~\ref{fig:single100}, we show the same ratio
as a function of the mass difference
for $m_{\chi^\pm}=100, 300, \mathrm{and}\, 1000$\,GeV.
The figures show that the
NLO correction 
reduces the decay rate from $\Gamma_\mathrm{LO}$ 
by about 2\%.

In Fig.~\ref{fig:single_error}, we show 
deviations of  
$\Gamma_\mathrm{tree}$~\eqref{eq: tree-level single pion rate},
$\Gamma_\mathrm{tree}^0$~\eqref{eq: Gamma0 single pion},
and 
$\Gamma_\mathrm{NLO}$~\eqref{eq: form of NLO single pion}
from $\Gamma_\mathrm{LO}$~\eqref{eq: pion decay at leading by BF}
together with the theoretical
uncertainties explained in subsubsection~\ref{sec: pion error} for the 300\,GeV Higgsino.
The blue band represents theoretical error
from the minimal resonance model described by Eqs.~\eqref{eq: MRM leading} and \eqref{eq: MRM subleading}. 
The orange band indicates uncertainty
due to contributions from operators 
with derivatives on the Higgsinos which are not included in our analysis.
The former error is dominated by the latter one
for $\mathit{\Delta}m_\pm\gtrsim0.4\,\mathrm{GeV}$.
The combined uncertainty of
the NLO analysis 
is a few percent
for the mass difference less than 1.5\,GeV.
Compared with the uncertainty in the ``tree-level'' estimation of about 9\%, i.e.
the difference between $\Gamma_\mathrm{tree}^0$
and $\Gamma_\mathrm{LO}$,
the NLO analysis substantially reduces 
the uncertainty of the chargino decay rate into the single pion.

\section{Decay into Single Kaon and Multi-Meson}
\label{sec: multimeson}
When the mass difference $\mathit{\Delta} m_\pm$ exceeds roughly 0.5\,GeV, 
the branching fractions 
of the chargino decay into a single Kaon
as well as multi-meson become significant.
For such a large mass difference, the hadron 
resonances such as the $\rho$-mesons play significant roles.
In the following analysis, 
instead of relying on hadron models, we make use of 
an observed differential decay rate of
the tau lepton to calculate the hadronic chargino decay width.
These methods are discussed in Refs.~\cite{Chen:1996ap}.
We revisit these estimations by considering short-distance corrections from electroweak interactions and by using updated data for the hadronic tau lepton decay.
These effects result in a difference of several tens of percent from the previous estimations.

\subsection{Inclusive Decay Rates}
Let us consider a generic decay process,
$\chi^-(p_1)\rightarrow\chi^0(p_2)+\mathrm{mesons}$.
Here, $p_{1(2)}$ is the momentum of the chargino (neutralino).
At tree level, 
the decay amplitude is given by
\begin{align}
    \mathcal{M}(\chi^-\rightarrow\chi^0+ \mathrm{mesons})
    &= -2G_F V^*_{uD}\,\bar{u}_{\chi^0}(p_2)\gamma^\mu\qty(O^W_LP_L+O^W_RP_R)u_{\chi^-}(p_1)
    \mel{\mathrm{mesons}}{U_\mu(0)}{0}\ ,
\end{align}
where $V_{uD}(D=d,s)$ is the CKM matrix element, and $U_\mu(x)$ is the vector 
$\qty(U_\mu = V_\mu = \overline{\Psi}_D\gamma^\mu \Psi_u)$
or axial 
$\qty(U_\mu = A_\mu = \overline{\Psi}_D\gamma^\mu \gamma_5 \Psi_u)$ color-singlet quark current.
With the spectral functions $v_J(s)$ and $a_J(s)\, (J=0,1)$~\cite{Tsai:1971vv},
we can write the decay rate into vector ($V^-_J$) or axial ($A^-_J$) channel as 
\begin{align}
    \label{eq: tree-level multimeson rate}
    \Gamma_\mathrm{tree}(\chi^-\rightarrow\chi^0 + (V/A)^-_J)
    =&\, \frac{G_F^2\abs{V_{uD}}^2}{(2\pi)^3}
    \int ds\,\frac{s^3}{m_{\chi^\pm}^3}
    \lambda_{\chi^\pm\chi^0}^{1/2}(s)
    \times (v/a)_J(s)K_J(s)\ ,
\end{align}
where $s = (p_1-p_2)^2$ is the invariant mass of the multi-meson
system in the final state. 
For the normalization of the spectral functions,
see Appendix~\ref{sec: Hadronic Decay of Tau}.
The kinematical function $\lambda_{AB}(s)$ 
is given by 
\begin{align}
    \label{eq: inclusive kallenlambda}
    \lambda_{AB}(s) &=
    \lambda(1,m_A^2/s, m_B^2/s)\ ,
\end{align}
with $\lambda(a, b, c) = a^2+b^2+c^2-2ab-2bc-2ca$, and
\begin{align}
    &K_{J=1}(s)
    =
    \frac{1}{2}
    \left\{-6(O^W_LO^{W*}_R+O^{W*}_LO^W_R)
    \frac{m_{\chi^0} m_{\chi^\pm}}{s}
    \phantom{\qty[-\qty(\frac{m_{\chi^\pm}^2-m_{\chi^0}^2}{s})^2]}
    \right.\cr
    &\left.\hspace{120pt}
    +\qty(\abs{O^W_L}^2+\abs{O^W_R}^2)
    \qty[\frac{m_{\chi^\pm}^2+m_{\chi^0}^2}{s}-2
    +\qty(\frac{m_{\chi^\pm}^2-m_{\chi^0}^2}{s})^2]
    \right\}\ ;\\
    &K_{J=0}(s)
    =
    \frac{1}{2}
    \left\{2(O^W_LO^{W*}_R+O^{W*}_LO^W_R)
    \frac{m_{\chi^0} m_{\chi^\pm}}{s}
    \phantom{
    \qty[-\frac{m_{\chi^\pm}^2+m_{\chi^0}^2}{s}+\qty(\frac{m_{\chi^\pm}^2-m_{\chi^0}^2}{s})^2]}
    \right.\cr
    &\left.\hspace{120pt}
    +\qty(\abs{O^W_L}^2+\abs{O^W_R}^2)
    \qty[-\frac{m_{\chi^\pm}^2+m_{\chi^0}^2}{s}+\qty(\frac{m_{\chi^\pm}^2-m_{\chi^0}^2}{s})^2]\right\}\ .
\end{align}
The spectral functions $v_J(s)$ and $a_J(s)$
encapsulate the hadronization effects.
In our analysis, we will use 
the spectral functions 
obtained from the data of
the tau lepton decay up to 
$s = m_\tau^2 \simeq 3\, \mathrm{GeV}^2$.
The specific extraction procedure
will be detailed in subsections~\ref{sec: non-strange multi-meson} and \ref{sec: strange multi-meson}.

We approximate the electroweak and short-distance QED corrections above $\mu_\mathrm{IR}$ 
by multiplying 
the tree-level decay rate~\eqref{eq: tree-level multimeson rate}
by the factor $S_\mathrm{EW}^{\chi}(\mu_\mathrm{IR})$~\eqref{eq: SEW}, 
with the central value of $\mu_\mathrm{IR}$ chosen to be the $\rho$ meson mass.
That is, the NLO decay rate
is given by
\begin{align}
    \label{eq: NLO multimeson rate}
    \Gamma_\mathrm{tree+EW}(\chi^-\rightarrow\chi^0 + (V/A)^-_J)
    =\Gamma_\mathrm{tree}(\chi^-\rightarrow\chi^0 + (V/A)^-_J)
    \times S_\mathrm{EW}^{\chi}(\mu_\mathrm{IR}=m_\rho)\ .
\end{align}
This approximation effectively captures the leading logarithmic enhancements 
in QED.
Additionally, it accounts for electroweak corrections that are enhanced 
by the inverse power of $s_W^2$. 
Relative to other sources of theoretical and experimental errors, 
such as those arising from spectral functions, 
the remaining non-logarithmic QED corrections are comparatively negligible.

\subsection{Single Kaon Mode}
\label{sec: single Kaon mode}
The spectral function $a_0(s)$ corresponds to $J^P=0^-$ transitions,
and therefore
includes the pole of a single meson $P^-=\pi^-,K^-$ as
\begin{align}
    a_0(s)\supset a_0^P(s) = \qty(2\pi F_P)^2\delta(s-m_P^2)\ .
\end{align}
Plugging this into Eq.~\eqref{eq: tree-level multimeson rate}, we obtain
the decay rate of the single meson mode, 
\begin{align}
    \label{eq: tree-level single meson rate}
    \Gamma_\mathrm{tree}\qty(\chi^-\to\chi^0+P^-)
    =
    &\frac{F_P^2 G_F^2 \abs{V_{uD}}^2}{8\pi m_{\chi^\pm}}
    \sqrt{\lambda\qty(1, m_{\chi^0}^2/m_{\chi^\pm}^2, m_P^2/m_{\chi^\pm}^2)}\cr
    &\times\Bigg\{\abs{O^W_L+O^W_R}^2
    \qty(m_{\chi^\pm} - m_{\chi^0})^2
    \qty[\qty(m_{\chi^\pm} + m_{\chi^0})^2-m_P^2]\cr
    &\hspace{40pt}
    +\abs{O^W_L-O^W_R}^2
    \qty(m_{\chi^\pm}+m_{\chi^0})^2
    \qty[\qty(m_{\chi^\pm}-m_{\chi^0})^2-m_P^2]
    \Bigg\}\ ,
\end{align}
which reproduces the result of the single pion mode~\eqref{eq: tree-level single pion rate}.

We define the leading-order rate of the single Kaon mode in terms of the Kaon partial decay rate to the muon as in the case of the single pion mode,
\begin{align}
    \label{eq: single Kaon LO}
    \Gamma_\mathrm{LO}(\chi^-\to\chi^0+K^-)
    := B\qty(K^-\to\mu^-+\overline{\nu_\mu}(+\gamma))
    \times \Gamma_K^\mathrm{tot}\times
    \frac{\Gamma_\mathrm{tree}(\chi^-\to\chi^0 + K^-)}{\Gamma_\mathrm{tree}\qty(K^-\to\mu^-+\overline{\nu_\mu})}\ ,
\end{align}
where $\Gamma^\mathrm{tot}_K$ is the total decay rate of the charged pion
and $B(K^-\to\mu^-+\overline{\nu_\mu}(+\gamma))$
is the branching fraction of the 
$K_{\mu2(\gamma)}$ decay.
We use measured values for both of $\Gamma_K^\mathrm{tot}=\tau_K^{-1}$ and $B(K^-\to\mu^-+\overline{\nu_\mu}(+\gamma))$
in Table~\ref{tab:input}.

We approximate
the electroweak and short-distance QED corrections
to $\Gamma_\mathrm{LO}$ as
\begin{align}
    \label{eq: NLO single kaon}
    \Gamma_\mathrm{LO+EW}(\chi^-\to\chi^0+K^-)
    =
    \Gamma_\mathrm{LO}(\chi^-\to\chi^0+K^-)\times S_\mathrm{EW}^{\chi\to K}(\mu_\mathrm{IR}=m_\rho)
    \ ,
\end{align}
where\,\footnote{Compare Eq.~\eqref{eq: single Kaon SEW} to
the first two lines in the NLO corrections to the single pion mode~\eqref{eq:analyticNLO}.}
\begin{align}
    \label{eq: single Kaon SEW}
  S_\mathrm{EW}^{\chi\to K}(\mu_\mathrm{IR})
  :=
  S_\mathrm{EW}^{\chi}(\mu_\mathrm{IR})
  -\frac{\alpha}{4\pi}\cdot 8\log\frac{m_Z}{\mu_\mathrm{IR}}\ .
\end{align}
In the single Kaon mode,
we do not compute 
the long-distance radiative correction, 
and hence it provides theoretical uncertainty
of the decay rate.
We take the difference 
$\Gamma_\mathrm{LO+EW}-\Gamma_\mathrm{tree+EW}$
as a crude estimate of the size of 
the long-distance correction to the single Kaon mode.

\subsection{Non-Strange Multi-Meson Modes}
\label{sec: non-strange multi-meson}
Let us consider multi-meson modes
without strangeness, 
e.g.,
$\pi^-\pi^0$, $K^-K^0$, $\pi^-2\pi^0$,
$\pi^+2\pi^-$, $4\pi$,\,\,$\cdots$
modes. Here, $K^0$ is the linear combination of the mass eigenstates $K_L$ and $K_S$, i.e., $\ket{K_0}  \simeq  
(\ket{K_L}+\ket{K_S})/\sqrt{2}$.
For these modes, the relevant spectral functions 
are $v_1(s)/a_1(s)$,
which can be extracted 
from the observation of the tau lepton decay
based on the relation
\begin{align}
    \label{eq: v_1 or a_1}
    v_1(s)/a_1(s)
    &= \frac{m_\tau^2}{6\abs{V_{ud}}^2S_\mathrm{EW}^\tau}
    \frac{B\qty(\tau^-\to V^-/A^-\nu_\tau)}{B\qty(\tau^-\to e^-\overline{\nu_e}\nu_\tau)}\frac{dN_{V/A}}{N_{V/A}ds}
    \qty[\qty(1-\frac{s}{m_\tau^2})^2\qty(1+\frac{2s}{m_\tau^2})]^{-1}\ .
\end{align}
Here, $S_\mathrm{EW}^\tau$ is the short-distance NLO correction
defined by Eq.~\eqref{eq: SEWtau},
$B(\cdots)$ is the branching fraction, and $(1/N_{V/A})dN_{V/A}/ds$ is
the normalized invariant mass-distribution
from the tau lepton decay. 
The effect of short- and long-distance corrections
to the extraction of these spectral functions will be also discussed below.

\subsubsection{Effects of Radiative Corrections to Tau Decay}
\label{sec: SEWtau}
The short-distance corrections to the hadronic tau lepton decay
introduce
about 2\% modification to the spectral functions. 
Therefore, it is important to exercise caution 
when deducing the spectral function from an experimentally measured mass distribution.

The factor 
$S_\mathrm{EW}^\tau$ in Eq.~\eqref{eq: v_1 or a_1}
is given by the ratio between the radiative corrections to the hadron mode 
and those to the lepton mode of the tau lepton decay, 
\begin{align}
    \label{eq: SEWtau}
    S_\mathrm{EW}^\tau = \frac{S^\tau(m_\tau, m_Z)}{\left.S^\tau_\mathrm{EW}\right|_\mathrm{lepton}}\ .
\end{align}
Here, the numerator is defined to be the short-distance correction to the hadronic tau lepton decay~\cite{Erler:2002mv},
\begin{align}
    \label{eq: resummed log with QCD}
    S^\tau(m_\tau, m_Z)
    = \,\,
    &\qty[\frac{\alpha(m_b)}{\alpha(m_\tau)}]^{9/19}\qty[\frac{\alpha(m_W)}{\alpha(m_b)}]^{9/20}
    \qty[\frac{\alpha(m_Z)}{\alpha(m_W)}]^{36/17}\cr
    &\times 
    \qty[\frac{\alpha_s(m_b)}{\alpha_s(m_\tau)}]^{(3/25)\qty(\alpha(m_\tau)/\pi)}
    \qty[\frac{\alpha_s(m_Z)}{\alpha_s(m_b)}]^{(3/23)\qty(\alpha(m_\tau)/\pi)} \cr
    \simeq \,\,
    &1.01907\ ,
\end{align}
where  leading-logarithmic corrections with perturbative QCD effects are resummed.
The denominator 
$\left.S^\tau_\mathrm{EW}\right|_\mathrm{lepton}$
represents
the short- and long-distance radiative corrections to the semi-leptonic decay of tau lepton~\cite{Marciano:1988vm, Braaten:1990ef},
\begin{align}
    \label{eq: lepton SEW}
    \left.S^\tau_\mathrm{EW}\right|_\mathrm{lepton}
    = 1+\frac{\alpha(m_\tau)}{4\pi}\qty(\frac{25}{2}-2\pi^2)\simeq 0.9957\ ,
\end{align}
where no logarithmic enhanced terms appear.
Hence, we obtain
\begin{align}
    \label{eq: Belle SEW}
    S_\mathrm{EW}^\tau = 
    \frac{S^\tau(m_\tau, m_Z)}{\left.S^\tau_\mathrm{EW}\right|_\mathrm{lepton}}
    \simeq 1.0235\pm 0.0003\ ,
\end{align}
where the error $\pm 0.0003$ incorporates the RG effects~\cite{Belle:2007goc}.

The formula for the spectral functions~\eqref{eq: v_1 or a_1}
also receives long-distance virtual/real photon corrections to the hadronic tau decay, 
which depend on what kind of hadrons are in the final state.
These effects can be encapsulated by a mode-dependent function
$G_\mathrm{EM}^\tau(s)$\,\footnote{For the two-pion mode
of the tau lepton, $G_\mathrm{EM}^\tau(s)$
has been estimated in Refs.\,\cite{Cirigliano:2001er,Cirigliano:2002pv,Flores-Baez:2006yiq, Flores-Tlalpa:2006snz, Miranda:2020wdg}.},
and the relation of the spectral functions to  observed mass distributions should be corrected as
\begin{align}
    \label{eq: v_1 or a_1 with GEM}
    v_1(s)/a_1(s)
    &= \frac{m_\tau^2}{6\abs{V_{ud}}^2S_\mathrm{EW}^\tau G_\mathrm{EM}^\tau(s)}
    \frac{B\qty(\tau^-\to V^-/A^-\nu_\tau)}{B\qty(\tau^-\to e^-\overline{\nu_e}\nu_\tau)}\frac{dN_{V/A}}{N_{V/A}ds}
    \qty[\qty(1-\frac{s}{m_\tau^2})^2\qty(1+\frac{2s}{m_\tau^2})]^{-1}\ .
\end{align}
When representing hadronic 
chargino decay rates with
the spectral functions 
defined by Eq.~\eqref{eq: v_1 or a_1 with GEM},
long-distance corrections to the chargino decay 
can also be factorized into 
a similar function $G_\mathrm{EM}^\chi(s)$.
The difference between $G_\mathrm{EM}^\chi(s)$
and $G_\mathrm{EM}^\tau(s)$ 
for each hadronic decay mode 
would be approximated as 
\begin{align}
    \label{eq: GEM error}
    \frac{G_\mathrm{EM}^\chi(s)}{G_\mathrm{EM}^\tau(s)}-1
    \simeq
    O\qty(\frac{\alpha}{\pi}\log\frac{m_\tau}{m_{\pi^\pm}})\ ,
\end{align}
where we expect that no enhanced corrections of $O((\alpha/\pi)\log m_{\chi^\pm})$ appear
as we have confirmed in the cases 
of the leptonic and the single pion modes.
In the following analysis, we define the spectral functions by Eq.~\eqref{eq: v_1 or a_1 with GEM}
and multiply the decay rate by $G_\mathrm{EM}^\chi(s)$ for each multi-meson  mode. 
We take $\pm(\alpha/\pi)\log(m_\tau/m_{\pi^\pm})=\pm0.6\%$
as the theoretical error for this approximation.
Note that the uncertainty from the choice of 
the IR cutoff $\mu_\mathrm{IR}$ in the correction
factor~\eqref{eq: SEW} is replaced by
the uncertainty from the long-distance correction $G_\mathrm{EM}^\chi(s)/G_\mathrm{EM}^\tau(s)$.

\subsubsection{Two-Pion Mode}
\begin{figure}[tb]
\centering
  \includegraphics[width=0.7\linewidth]{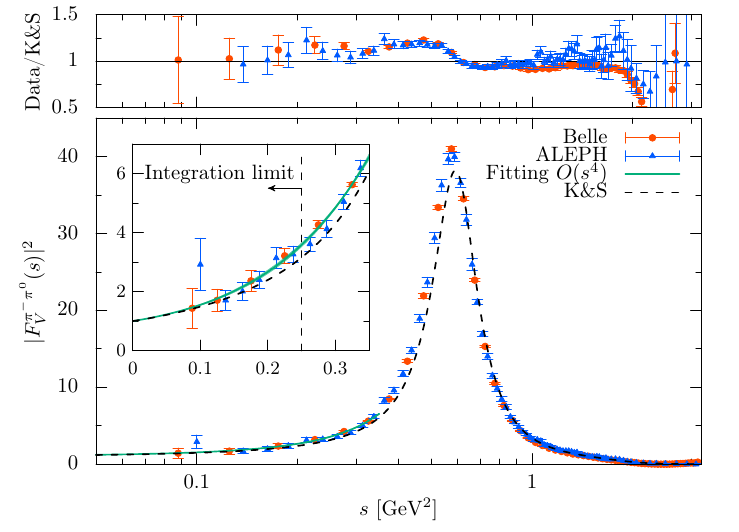}
\caption{The pion form factor $|F^{\pi^-\pi^0}_V(s)|^2$ measured by ALEPH and Belle experiments.
The black-dashed line represents the fitting given by K\"uhn\&Santamaria (K\&S)\,\cite{Kuhn:1990ad}.
The green band in the close-up window
is our fitting with Eq.~\eqref{eq: Fpi cubic}.
In estimation of the decay rate,
we apply 
the cubic polynomial approximation~\eqref{eq: Fpi cubic} 
up to the integration limit, $s=0.25\,\mathrm{GeV}^2$.
Above this limit,
we use the linear interpolation of 
the ALEPH data.
The upper panel shows the ratio between
the observed data and the K\&S fitting curve.
}\label{fig:Fpi}
\end{figure}

Let us consider the two-pion mode, which
is dominant among the multi-meson modes
and mediated only by the vector-current.
The axial current contribution to the two-pion mode
is suppressed by the conservation of $G$-parity.
The two-pion part of the spectral function $v_1(s)$~\eqref{eq: v_1 or a_1} is related to
the pion form factor $F_V^{\pi^-\pi^0}(s)$
\begin{align}
     v_1(s)
    \supset
    v^{\pi^-\pi^0}_1(s) 
    &= 
    \frac{1}{12}\lambda^{3/2}_{\pi^-\pi^0}(s)
    \abs{F^{\pi^-\pi^0}_V(s)}^2\ .
\end{align}

In our analysis, the numerical estimation of the pion form factor $\abs{F_V^{\pi^-\pi^0}(s)}^2$ 
is obtained as follows.
The pion form factor can be measured by 
decay of the tau lepton. 
Following Ref.~\cite{Davier:2002dy},
we fit the experimental data
on the pion form factor
obtained by 
the ALEPH~\cite{ALEPH:2005qgp, Davier:2008sk, Davier:2013sfa}
and
Belle~\cite{Belle:2007goc}
in the range of $s\in[(m_{\pi^0} + m_{\pi^\pm})^2, 0.35\,\mathrm{GeV}^2]$
with 
\begin{align}
    \label{eq: Fpi cubic}
    F^{\pi^-\pi^0}_V(s)
    = 1 + \frac{1}{6}\langle r^2 \rangle_\pi s
    + c_1 s^2 +c_2 s^3 + O(s^4)\ .
\end{align}
Here,  
we use 
the pion charge radius
$\langle r^2\rangle_\pi$
given by the PDG value~\cite{Workman:2022ynf},
$\langle r^2 \rangle_\pi=0.434\pm0.005\,\mathrm{fm}^2$,
by assuming the isospin symmetry.
In the integration in Eq.~\eqref{eq: tree-level multimeson rate}, we apply 
this cubic polynomial approximation to the form factor up to the integration limit, $s=0.25\,\mathrm{GeV}^2$. 
Above this limit, 
we use the linear interpolation of 
the ALEPH data\,\footnote{We have used the 2013 data distributed at~\url{http://aleph.web.lal.in2p3.fr/tau/specfun13.html}\ , which is based on Ref.~\cite{Davier:2013sfa}. 
For the other non-strange modes, we use the 2013 data provided
at the same website.
} in Eq.~\eqref{eq: tree-level multimeson rate}.

In Fig.~\ref{fig:Fpi}, we show the observed data~\cite{Belle:2007goc, Davier:2013sfa} of 
the pion form factor $\abs{F_V^{\pi^-\pi^0}(s)}^2$ together with
the fitting curve by the K\"uhn \& Santamaria (K\&S) model~\cite{Kuhn:1990ad} in the black-dashed line. 
Here, the parameters of the K\&S model are chosen
according to the analysis by CDG.
The figure shows
the observed data is about 10\%--20\% larger than the K\&S fitting curve around
$s = 0.2$--$0.6$\,GeV$^2$.

\begin{figure}[t!]
	\centering
  	\subcaptionbox{\label{fig:twopi_rate_error}}
	{\includegraphics[width=0.47\textwidth]{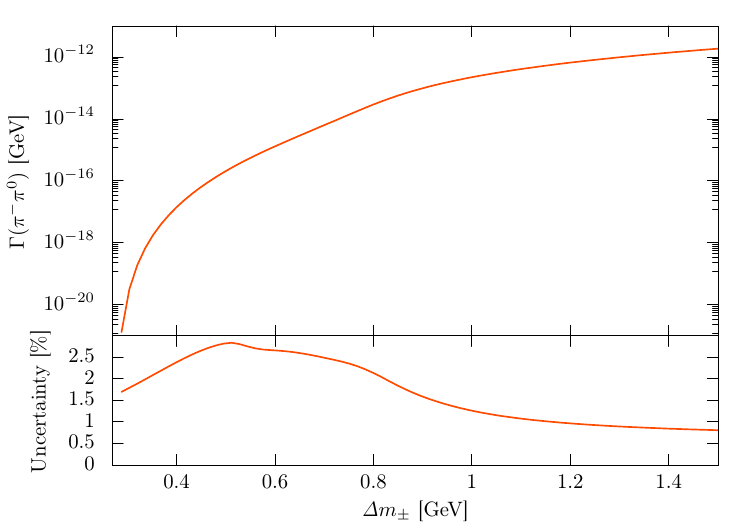}} 
  	\subcaptionbox{ \label{fig:twopi_ratio}}
	{\includegraphics[width=0.47\textwidth]{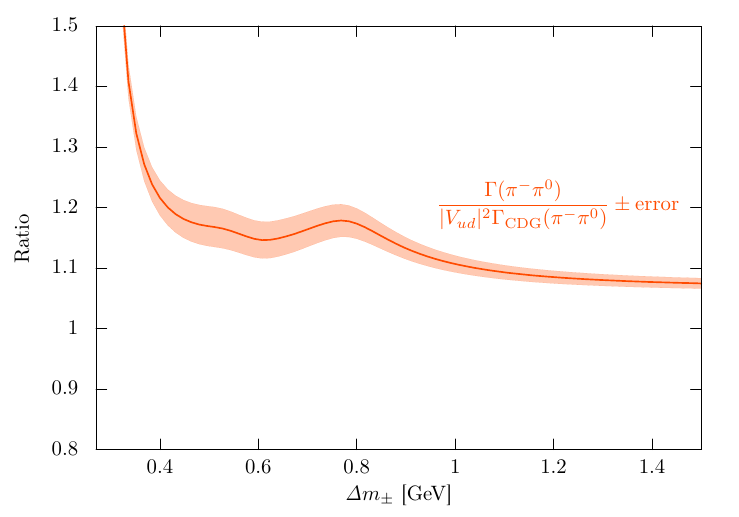}}  
\caption{
(a) The 
partial decay rate and the uncertainty of the two-pion mode
for $300\,$GeV Higgsino.
(b) The ratio of 
the decay rates of the two pion mode
in the present analysis and that 
of CDG for $300\,$GeV Higgsino multiplied by $|V_{ud}|^2$.
}\label{fig:twpi}
\end{figure}

In Fig.~\ref{fig:twopi_rate_error}, we show the 
partial decay rate and its uncertainty of the two-pion mode
for the $300\,$GeV Higgsino.
The uncertainty comprises errors in data of
the pion form factor and the errors from the long-distance NLO correction 
to the chargino decay,
$G_\mathrm{EM}^\chi(s)/G_\mathrm{EM}^\tau(s)$.
The figure shows that
for $\mathit{\Delta}m_\pm\simeq 0.5$\,GeV,
the error of the measured data dominates
and provides about 3\% uncertainty to the decay rate.
For a larger mass difference, e.g. $\mathit{\Delta}m_\pm\simeq1.4$\,GeV,
the error is dominated by that of the long-distance NLO corrections.

In Fig.~\ref{fig:twopi_ratio}, we compare our numerical result of the chargino decay rate into the two pions with
the previous estimate by CDG.
Here, we have multiplied the CDG decay rate by
the CKM factor $\abs{V_{ud}}^2$, which has not been
included in the original expression by CDG.
The red band around the ratio
corresponds to the error of our analysis.
Typically, our estimate is $O(10\%)$
larger than the previous one.

\subsubsection{Other Non-Strange Multi-Meson Modes}

\begin{figure}[t!]
	\centering
  	\subcaptionbox{\label{fig:threepi_rate_error}}
	{\includegraphics[width=0.47\textwidth]{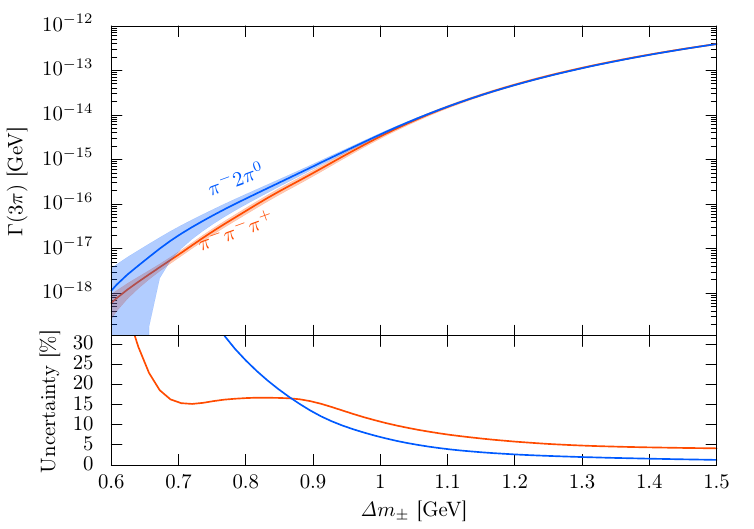}} 
  	\subcaptionbox{\label{fig:threepi_ratio}}
	{\includegraphics[width=0.47\textwidth]{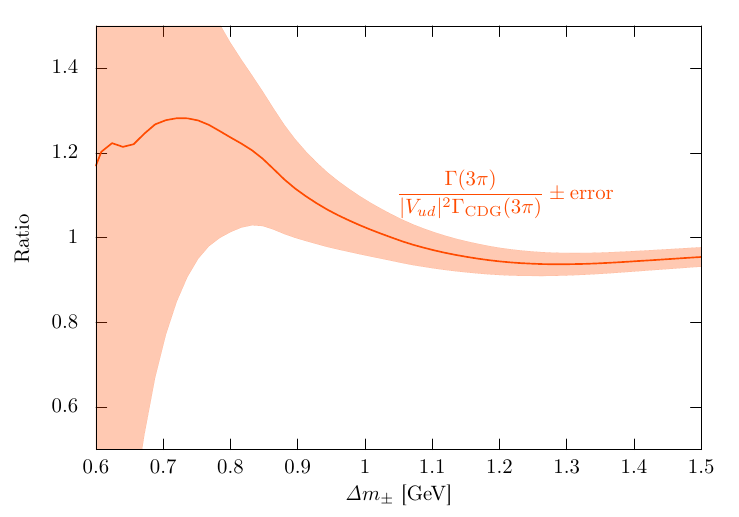}}  
\caption{
(a) The 
partial decay rate and the uncertainty of the three-pion modes
for $300\,$GeV Higgsino.
(b) The ratio of 
the decay rates of the three pion modes
in the present analysis and that 
of CDG for $300\,$GeV Higgsino multiplied by $|V_{ud}|^2$.
}\label{fig:threepi}
\end{figure}

\begin{figure}[t!]
	\centering
  	\subcaptionbox{ \label{fig:multipi_rate_ratio}}
	{\includegraphics[width=0.47\textwidth]{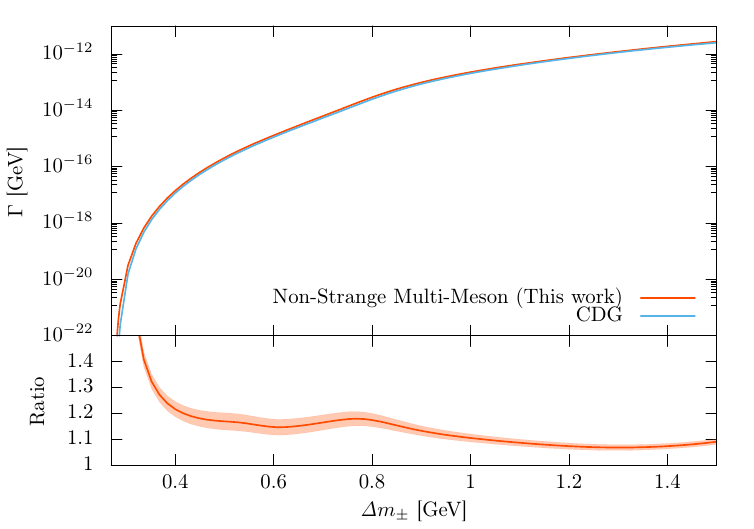}} 
   	\subcaptionbox{\label{fig:multipi_BF}}
	{\includegraphics[width=0.47\textwidth]{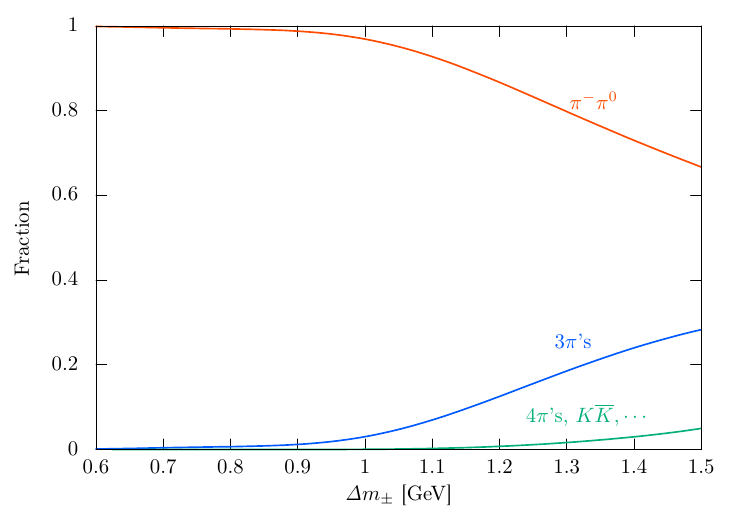}} 
\caption{
(a) The decay rate and the uncertainty of the non-strange multi-meson modes
for the $300\,$GeV Higgsino.
The bottom panel shows the ratio of 
the decay rates of the 
non-strange multi-meson modes
in the present analysis and that 
of CDG including $|V_{ud}|^2$.
(b) The branching fractions of each mode
for the 300\,GeV Higgsino.
}\label{fig:multi}
\end{figure}

When the mass difference is greater than about 1\,GeV,
the three-pion modes also provide significant contribution
to the chargino decay.
We use the linear interpolation of the ALEPH data
of the three pion modes
to obtain the spectral functions, $a_1^{\pi^+2\pi^-}(s)$ and $a_1^{\pi^-2\pi^0}(s)$.
Again, the contribution of $v_1(s)$ to the three-pion modes is negligible due to the $G$-parity conservation.
In Fig.~\ref{fig:threepi_rate_error},
we show the 
chargino decay rates into three pions 
and their uncertainties.
These uncertainties include both experimental error of the
ALEPH data and theoretical error from $G_\mathrm{EM}$~\eqref{eq: GEM error}.
In the low mass difference, 
the error of the three-pion modes is dominated by the experimental uncertainty of the ALEPH data.
This large uncertainty scarcely affects the estimation
of the total chargino decay rate due to their small 
branching fractions.
In Fig.~\ref{fig:threepi_ratio}, we compare our numerical result of the chargino decay rate into three pions with
the previous estimate by CDG.

We also include other minor
non-strange multi-meson modes,
$K^-K^0$, $4\pi$, $\cdots$,
which have not been included in the CDG analysis.
In Fig.~\ref{fig:multi},
we summarize non-strange multi-meson modes of the chargino decay with comparison to the previous CDG decay rate.
Compared to the previous estimate by CDG, the total 
rate of the non-strange multi-meson modes
increases by about 10\%.

\subsection{Multi-Meson Modes with Strangeness}
\label{sec: strange multi-meson}

There are multi-meson modes with 
the strangeness final states $\mathit{\Delta}S = -1$.
Those modes are suppressed by
the Cabbibo angle $V_{us}$, 
and they provide a few percent contribution
to the total decay rate
for a larger mass difference.

\subsubsection{$\qty(K\pi)^-$ Mode}
The form factors for the $\pi^-{K}_S$ 
mode are provided by Ref.~\cite{Belle:2007goc}
in the Breit-Wigner approximation,
\begin{align}
    F_V^{\pi^- K_S}(s)
    &= \frac{1}{1+\beta}\qty[BW_{K^*(892)}^{J=1}(s)
    +\beta BW_{K^*(1410)}^{J=1}(s)]\ ; \\
    F_S^{\pi^- K_S}(s)
    &= \kappa\frac{s}{M_{K^*_0(800)}}BW_{K^*_0(800)}^{J=0}(s)\ .
\end{align}
Here, $BW_R^J(s)$ is the Breit-Wigner function for
resonance $R$ with mass $M_R$,
\begin{align}
    BW_R^J(s) = \frac{M_R^2}{s-M_R^2+i\sqrt{s}\,\Gamma_R^J(s)}\ .
\end{align}
The $s$-dependent total width $\Gamma_R(s)$ is given by
\begin{align}
    \Gamma_R^J(s) = \Gamma_{0R}\frac{M_R^2}{s}
    \qty(\frac{P(s)}{P(M_R^2)})^{2J+1}\ ,
\end{align}
where $\Gamma_{0R}$ is the resonance width at its peak,
$P(x)$ is given by
\begin{align}
    P(x) = \frac{1}{2\sqrt{x}}\sqrt{\qty[x-(m_K+m_{\pi^\pm})^2]\qty[x-(m_K-m_{\pi^\pm})^2]}\ ,
\end{align}
and $J = 0, 1$ is the angular momentum of the $(K\pi)^-$
system. 
Following Ref.~\cite{Belle:2007goc},
we adopt the parameters such as $\beta$ and $\kappa$
fitted in the model including $K_0^*(800)$, $K^*(892)$, and $K^*(1410)$ resonances.

These form factors can be translated into 
the spectral functions in the following relations:
\begin{align}
    v^{\pi^-K_S}_1(s)
    &= \frac{1}{48}\qty{\frac{2P(s)}{\sqrt{s}}}^3
    \abs{F^{\pi^- K_S}_V(s)}^2\ ; \\
    v^{\pi^-K_S}_0(s)
    &= \frac{1}{16}\qty(\frac{m_K^2-m_{\pi^\pm}^2}{s})^2\,
    \frac{2P(s)}{\sqrt{s}}
    \abs{F^{\pi^- K_S}_S(s)}^2\ ; \\
    v^{\pi^0K^-}_1(s)
    &=v^{\pi^-K_L}_1(s)
    =v^{\pi^-K_S}_1(s) \ ; \quad
    v^{\pi^0K^-}_0(s)
    =v^{\pi^-K_L}_0(s)
    =v^{\pi^-K_S}_0(s) \ ,
\end{align}
where we have used the 
isospin symmetry.
By substituting these spectral functions into 
Eq.~\eqref{eq: tree-level multimeson rate},
we obtain the decay rate into the final state
indicated by the superscript of the vector spectral functions.

\subsubsection{Other Multi-Meson Modes with Strangeness}
\begin{figure}[t!]
	\centering
  	\subcaptionbox{\label{fig:K_rate_error}}
	{\includegraphics[width=0.47\textwidth]{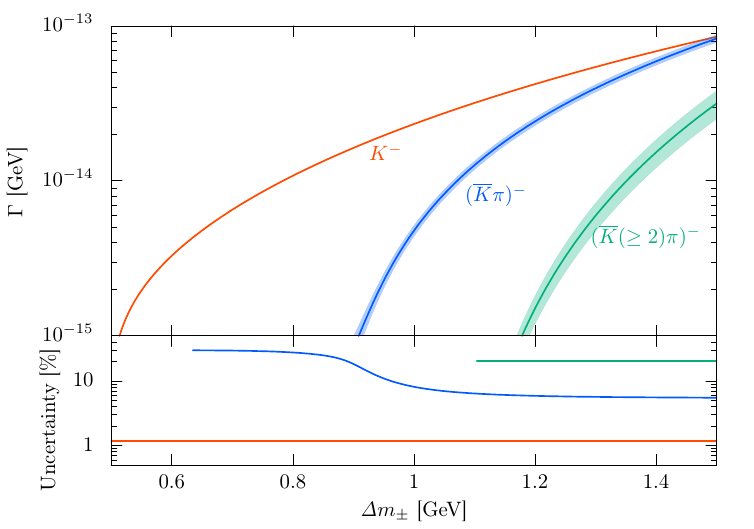}} 
   	\subcaptionbox{\label{fig:K_ratio}}
	{\includegraphics[width=0.47\textwidth]{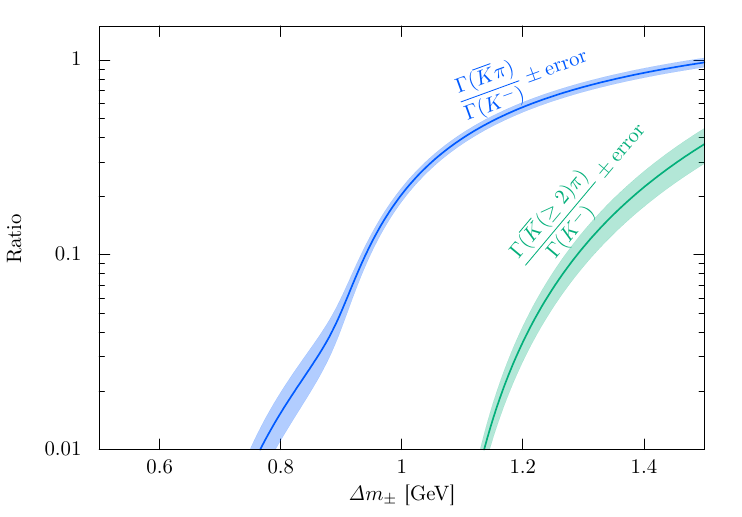}} 
\caption{
The strange hadronic decay modes for the 300\,GeV Higgsino.
(a) The decay rates and uncertainties of the 
single $K^-$, $\qty(\overline{K}\pi)^-$
and $\qty(\overline{K}(\geq2)\pi)^-$ modes.
(b) The branching
ratio of the strange multi-meson modes
to the single Kaon mode together with
their uncertainties.
}\label{fig:strange}
\end{figure}

The spectral functions for 
other strange modes such as
$\qty(\overline{K}(\geq2)\pi)^-$
are contained in $v+a$ with $\mathit{\Delta}S = -1$.
We approximate the inclusive spectral function by
$\left.v+a\right|_{\mathit{\Delta}S = -1} \simeq 1$
which is obtained by the parton model.
This approximation is consistent with 
the observed data of the spectral functions 
of the strangeness final states.
See Fig.~9 in Ref.~\cite{ALEPH:1999uux}.
To avoid over-counting with 
the $(K\pi)^-$ mode, we include
$\left.v+a\right|_{\mathit{\Delta}S = -1} \simeq 1$
contribution only for $s \gtrsim 1.1\,\mathrm{GeV}^2$.
This treatment reproduces the
tau lepton decay rate to
$K\,\qty(\geq2)\pi$ to accuracy
of about $20\%$, and hence the 
$K\qty(2\geq)\pi$ modes of
the chargino can be predicted to precision of 
about $20\%$.
Since the contributions of the
multi-meson modes with the strangeness final states to the total decay rate are minor, 
this approach provides the level of accuracy necessary for determining the total chargino decay rate as aimed for in this paper.

In Fig.~\ref{fig:strange},
we summarize strange hadronic decay modes
for the 300\,GeV Higgsino.
In Fig.~\ref{fig:K_rate_error},
we show the decay rates and uncertainties
of the 
single $K^-$, $\qty(\overline{K}\pi)^-$
and $\qty(\overline{K}(\geq2)\pi)^-$ modes.
The error of the single Kaon mode
is addressed 
in subsection~\ref{sec: single Kaon mode}.
The uncertainty of the $\qty(\overline{K}\pi)^-$
mode comes from that of the experimental data.
We estimate the error of $\qty(\overline{K}(\geq2)\pi)^-$
mode as explained in the previous paragraph.
In Fig.~\ref{fig:K_ratio}, we show the branching
ratio of the strange multi-meson modes
to the single Kaon mode together with
their uncertainties.

\section{Conclusions}
\label{sec: Conclusions}

\begin{figure}[tb]
\centering
  \includegraphics[width=0.7\linewidth]{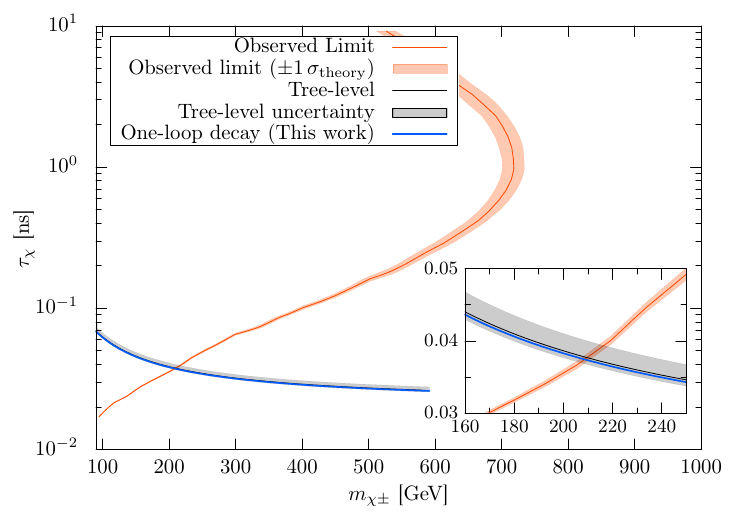}
\caption{The LHC constraint on the charged Higgsino 
mass and lifetime based on the disappearing charged track search,
where $\mathit{\Delta}m_\pm \sim 300$\,MeV.
The red solid line shows the 95\%. limit by the ATLAS with data of $\int  \mathcal{L} dt =  136\,\mathrm{fb}^{-1}$ and $\sqrt{s} = 13$\,TeV\,\cite{ATLAS:2022rme}.
The red band shows $\pm 1\,\sigma$ uncertainty of the production cross section of the Higgsino.
The upper-left region of the red line has been excluded.
The blue solid line is the lifetime  
based on the NLO decay rate for the pure Higgsino, which decays into $\chi^0_1$ and $\chi^0_2$.
The gray band shows the tree-level prediction with its uncertainty.
}\label{fig:atlas_higgsino}
\end{figure}

In this paper, we performed precise estimate of the decay rate of the chargino.
We extend the previous analysis on the charged Wino decay~\cite{Ibe:2022lkl} so as to apply our analysis to larger mass difference between the 
chargino and the neutralino.
We have updated $O(\alpha)$ correction to the single pion mode following the D\&M analysis~\cite{Descotes-Genon:2005wrq} as in the Wino case.
Other difference from our previous analysis is in the inclusion of the leptonic mode,
the single Kaon mode, and other multi-meson modes.
In the leptonic mode,
we have computed 
 $O(\alpha)$ corrections
with the real emission effects,
and predicted the decay rate 
neglecting 
of $O(\alpha\mathit{\Delta}m_\pm/m_{\chi^\pm})$
contributions.
We also estimated the single Kaon mode, including the dominant short-distance $O(\alpha)$ corrections, and evaluated the remaining uncertainty by comparison with the single pion case.
We also utilized the latest data on hadronic tau lepton decays to estimate the multi-meson decay rates and their associated uncertainties. Through these updates, we achieved the NLO chargino decay rate with a precision of less than 1\% for $\mathit{\Delta}m_\pm \lesssim 1.5$\,GeV.
We also emphasize that 
the NLO results can be applied to the case of more general fermionic electroweak multiplets, e.g., quintuplet dark matter.

As emphasised in the text, the NLO calculation is mandatory to obtain the decay rate with this high precision.
In fact, ``tree-level'' analysis is always subject to indeterminacy due to the choice of tree-level coupling. 
As we have seen, for example, the choice of either $G_F$ or $G_F^0$ results in the uncertainty of the decay rate about $8$\%.
Besides, the use of the running gauge coupling constant as the tree-level coupling makes the uncertainty larger.

The reduction of this uncertainty by our NLO analysis plays a critical role in the search for metastable particles at colliders such as the LHC. Fig\,.\ref{fig:atlas_higgsino} presents the results from a search using the disappearing charged track technique at the current ATLAS setup. The gray area in the figure underscores the tree-level uncertainty in the pure-Higgsino decay rate, which markedly influences the search for the Higgsinos. As a consequence of this uncertainty, the established mass limit for the Higgsino is subject to a margin of error of approximately 10\%.
Our one-loop analysis resolved 
those problems.
Currently, a comprehensive two-loop calculation to determine the mass difference of the pure Higgsino is not available. 
It is anticipated that this gap contributes to an uncertainty in the sub-percent range regarding the mass difference. 
This uncertainty, in turn, leads to a theoretical prediction of the charged pure Higgsino lifetime with an approximate 1\% margin of error. 
While these theoretical calculations is beyond the scope of this paper, we aim to conduct precise calculations of the mass difference in the near future.

Finally, it should be emphasised that we confirmed that the decay rate
becomes independent of $m_{\chi^\pm}$ 
in the limit of $m_{\chi^\pm} \to \infty$ at the NLO level.
This confirms that the decoupling theorem similar to the Appelquist-Carazzone theorem holds
at one-loop level in the chargino decay, although 
the theorem is not directly applicable to an amplitude where the external lines of the diagrams include heavy particles.
We also confirmed that the decay rates of the leptonic and the single pion modes are free from collinear divergences, in accordance with the KLN theorem.

\section*{Acknowledgements}
This work is supported by Grant-in-Aid for Scientific Research from the Ministry of Education, Culture, Sports, Science, and Technology (MEXT), Japan, 21H04471, 22K03615 (M.I.), 20H01895, 20H05860 and 21H00067 (S.S.) and by World Premier International Research Center Initiative (WPI), MEXT, Japan. This work is also supported by Grant-in-Aid for JSPS Research Fellow 
JP21J20421 and JP22KJ0556 (Y.N.), 
and by International Graduate Program for Excellence in Earth-Space Science (Y.N.).

\appendix

\section{Details of Electroweak Corrections}
\label{app: explicit electroweak}
\renewcommand{\theequation}{\thesection.\arabic{equation}}
Here, we present the detailed computation of the electroweak corrections 
to the four-fermion interaction
based on the pure-Higgsino Lagrangian~\eqref{eq: pure-Higgsino Lagrangian}. 
To see the generalization to other representations
of $\mathrm{SU}(2)_L\times\mathrm{U}(1)_Y$,
we introduce a redundant notation of 
the coupling constants of the fermion $F$ to the $Z$-boson,
\begin{align}
    g_{F}^Z = T_{F}^3-s_W^2Q_{F}\ ,
\end{align}
where $T^3_F$ and $Q_F$ 
represent the eigenvalues of the diagonal generator of $\mathrm{SU}(2)_L$ and the charge operator with respect to $F$.

\subsection{Corrections to Wave-Functions and Vertices}
\label{sec: wave-function and vertices}
\subsubsection{Wave-Function Renormalization of Higgsinos}
To begin with, we show the self-energy
of the fermions which appear in our problem at one-loop level.
We define the self-energy of a fermion
with a mass parameter $m_\mathrm{fermion}$, which we denote by $\Sigma_\mathrm{fermion}(\slashed{p})$,
by the following relation:
\begin{align}
    \Gamma_{\mathrm{fermion}}^{(2)}(p) 
    = (\slashed{p}-m_{\mathrm{fermion}}) - \Sigma_{\mathrm{fermion}}(\slashed{p}) + (\mathrm{CT})\ .
\end{align}
Here, $p$ is the momentum of the fermion,  $\Gamma^{(2)}_\mathrm{fermion}(p)$
is the coefficient of the quadratic term in the quantum effective action,
and $(\mathrm{CT})$ represents the wavefunction and the mass counterterm.
In the following, we use the dimensional regularization
with $d=4-2\epsilon_\mathrm{EW}$.

The self-energies provide corrections
to a decay amplitude as a residue of
external legs. At $O(\alpha)$,
the derivatives of $\Sigma(\slashed{p})$ at the $\mathrm{SU}(2)_L$ symmetric mass
appear in the amplitude.
The virtual $\gamma/Z/W$ contributions to the wave-function of the charged Higgsino
are given by
\begin{align}
    \label{eq: QED wave-function of chargino}
	&\frac{d}{d\slashed{p}}\Sigma_-^\mathrm{\gamma}(m_\chi)
     = - Q_{\chi^-}^2\frac{\alpha}{4\pi}
     \left(\frac{1}{\bar{\epsilon}_{\mathrm{EW}}} + \log \frac{\mu_{\mathrm{EW}}^2}{m_\chi^2} 
     -2 \log \frac{m_\chi^2}{m_\gamma^2} + 4\right) \ ; \\
    &\frac{d}{d\slashed{p}}\Sigma_-^{Z\mathrm{(EW)}}(m_\chi)
	= - \frac{1}{c_W^2s_W^2}\qty(g^Z_{\chi^-})^2
	 \frac{\alpha}{4\pi}\left[\frac{1}{\bar{\epsilon}_{\mathrm{EW}}} 
    + \log \frac{\mu_{\mathrm{EW}}^2}{m_\chi^2} 
    +F_\Sigma(r_Z)\right] \ ; \\
    &\frac{d}{d\slashed{p}}\Sigma_-^{W\mathrm{(EW)}}(m_\chi)
    = - \qty(\frac{1}{\sqrt{2}s_W})^2\frac{\alpha}{4\pi}
    \left[\frac{1}{\bar{\epsilon}_{\mathrm{EW}}} + \log \frac{\mu_{\mathrm{EW}}^2}{m_\chi^2} 
    +F_\Sigma(r_W)
    \right]\ .
\end{align}
Here,
$\mu_{\mathrm{EW}}$ is an arbitrary 
mass parameter, $\bar{\epsilon}_{\mathrm{EW}}^{-1}=\epsilon_{\mathrm{EW}}^{-1}-\gamma_E + \log4\pi$ with $\gamma_E$ being the Euler-Mascheroni constant, 
and $m_\gamma$ a photon mass to regulate the infrared singularity. 
The function $F_\Sigma$ is given by
\begin{align}
    F_\Sigma(r)
    = 4\log r + 4+3r^2 -3r^4\log r
    - \frac{3r(4+2r^2-r^4)}{\sqrt{4-r^2}}\cos^{-1}\frac{r}{2}\ .
\end{align}
The arguments $r_W$ and $r_Z$ are given by $r_W = m_W/m_\chi$
and $r_Z = m_Z/m_\chi$, respectively.

To avoid confusion in the matching
procedure, we label the UV pole and the arbitrary scale with
the name of theory which provides
them; The pole $\epsilon^{-1}_\mathrm{EW}$ means
the UV divergence arising from loops in the electroweak theory. Similarly, the subscript of $\mu_\mathrm{EW}$ shows that the scale is generated by the dimensional regularization in the electroweak theory.

The normalization of the wave-function of the neutral Higgsino is contributed by the virtual $Z/W$ loops:\,\footnote{In our computation the neutral Higgsinos are treated as a Dirac fermion,
and therefore we should not include Majorana factors.}
\begin{align}
    & \frac{d}{d\slashed{p}}\Sigma_0^{Z\mathrm{(EW)}}(m_\chi)
    = -\frac{1}{c_W^2s_W^2}\qty(g^Z_{\chi^0})^2
	 \frac{\alpha}{4\pi}
    \left[\frac{1}{\bar{\epsilon}_{\mathrm{EW}}} 
    + \log \frac{\mu_{\mathrm{EW}}^2}{m_\chi^2} 
    +F_\Sigma(r_Z)
    \right]\ ; \\
    & \frac{d}{d\slashed{p}}\Sigma_0^{W\mathrm{(EW)}}(m_\chi)
    = -\qty(\frac{1}{\sqrt{2}s_W})^2\frac{\alpha}{4\pi}
    \left[\frac{1}{\bar{\epsilon}_{\mathrm{EW}}} 
    + \log \frac{\mu_{\mathrm{EW}}^2}{m_\chi^2} 
    +F_\Sigma(r_W)
    \right]\ .
\end{align}
The contribution from the wave-function renormalizations 
of the Higgsinos is given by\,\footnote{The counterterm contributions to the wave-function renormalizations will be canceled
by the counterterm contribution
to the vertex corrections.}
\begin{align}
    \mathcal{M}^{\chi,0}_{\mathrm{WF(EW)}}
    = \mathcal{M}^0_{\mathrm{tree}} 
    \times\qty[
    \frac{1}{2}\frac{d}{d\slashed{p}}\Sigma_{0}^{(\mathrm{EW})}(m_\chi)
    +\frac{1}{2}\frac{d}{d\slashed{p}}\Sigma_{-}^{(\mathrm{EW})}(m_\chi)]\ ,
\end{align}
where
\begin{align}
    \frac{d}{d\slashed{p}}\Sigma_{0}^{(\mathrm{EW})}
    &=\frac{d}{d\slashed{p}}\Sigma_{0}^{Z(\mathrm{EW})}
    +\frac{d}{d\slashed{p}}\Sigma_{0}^{W(\mathrm{EW})}\ ; \cr
    \frac{d}{d\slashed{p}}\Sigma_{-}^{(\mathrm{EW})}
    &=
    \frac{d}{d\slashed{p}}\Sigma_{-}^{\gamma(\mathrm{EW})}
    +\frac{d}{d\slashed{p}}\Sigma_{-}^{Z(\mathrm{EW})}
    +\frac{d}{d\slashed{p}}\Sigma_{-}^{W(\mathrm{EW})}\ .
\end{align}

\subsubsection{Corrections to Vertex of Higgsinos}
\begin{figure}[t]
    \centering
    \subcaptionbox{\label{fig:HisssinoVertexCorrectionA}}
    {\includegraphics[width=0.23\textwidth]{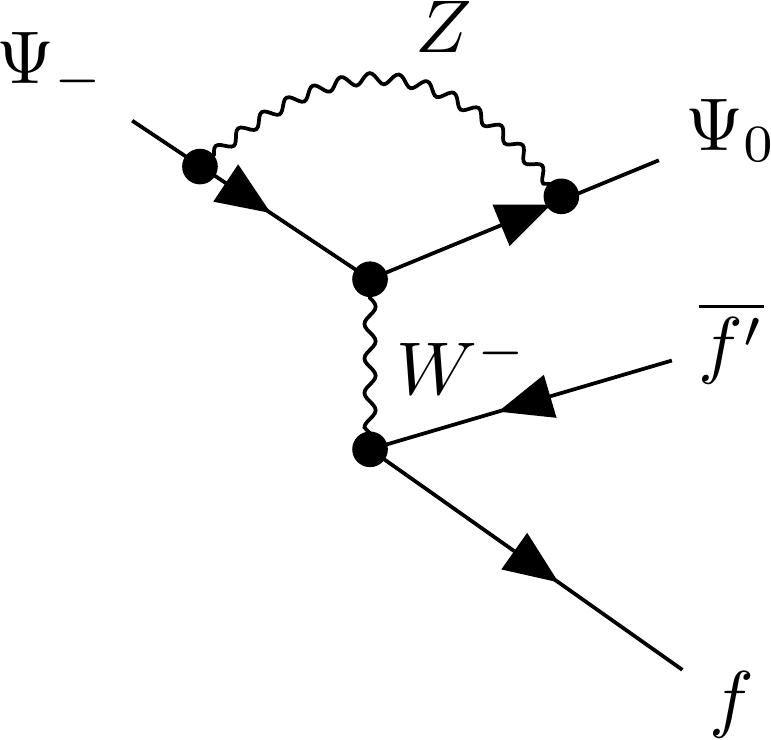}}
    \hspace{20pt}
    \subcaptionbox{\label{fig:HiggsinoVertexCorrectionB}}
    {\includegraphics[width=0.23\textwidth]{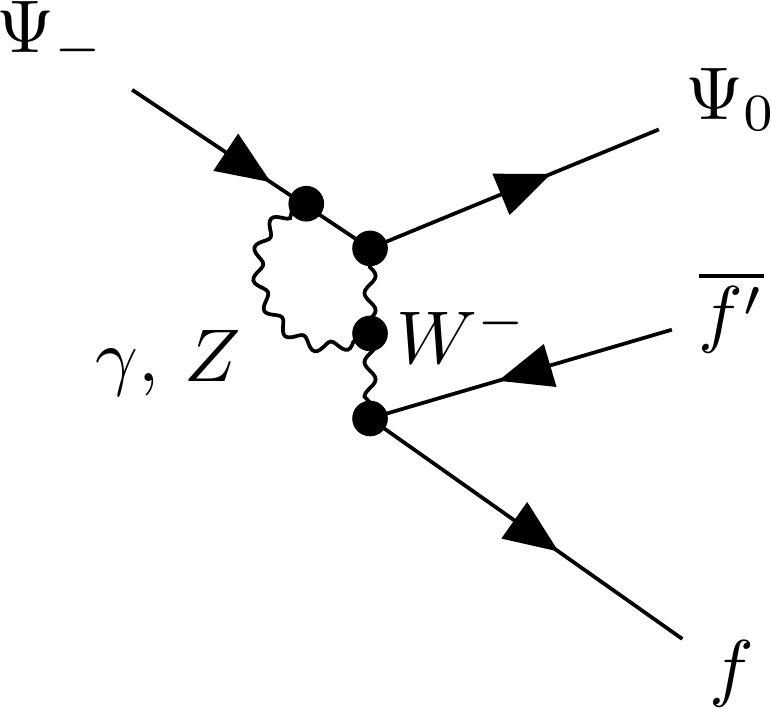}}
    \subcaptionbox{\label{fig:HiggsinoVertexCorrectionC}}
    {\includegraphics[width=0.23\textwidth]{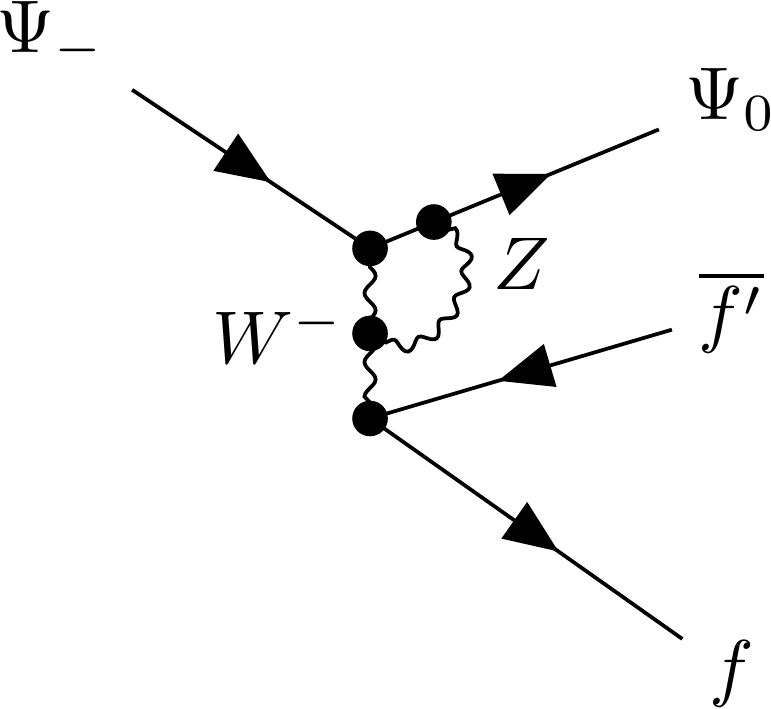}}
    \hspace{20pt}
    \caption{Radiative corrections to the Higgsino's vertex.}
    \hspace{20pt}
\end{figure}
The Higgsino's vertex receives the radiative corrections
shown in Figs.~\ref{fig:HisssinoVertexCorrectionA}-\ref{fig:HiggsinoVertexCorrectionC}, 
which can be written as multiplicative factors to
the tree-level amplitude:
\begin{align}
    \mathcal{M}_\chi^{Z,0} &= \mathcal{M}_\mathrm{tree}^0
    \times \frac{g^Z_{\chi^0}g^Z_{\chi^-}}{c_W^2 s_W^2}\frac{\alpha}{4\pi}
    \qty[\frac{1}{\bar{\epsilon}_\mathrm{EW}} + \log\frac{\mu_\mathrm{EW}^2}{m_\chi^2}
    +F_\Sigma\qty(r_Z)]\ ; \\
    \label{eq: gamW to chi vertex}
    \mathcal{M}_\chi^{\gamma W, 0} &= \mathcal{M}_\mathrm{tree}^0
    \times \mathit{\Delta}Q_\chi\frac{\alpha}{4\pi}
    \qty[\frac{3}{\bar{\epsilon}_\mathrm{EW}} + 3\log\frac{\mu_\mathrm{EW}^2}{m_\chi^2}
    +F_\chi^{\gamma W}(r_W)]\ ; \\
    \label{eq: WZ to chi vertex}
    \mathcal{M}_\chi^{WZ,0} &= \mathcal{M}_\mathrm{tree}^0
    \times \frac{g^Z_{\chi^0}-g^Z_{\chi^-}}{s_W^2}\frac{\alpha}{4\pi}
    \qty[\frac{3}{\bar{\epsilon}_\mathrm{EW}} + 3\log\frac{\mu_\mathrm{EW}^2}{m_\chi^2} + F_\chi^{WZ}(r_W)]\ ,
\end{align}
where the $\mathit{\Delta}Q_\chi = Q_{\chi^0}-Q_{\chi^-} = 1$ 
is the charge difference between
the neutralino and the chargino, and
\begin{align}
    F_\chi^{\gamma W}(r_W) =&\, 
    4+r_W^2-r_W^4\log(r_W)
    -r_W(2+r_W^2)\sqrt{4-r_W^2}\cot^{-1}\qty(\frac{r_W}{\sqrt{4-r_W^2}}) \, ,\\
    F_\chi^{WZ}(r_W) =&\, 
    4 + \qty(1+\frac{1}{c_W^2})r_W^2 + \frac{\log(c_W)}{c_W^4s_W^2}r_W^4 
    - \qty(1+\frac{1}{c_W^2}+\frac{1}{c_W^4})r_W^4\log(r_W) \notag \\
    &+\frac{c_W^2}{s_W^2}(2+r_W^2)r_W\sqrt{4-r_W^2}\cot^{-1}\qty(\frac{r_W}{\sqrt{4-r_W^2}}) \notag \\
    &-\frac{2c_W^2+r_W^2}{c_W^4s_W^2}r_W\sqrt{4c_W^2-r_W^2}\cot^{-1}\qty(\frac{r_W}{\sqrt{4c_W^2-r_W^2}})\ .
\end{align}
Note that the virtual $W$-boson loop does not exist in
the corrections to the charged-current operator of the Higgsinos, 
since we treat the neutral Higgsino as a Dirac fermion.

\subsubsection{Renormalization}
The virtual $Z$-boson contribution
to the charged-current operator of the Higgsinos
is given by 
\begin{align}
    \label{eq: Z to Jchi}
    \mathcal{M}^{Z,0}_{J_\chi^-}
    &=
    \mathcal{M}_\chi^{Z,0}
    +\mathcal{M}^0_\mathrm{tree}\times
   \qty[
    \frac{1}{2}\frac{d}{d\slashed{p}}\Sigma_{0}^{Z(\mathrm{EW})}(m_\chi)
    +\frac{1}{2}\frac{d}{d\slashed{p}}\Sigma_{-}^{Z(\mathrm{EW})}(m_\chi)] \cr
    &=
    \mathcal{M}^0_\mathrm{tree}
    \times
    \frac{-\qty(g^Z_{\chi^0}-g^Z_{\chi^-})^2}{2c_W^2s_W^2}
    \frac{\alpha}{4\pi}
    \qty[\frac{1}{\bar{\epsilon}_\mathrm{EW}} + \log\frac{\mu_\mathrm{EW}^2}{m_\chi^2}
    +F_\Sigma(r_Z)]\cr
    &=
    \mathcal{M}^0_\mathrm{tree}
    \times
    \frac{-\qty(\mathit{\Delta}Q_\chi)^2}{2}
    \frac{c_W^2}{s_W^2}
    \frac{\alpha}{4\pi}
    \qty[\frac{1}{\bar{\epsilon}_\mathrm{EW}} + \log\frac{\mu_\mathrm{EW}^2}{m_\chi^2}
    +F_\Sigma(r_Z)]\ .
\end{align}
Note that not only the divergent part but also the finite part
is independent of the hypercharge thanks to neglecting the momentum transfer. 
Similarly, the virtual photon/$W$
contribution to the charged-current operator of the Higgsinos 
is given by
\begin{align}
    \label{eq: gam to Jchi}
    \mathcal{M}^{\gamma,0}_{J_\chi^-}
    &=
    \mathcal{M}^0_\mathrm{tree}
    \times
    \frac{-\qty(\mathit{\Delta}Q_\chi)^2}{2}
    \frac{\alpha}{4\pi}
    \qty(\frac{1}{\bar{\epsilon}_\mathrm{EW}} + \log\frac{\mu_\mathrm{EW}^2}{m_\chi^2}
    +4-2\log\frac{m_\chi^2}{m_\gamma^2})\ ; \\
    \label{eq: W to Jchi}
    \mathcal{M}^{W,0}_{J_\chi^-}
    &=
    \mathcal{M}^0_\mathrm{tree}
    \times
    \frac{-1}{2s_W^2}
    \frac{\alpha}{4\pi}
    \qty[\frac{1}{\bar{\epsilon}_\mathrm{EW}} + \log\frac{\mu_\mathrm{EW}^2}{m_\chi^2}
    +F_\Sigma(r_W)]\ .
\end{align}

The total of virtual contributions from the wave-function renormalization
and the vertex corrections can be simplified as
\begin{align}
    \label{eq: WF+Vertex chi w/ UV}
    \left.\mathcal{M}^{\chi,0}_\mathrm{WF+Vertex(EW)}\right|_\mathrm{Virtual}
    = \,\,
    &\sum_{V=\gamma,Z,W}\mathcal{M}_{J_\chi^-}^{V,0}+\mathcal{M}_\chi^{\gamma W,0}
    +\mathcal{M}_\chi^{WZ,0}\cr
    =\,\,
    &\frac{\alpha}{4\pi}\left\{\qty[\frac{2}{s_W^2}-\frac{(5-\mathit{\Delta}Q_\chi)(1-\mathit{\Delta}Q_\chi)}{2s_W^2}]
    \qty(\frac{1}{\bar{\epsilon}_\mathrm{EW}} + \log\frac{\mu_\mathrm{EW}^2}{m_\chi^2})
    \right.\cr
    &\left.-\frac{1}{2s_W^2}F_\Sigma(r_W)
    +\mathit{\Delta}Q_\chi\qty[F_\chi^{\gamma W}(r_W)+
    \frac{c_W^2}{s_W^2}F_\chi^{WZ}(r_W)]\right.\cr
    &\left.-\frac{1}{2}\qty(\mathit{\Delta}Q_\chi)^2
    \qty[4-2\log\frac{m_\chi^2}{m_\gamma^2}+F_\Sigma(r_Z)]\right\}
    \times\mathcal{M}^0_\mathrm{tree}\ .
\end{align}
As long as we consider transitions with
$\mathit{\Delta}Q_\chi = 1$,
the UV divergence must be canceled with
the counterterm determined by
the renormalization of the weak gauge coupling and
the wave-function of the weak gauge boson
because of the gauge invariance.
The counterterm contribution is given by
\begin{align}
\label{eq: EWcounter}
    \left.\mathcal{M}^0_{\mathrm{Vertex}(\mathrm{EW})}\right|_{\mathrm{CT}}
    = \mathcal{M}^0_{\mathrm{tree}}\times (\delta Z_1^W-\delta Z_2^W)\ ,
\end{align}
where $\delta Z_{1}^{W}$
is the renormalization constants of the weak gauge coupling and
$\delta Z_{2}^{W}$ is the renormalization factor of 
the wave function of the weak gauge boson.
In the on-shell renormalization scheme of the electroweak theory,
this counterterm is set to satisfy~\cite{Bohm:1986rj}
\begin{align}
    \delta Z_1^W-\delta Z_2^W= - \frac{2}{s_W^2}\frac{\alpha}{4\pi}
    \left(\frac{1}{\bar{\epsilon}_{\mathrm{EW}}}+\log \frac{\mu_{\mathrm{EW}}^2}{m_W^2}\right)\ ,
\end{align}
which successfully subtract the UV pole in Eq.~\eqref{eq: WF+Vertex chi w/ UV} when $\mathit{\Delta}Q_\chi=1$.
By combining the counterterm contribution~\eqref{eq: EWcounter} 
with Eq.~\eqref{eq: WF+Vertex chi w/ UV}, we obtain
\begin{align}
    \label{eq: WF+Vertex chi}
    \mathcal{M}^{\chi,0}_\mathrm{WF+Vertex(EW)}
    &=
    \left.\mathcal{M}^{\chi,0}_\mathrm{WF+Vertex(EW)}\right|_\mathrm{Virtual}
    +\left.\mathcal{M}^{0}_\mathrm{Vertex(EW)}\right|_\mathrm{CT} \cr
    &=\frac{\alpha}{4\pi}\mathcal{M}_\mathrm{tree}^0 
    \times\qty[2\log\frac{m_W}{m_\gamma}+F^\mathrm{WF+Vertex(EW)}_\chi(r_W)]\ ,
\end{align}
where we have set $\mathit{\Delta}Q_\chi=1$ and defined the functions
\begin{align}
    F^\mathrm{WF+Vertex(EW)}_\chi(r_W)
    = \,\,
    &-2
    +\frac{2c_W^2}{s_W^2}\log c_W\cr
    &-\frac{1}{2s_W^2}\bar{F}_\Sigma(r_W)
    -\frac{c_W^2}{2s_W^2}\bar{F}_\Sigma(r_Z)
    +F_\chi^{\gamma W}(r_W)+
    \frac{c_W^2}{s_W^2}F_\chi^{WZ}(r_W)\ ;\cr
    \bar{F}_\Sigma(r)
    =\,\,
    &4+3r^2 -3r^4\log r
    - \frac{3r(4+2r^2-r^4)}{\sqrt{4-r^2}}\cos^{-1}\frac{r}{2}\ ,
\end{align}
which remain finite in the limit of $r\to0$.

We would like to stress that
the coefficients of the
amplitudes $\mathcal{M}_\chi^{\gamma W,0}$~\eqref{eq: gamW to chi vertex},
$\mathcal{M}_\chi^{WZ,0}$~\eqref{eq: WZ to chi vertex}, $\mathcal{M}_{J_\chi^-}^{Z,0}$~\eqref{eq: Z to Jchi}, and $\mathcal{M}_{J_\chi^-}^{\gamma,0}$~\eqref{eq: gam to Jchi}
are determined only by the 
charge difference.
Given that the cancellation of
the UV divergence in Eq.~\eqref{eq: WF+Vertex chi w/ UV}
by the counterterm contribution~\eqref{eq: EWcounter},
which is
independent of the $\mathrm{SU}(2)_L\times\mathrm{U}(1)_Y$ representation of the chargino/neutralino multiplet,
the coefficient of 
Eq.~\eqref{eq: W to Jchi} is also independent of
the representation of the decaying particle 
as long as $\mathit{\Delta}Q_\chi = 1$.\,\footnote{In the case of the multiplet with $Y=0$ such as the Wino or the quintuplet,
the virtual $Z$-boson exchange in Fig.~\ref{fig:HisssinoVertexCorrectionA} does not exist
in the weak transition into the neutral component. 
Our discussion can include 
such a situation by considering 
the virtual $Z$-boson exchange with a zero coupling
to the neutral Majorana fermion.}
Consequently, the one-loop contribution from the wave-function and vertex corrections~\eqref{eq: WF+Vertex chi} does not depend on the representation of the external fermion for the
weak transition with $\mathit{\Delta}Q_\chi=1$.

\subsubsection{Corrections to SM Fermions}
The contribution from the wave-function renormalizations
and the vertex corrections to the SM fermions 
can be computed in a similar way.
The wave-function renormalization in the SM fermions gives rise to 
the amplitude,
\begin{align}
    \mathcal{M}_{\mathrm{WF(EW)}}^{f,0}
    = \mathcal{M}_{\mathrm{tree}}^0
    \times\qty[
    \frac{1}{2}\frac{d}{d\slashed{p}}\Sigma_{f_L}^{(\mathrm{EW})}(0)
    +\frac{1}{2}\frac{d}{d\slashed{p}}\Sigma_{f'_L}^{(\mathrm{EW})}(0)]\ ,
\end{align}
where for $F = f$ or $f'$,
\begin{align}
    \frac{d}{d\slashed{p}}\Sigma_F^\mathrm{(EW)}(0)
    =
    \frac{d}{d\slashed{p}}\Sigma_F^{\gamma\mathrm{(EW)}}(0)
    +\frac{d}{d\slashed{p}}\Sigma_{F_L}^{Z\mathrm{(EW)}}(0)
    +\frac{d}{d\slashed{p}}\Sigma_{F_L}^{W\mathrm{(EW)}}(0)\ ,
\end{align}
and
\begin{align}
    & \frac{d}{d\slashed{p}}\Sigma_F^\mathrm{\gamma(EW)}(0) 
    = -Q_F^2\frac{\alpha}{4\pi}
    \qty(\frac{1}{\bar{\epsilon}_{\mathrm{EW}}} 
    + \log \frac{\mu_{\mathrm{EW}}^2}{m_\gamma^2} 
    -\frac{1}{2})\ ;\\
	& \frac{d}{d\slashed{p}}\Sigma_{F_L}^{Z\mathrm{(EW)}}(0) 
    = -\frac{1}{c_W^2s_W^2}\qty(g_F^Z)^2
    \frac{\alpha}{4\pi}
    \qty(\frac{1}{\bar{\epsilon}_{\mathrm{EW}}} 
    + \log \frac{\mu_{\mathrm{EW}}^2}{m_Z^2} 
    -\frac{1}{2})\ ; \\
    & \frac{d}{d\slashed{p}}\Sigma_{F_L}^{W\mathrm{(EW)}}(0) 
    = -\qty(\frac{1}{\sqrt{2}s_W})^2
    \frac{\alpha}{4\pi}
    \qty(\frac{1}{\bar{\epsilon}_{\mathrm{EW}}} 
    + \log \frac{\mu_{\mathrm{EW}}^2}{m_W^2} 
    -\frac{1}{2})\ .
\end{align}
\begin{figure}[t]
    \centering
    \subcaptionbox{\label{fig:SMfermionVertexCorrectionA}}
    {\includegraphics[width=0.23\textwidth]{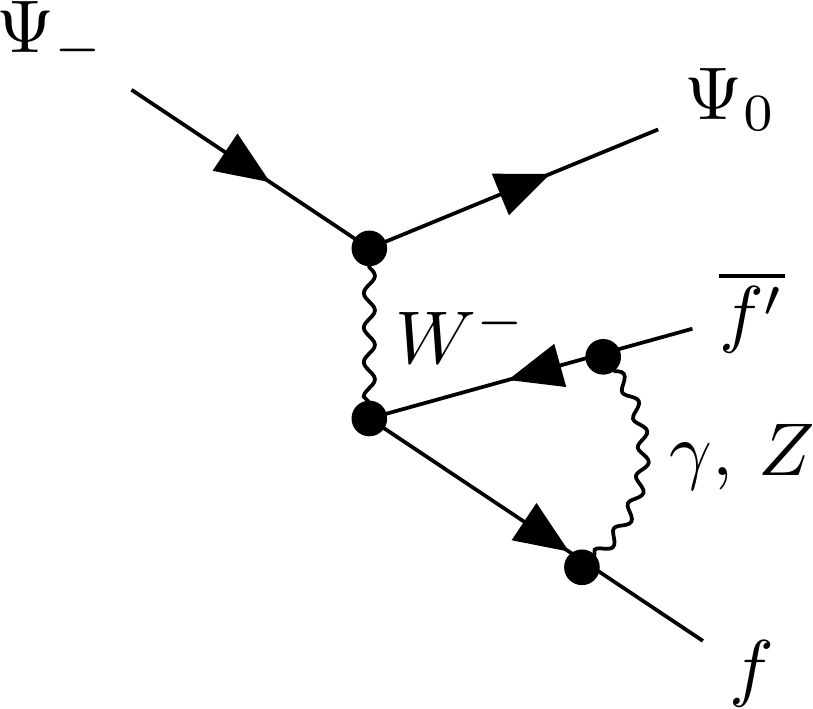}}
    \hspace{20pt}
    \subcaptionbox{\label{fig:SMfermionVertexCorrectionB}}
    {\includegraphics[width=0.23\textwidth]{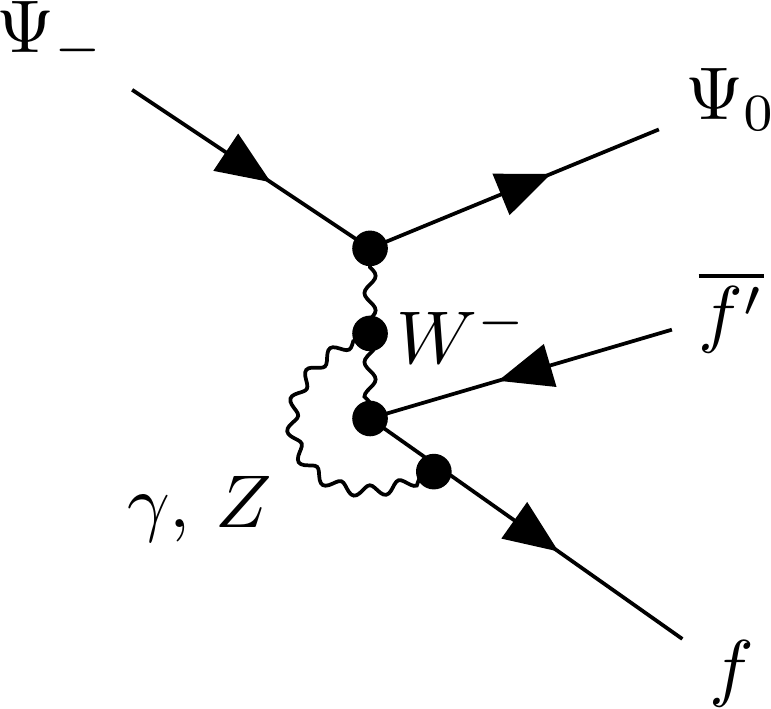}}
    \subcaptionbox{\label{fig:SMfermionVertexCorrectionC}}
    {\includegraphics[width=0.23\textwidth]{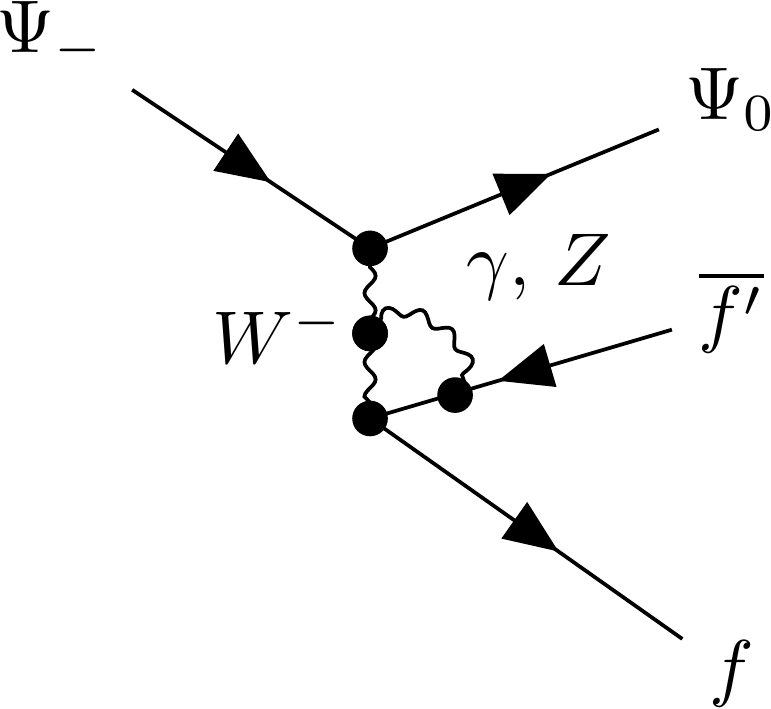}}
    \hspace{20pt}
    \caption{Radiative corrections to SM fermion's vertex.}
    \hspace{20pt}
\end{figure}
The radiative corrections to the SM fermion's vertex
(Figs.~\ref{fig:SMfermionVertexCorrectionA}-\ref{fig:SMfermionVertexCorrectionC})
are given by
\begin{align}
    \mathcal{M}^{\gamma,0}_{ff'} 
    &=  
    \mathcal{M}_{\mathrm{tree}}^0
      \times 
      Q_{f}Q_{f'}\frac{\alpha}{4\pi}
      \left(\frac{1}{\bar{\epsilon}_{\mathrm{EW}}} 
      + \log\frac{\mu_{\mathrm{EW}}^2}{m_\gamma^2}-\frac{1}{2}\right)\ ; 
      \\
    \mathcal{M}^{Z,0}_{ff'} 
    &=  
    \mathcal{M}_{\mathrm{tree}}^0
      \times 
      \frac{g_{f}^Zg_{f'}^Z}{c_W^2 s_W^2}\frac{\alpha}{4\pi}
      \left(\frac{1}{\bar{\epsilon}_{\mathrm{EW}}} 
      + \log\frac{\mu_{\mathrm{EW}}^2}{m_Z^2}-\frac{1}{2}\right)\ ; 
      \\
      \mathcal{M}^{\gamma W,0}_{ff'} 
      &=  
      \mathcal{M}_{\mathrm{tree}}^0
      \times \mathit{\Delta}Q_f\frac{\alpha}{4\pi}
      \left(\frac{3}{\bar{\epsilon}_{\mathrm{EW}}} 
      +3\log\frac{\mu_{\mathrm{EW}}^2}{m_W^2}+\frac{5}{2}\right)\ ; 
     \\
      \mathcal{M}^{WZ,0}_{ff'} 
      &= 
      \mathcal{M}_{\mathrm{tree}}^0
      \times \frac{g_{f'}^Z-g_{f}^Z}{s_W^2}\frac{\alpha}{4\pi}
      \left(\frac{3}{\bar{\epsilon}_{\mathrm{EW}}} 
      +3\log\frac{\mu_{\mathrm{EW}}^2}{m_W^2}+\frac{5}{2}
      + \frac{3}{s_W^2}\log c_W^2\right)\ ,
\end{align}
where we have defined the charge difference $\mathit{\Delta}Q_f = Q_{f'}-Q_{f} = 1$.

From these amplitudes, we obtain 
the virtual $\gamma/Z/W$ corrections to the charged-current operator 
composed of the SM fermions,
\begin{align}
    \mathcal{M}_{J_f^-}^{\gamma,0}
    &= -\frac{\qty(\mathit{\Delta}Q_f)^2}{2}
    \frac{\alpha}{4\pi}
      \left(\frac{1}{\bar{\epsilon}_{\mathrm{EW}}} 
      + \log\frac{\mu_{\mathrm{EW}}^2}{m_\gamma^2}-\frac{1}{2}\right)
      \times\mathcal{M}_\mathrm{tree}^0\ ;\\
      \mathcal{M}_{J_f^-}^{Z,0}
      &= -\frac{\qty(g_{f'Z}-g_{fZ})^2}{2c_W^2s_W^2}
    \frac{\alpha}{4\pi}
      \left(\frac{1}{\bar{\epsilon}_{\mathrm{EW}}} 
      + \log\frac{\mu_{\mathrm{EW}}^2}{m_Z^2}-\frac{1}{2}\right)
      \times\mathcal{M}_\mathrm{tree}^0 \cr
    &= -\frac{c_W^2}{s_W^2}
    \frac{\qty(\mathit{\Delta}Q_f)^2}{2}
    \frac{\alpha}{4\pi}
      \left(\frac{1}{\bar{\epsilon}_{\mathrm{EW}}} 
      + \log\frac{\mu_{\mathrm{EW}}^2}{m_Z^2}-\frac{1}{2}\right)
      \times\mathcal{M}_\mathrm{tree}^0\ ;\\
    \mathcal{M}_{J_f^-}^{W,0}
    &= -\frac{1}{2s_W^2}
    \frac{\alpha}{4\pi}
      \left(\frac{1}{\bar{\epsilon}_{\mathrm{EW}}}
      + \log\frac{\mu_{\mathrm{EW}}^2}{m_W^2}-\frac{1}{2}\right)
      \times\mathcal{M}_\mathrm{tree}^0\ .
\end{align}
The total of virtual contributions from the wave-function renormalization
and the vertex corrections to the SM fermions
is given by
\begin{align}
    \label{eq: WF+Vertex F w/ UV}
    \left.\mathcal{M}^{f,0}_\mathrm{WF+Vertex(EW)}\right|_\mathrm{Virtual}
    =\,\,
    &\sum_{V=\gamma,Z,W}\mathcal{M}_{J_f^-}^{V,0} 
    + \mathcal{M}_{ff'}^{\gamma W,0}+\mathcal{M}_{ff'}^{WZ,0}
    \cr
    =\,\,
    &\frac{\alpha}{4\pi}\left\{\qty[\frac{2}{s_W^2}-\frac{(5-\mathit{\Delta}Q_f)(1-\mathit{\Delta}Q_f)}{2s_W^2}]
    \qty(\frac{1}{\bar{\epsilon}_\mathrm{EW}} + \log\frac{\mu_\mathrm{EW}^2}{m_W^2})\right.\cr
    &\left.-\frac{1}{2s_W^2}\qty(-\frac{1}{2})
    +\mathit{\Delta}Q_f\qty[\frac{5}{2}+
    \frac{c_W^2}{s_W^2}\qty(\frac{5}{2}+\frac{6}{s_W^2}\log c_W)]\right.\cr
    &\left.-\frac{1}{2}\qty(\mathit{\Delta}Q_f)^2
    \qty[\log\frac{m_W^2}{m_\gamma^2}-\frac{1}{2}
    +\frac{c_W^2}{s_W^2}\qty(\log c_W -\frac{1}{2})]\right\}
    \times\mathcal{M}_\mathrm{tree}^0\ .
\end{align}
By combining the counterterm contribution~\eqref{eq: EWcounter} 
with Eq.~\eqref{eq: WF+Vertex F w/ UV}, we obtain
the UV finite amplitude,
\begin{align}
    \mathcal{M}^{f,0}_\mathrm{WF+Vertex(EW)}
    &= \left.\mathcal{M}^{f,0}_\mathrm{WF+Vertex}\right|_\mathrm{Virtual}
    +\left.\mathcal{M}^0_\mathrm{Vertex(EW)}\right|_\mathrm{CT}\cr
    &= \frac{\alpha}{4\pi}\mathcal{M}_\mathrm{tree}^0
    \times\qty[-\log\frac{m_W}{m_\gamma}
    +\frac{3}{s_W^2}+\frac{(6-s_W^2)c_W^2}{s_W^4}\log c_W]\ ,
\end{align}
where we have set $\mathit{\Delta}Q_f = 1$.

\subsection{Box Contribution}
\label{sec: box contributions}
\begin{figure}[t]
    \centering
    \subcaptionbox{\label{fig:general photonboxa}}{\includegraphics[width=0.23\textwidth]{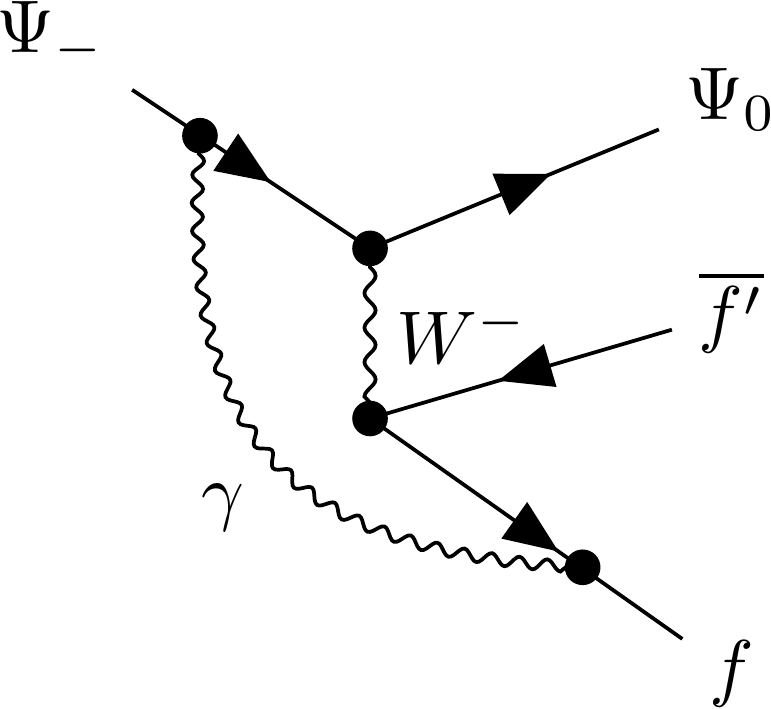}}
    \subcaptionbox{\label{fig:general photonboxb}}{\includegraphics[width=0.23\textwidth]{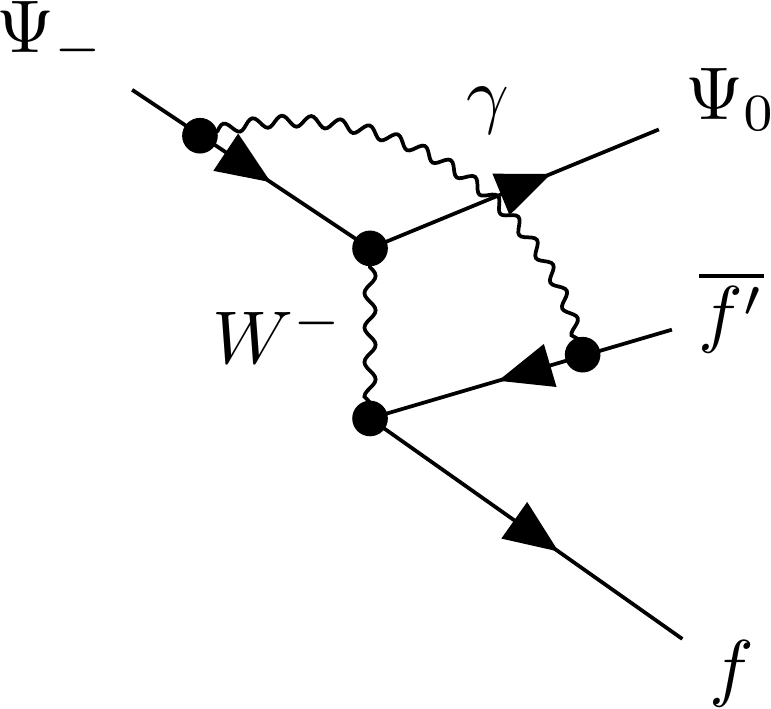}}
    \caption{Box contribution from virtual photon.}
\end{figure}
\begin{figure}[t]
    \centering
    \subcaptionbox{\label{fig:general Z boson box a}}
    {\includegraphics[width=0.23\textwidth]{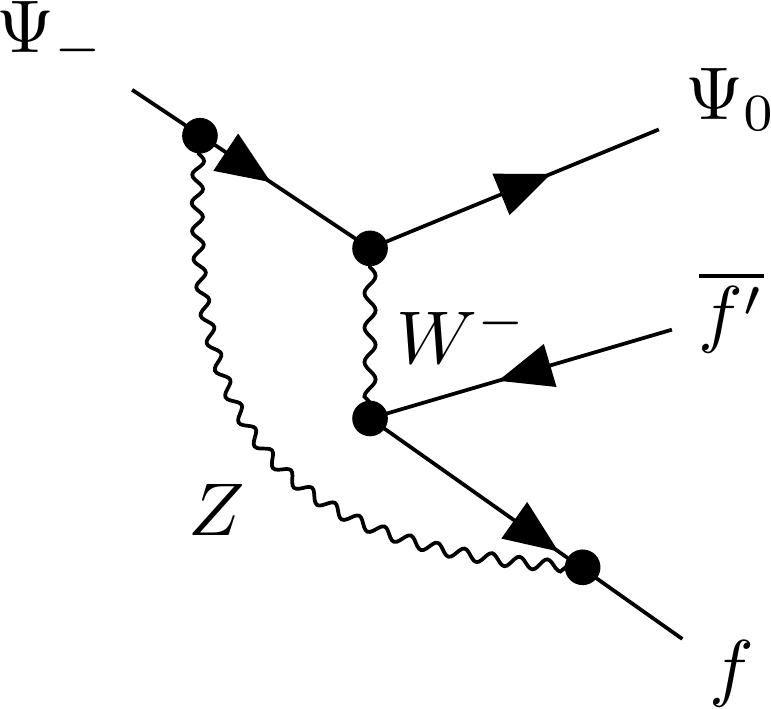}}
    \hspace{20pt}
    \subcaptionbox{\label{fig:general Z boson box b}}
    {\includegraphics[width=0.23\textwidth]{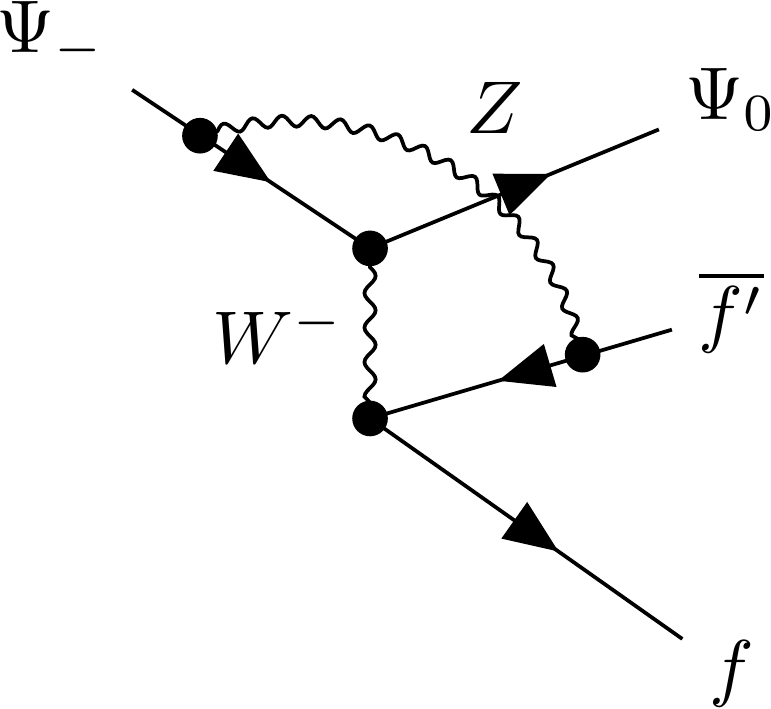}}
    \subcaptionbox{\label{fig:general Z boson box c}}
    {\includegraphics[width=0.23\textwidth]{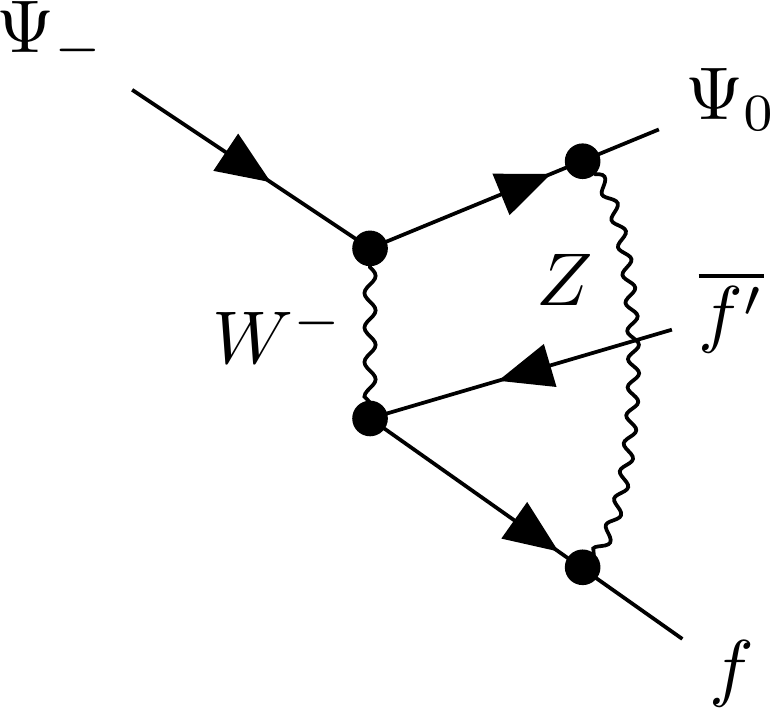}}
    \hspace{20pt}
    \subcaptionbox{\label{fig:general Z boson box d}}
    {\includegraphics[width=0.23\textwidth]{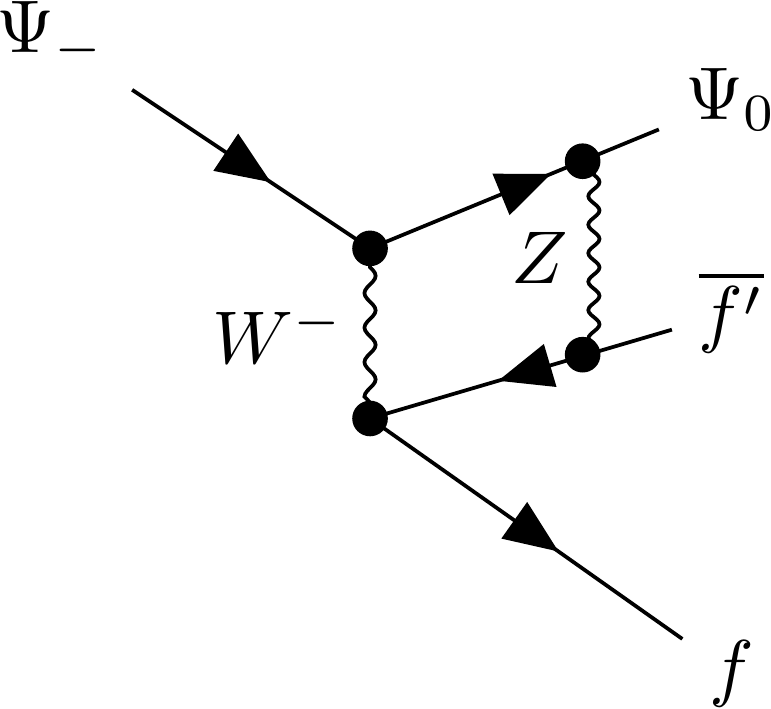}}
    \hspace{20pt}
    \caption{Box contributions from virtual $Z$ boson.}
\end{figure}

The box contributions generated by the virtual photon
exchange (Fig~\ref{fig:general photonboxa} and \ref{fig:general photonboxb}) can be written in the form:
\begin{align}
    \mathcal{M}_{\mathrm{box}, a}^{\gamma\mathrm{(EW)},0}
    &= Q_{f}Q_{\chi^-} \frac{\alpha}{4\pi}
        \qty{\mathcal{M}_\mathrm{tree}^0\times 
        \qty[\log\frac{m_W^2}{m_\gamma^2}+F^{\gamma\mathrm{(EW)}}_V(r_W)]
    + \mathcal{M}_A^0\times F^{\gamma\mathrm{(EW)}}_A(r_W)}\ ; \cr
    \mathcal{M}_{\mathrm{box}, b}^{\gamma\mathrm{(EW)},0}
    &=  Q_{f'}Q_{\chi^-}\frac{\alpha}{4\pi}\qty{-\mathcal{M}_\mathrm{tree}^0
    \times \qty[\log\frac{m_W^2}{m_\gamma^2}+F^{\gamma\mathrm{(EW)}}_V(r_W)]
    + \mathcal{M}_A^0\times F^{\gamma\mathrm{(EW)}}_A(r_W)}\ ,
\end{align}
where
$\mathcal{M}_A^0$ is the axial current 
contribution given by Eq.~\eqref{eq: axial amplitude}
and
\begin{align}
    \label{eq: photon box vector}
    F^{\gamma\mathrm{(EW)}}_V(r_W) 
    &= -\frac{r_W^2}{2} + \frac{1}{2}r_W^4\log(r_W)
    +r_W\qty(1+\frac{r_W^2}{2})\sqrt{4-r_W^2}\cot^{-1}\qty(\frac{r_W}{\sqrt{4-r_W^2}})\ ;
    \\
    \label{eq: photon box axial}
    F^{\gamma\mathrm{(EW)}}_A(r_W) &=
    \frac{r_W}{6}
    \left[-r_W+r_W(6+r_W^2)\log(r_W)
    +(8+r_W^2)\sqrt{4-r_W^2}\cot^{-1}\qty(\frac{r_W}{\sqrt{4-r_W^2}})
    \right]\ .
\end{align}
Hence, the photon box contribution can be summarized as
\begin{align}
    \mathcal{M}_\mathrm{box}^{\gamma,0}
    =
    \mathcal{M}_\mathrm{tree}^0\times 
    \mathit{\Delta}Q_\chi\mathit{\Delta}Q_f
    \frac{\alpha}{4\pi}
        \qty[\log\frac{m_W^2}{m_\gamma^2}+F^{\gamma\mathrm{(EW)}}_V(r_W)]
        -\mathcal{M}_A^0\times 
        \overline{Q}\frac{\alpha}{4\pi}
        F^{\gamma\mathrm{(EW)}}_A(r_W)\ ,
\end{align}
where $\overline{Q} = Q_f+Q_{f'}$. The coefficient of the vector-like part
depends only on the charge differences, and hence, is independent
of the representation of the chargino/neutralino multiplet.

$Z$-boson exchanges generate 
the box diagrams depicted in 
Figs.~\ref{fig:general Z boson box a}$-$\ref{fig:general Z boson box d},
which have contributions given by
\begin{align}
    \mathcal{M}^{Z\mathrm{(EW)},0}_{\mathrm{box},a}
    =\, &\frac{g_{\chi^-}^Zg_{f}^Z}{4 c_W^4 s_W^4}
    \frac{\alpha}{4\pi}
    \qty[\mathcal{M}_\mathrm{tree}^0
    \times F^{Z\mathrm{(EW)}}_V(r_W)
    + \mathcal{M}_A^0
    \times F^{Z\mathrm{(EW)}}_A(r_W)]\ ; \\
    \mathcal{M}^{Z\mathrm{(EW)},0}_{\mathrm{box},b}
    =\, &\frac{g_{\chi^-}^Zg_{f'}^Z}{4 c_W^4 s_W^4}
    \frac{\alpha}{4\pi}
    \qty[-\mathcal{M}_\mathrm{tree}^0
    \times F^{Z\mathrm{(EW)}}_V(r_W)
    + \mathcal{M}_A^0
    \times F^{Z\mathrm{(EW)}}_A(r_W)]\ ; \\
    \mathcal{M}^{Z\mathrm{(EW)},0}_{\mathrm{box},c}
    =\, &
    \frac{g_{\chi^0}^Zg_{f}^Z}{4 c_W^4 s_W^4}
    \frac{\alpha}{4\pi}
    \qty[-\mathcal{M}_\mathrm{tree}^0
    \times F^{Z\mathrm{(EW)}}_V(r_W)
    +\mathcal{M}_A^0
    \times F^{Z\mathrm{(EW)}}_A(r_W)]\ ; \\
    \mathcal{M}^{Z\mathrm{(EW)},0}_{\mathrm{box},d}
    =\, &
    \frac{g_{\chi^0}^Zg_{f'}^Z}{4 c_W^4 s_W^4}
    \frac{\alpha}{4\pi}
    \qty[\mathcal{M}_\mathrm{tree}^0
    \times F^{Z\mathrm{(EW)}}_V(r_W)
    +\mathcal{M}_A^0
    \times F^{Z\mathrm{(EW)}}_A(r_W)]\ ,
\end{align}
where 
\begin{align}
    \label{eq: Z box vector}
    F^{Z\mathrm{(EW)}}_V(r_W) 
    =\, &
    -2c_W^4(4+r_W^4)\log(c_W)-2c_W^2r_W^2
    +2c_W^4r_W^2-(1-c_W^4)r_W^4\log(\frac{c_W^2}{r_W^2})
    \notag \\
    &\hspace{20pt}
    -2c_W^4r_W(2+r_W^2)\sqrt{4-r_W^2}\cot^{-1}\qty(\frac{r_W}{\sqrt{4-r_W^2}})
    \notag \\
    &\hspace{20pt}
    +2r_W(2c_W^2+r_W^2)\sqrt{4c_W^2-r_W^2}\cot^{-1}\qty(\frac{r_W}{\sqrt{4c_W^2-r_W^2}})\ ;
    \\
    \label{eq: Z box axial}
    F^{Z\mathrm{(EW)}}_A(r_W) =\, 
    &\frac{r_W}{3}
    \left[-2c_W^4r_W(6+r_W^2)\log(c_W)
    -r_W\qty(6s_W^2 c_W^2+s_W^2 (1+c_W^2)r_W^2)
    \log(\frac{c_W^2}{r_W^2})
    \right. \notag \\
    &\left.\hspace{20pt}
    -2c_W^2s_W^2r_W-2c_W^4(8+r_W^2)\sqrt{4-r_W^2}
    \cot^{-1}\qty(\frac{r_W}{\sqrt{4-r_W^2}})
    \right. \notag \\
    &\left.\hspace{20pt}
    + 2(8c_W^2+r_W^2)\sqrt{4c_W^2-r_W^2}\cot^{-1}\qty(\frac{r_W}{\sqrt{4c_W^2-r_W^2}})\right]\ .
\end{align}
Hence, the $Z$-boson box contributions can be summarized as
\begin{align}
    \mathcal{M}_\mathrm{box}^{Z\mathrm{(EW)},0}
    =\,\,
    &\frac{\qty(g_{\chi^-}^Z-g_{\chi^0}^Z)\qty(g_{f}^Z-g_{f'}^Z)}{4c_W^4s_W^4}
    \frac{\alpha}{4\pi}
        \mathcal{M}_\mathrm{tree}^0\times 
        F^{Z\mathrm{(EW)}}_V(r_W)\cr
        &+\frac{\qty(g_{\chi^-}^Z+g_{\chi^0}^Z)\qty(g_{f}^Z+g_{f'}^Z)}{4c_W^4s_W^4}
        \frac{\alpha}{4\pi}
        \mathcal{M}_A^0\times F^{Z\mathrm{(EW)}}_A(r_W) \cr
    =\,\,
    &\frac{\mathit{\Delta}Q_\chi\mathit{\Delta}Q_f}{4s_W^4}
    \frac{\alpha}{4\pi}
        \mathcal{M}_\mathrm{tree}^0\times 
        F^{Z\mathrm{(EW)}}_V(r_W)\cr
        &+\frac{\qty(c_W^2\overline{Q}_\chi-2Y_\chi)\qty(c_W^2\overline{Q}-2Y)}{4c_W^4s_W^4}
        \frac{\alpha}{4\pi}
        \mathcal{M}_A^0\times F^{Z\mathrm{(EW)}}_A(r_W)\ .
\end{align}
where $\overline{Q}_\chi = Q_{\chi^-}+Q_{\chi^0} = -1$ and 
$Y_\chi(Y)$ is the hypercharge of the Higgsino doublet (the SM fermion doublet). 
The coefficient of the vector-like part
depends only on the charge differences, and hence, is independent
of the representation of the chargino/neutralino multiplet.

By using $2Y_\chi = \overline{Q}_\chi = -1$, $2Y = \overline{Q}$,
and $\mathit{\Delta}Q_\chi = \mathit{\Delta}Q_f = 1$,
the total box contribution can be reduced as
\begin{align}
    \mathcal{M}_\mathrm{Box(EW)}^0
    &= \mathcal{M}^{\gamma\mathrm{(EW)},0}_{\mathrm{box}}
    +\mathcal{M}^{Z\mathrm{(EW)},0}_{\mathrm{box}}\cr
    =\, &\mathcal{M}_\mathrm{tree}^0
    \times\frac{\alpha}{4\pi}
    \qty[\log\frac{m_W^2}{m_\gamma^2}+F^{\mathrm{Box(EW)}}_V(r_W)]
    +\mathcal{M}_A^0
    \times\frac{\alpha}{4\pi}F_A^{\mathrm{Box(EW)}}(r_W)\ , 
\end{align}
where 
\begin{align}
    F^{\mathrm{Box(EW)}}_V(r_W)
    &= F^{\gamma\mathrm{(EW)}}_V(r_W)
    +\frac{1}{4s_W^4}
    F_V^{Z\mathrm{(EW)}}(r_W)\ ;\\
    F^{\mathrm{Box(EW)}}_A(r_W)
    &= -\overline{Q}\qty[F_A^{\gamma\mathrm{(EW)}}(r_W)
    +\frac{1}{4c_W^4}F_A^{Z\mathrm{(EW)}}(r_W)]\ .
\end{align}

\subsection{Total Contribution}
\label{sec: EW total contribution}
In addition to the above contributions, 
we have to include the amplitude
coming from the vacuum polarization of the $W$-boson:
\begin{align}
    \mathcal{M}_\mathrm{VP(EW)}^0
    = \mathcal{M}_\mathrm{tree}^0
    \times\qty[1-\frac{\alpha}{4\pi}
    \frac{\hat{\Sigma}_{WW}^\mathrm{1PI}(0)}{m_W^2}]^{-1}
    \simeq
    \mathcal{M}_\mathrm{tree}^0
    \times\qty[1+\frac{\alpha}{4\pi}
    \frac{\hat{\Sigma}_{WW}^\mathrm{1PI}(0)}{m_W^2}]\ .
\end{align}
Putting all the contributions together, 
we can write down the electroweak corrections in the form of
\begin{align}
    \mathcal{M}_\mathrm{Virtual(EW)}^0 
    &= \mathcal{M}_\mathrm{WF+Vertex(EW)}^{\chi,0}
    +\mathcal{M}_\mathrm{WF+Vertex(EW)}^{f,0}
    + \mathcal{M}_\mathrm{Box(EW)}^0 
    +\mathcal{M}_\mathrm{VP(EW)}^0\cr
    &= \mathcal{M}_\mathrm{tree}^0
    \times \frac{\alpha}{4\pi}\qty[3\log\frac{m_W}{m_\gamma}
    + \frac{\hat{\Sigma}_{WW}^\mathrm{1PI}(0)}{m_W^2}
    + F^\mathrm{Virtual(EW)}_V(r_W)]
    \notag \\
    &\hspace{20pt}
    +\mathcal{M}_A^0\times \frac{\alpha}{4\pi}F^\mathrm{Virtual(EW)}_A(r_W)\ ,
\end{align}
where
\begin{align}
    F^\mathrm{Virtual(EW)}_V(r_W)
    & = F_\chi^\mathrm{WF+Vertex(EW)}(r_W) 
    + F_f^{\mathrm{WF+Vertex(EW)}}
    + F_V^{\mathrm{Box(EW)}}(r_W)\ ; \\
    F^\mathrm{Virtual(EW)}_A(r_W) & = F^\mathrm{Box(EW)}_A(r_W)\ .
\end{align}
The finite correction to the vector-like part, $F_V^{\mathrm{Virtual(EW)}}(r_W)$, 
does not depend on 
both the charge difference $\overline{Q}$ and the representation of the decaying particle.
Here, the analytical expression of $F_{V}^\mathrm{Virtual(EW)}(r_W)$
is given by Eq.~\eqref{eq: vector-like form factor}.

\section{Renormalization Group Analysis to Fermi Interaction}
\label{sec: RG analysis}
\renewcommand{\theequation}{\thesection.\arabic{equation}}
The vector-like part of the one-loop level amplitude~\eqref{eq: EW virtual corrections} does not depend on
the charge difference in the final state, $\overline{Q} = Q_{f} - \qty(-Q_{f'})$. 
This means that 
the short-distance correction is insensitive to the choice of 
the SM particles. 
The situation is different from the decay of the tau lepton;
The hadronic and the leptonic decay have different the leading logs 
in their short-distance corrections in the tau-lepton case.
In the following, we clarify the leading-log structure
in the case of the Higgsino by the language of the Wilson coefficient.

Let us introduce the operator basis
\begin{align}
   \mathcal{O}_V = \qty(\overline{\Psi}_{\chi^0}\gamma^\mu \Psi_{\chi^-})
        \qty(\overline{\Psi}_f\gamma_\mu P_L\Psi_{f'})\ ;\quad
    \mathcal{O}_A = \qty(\overline{\Psi}_{\chi^0}\gamma^\mu\gamma_5 \Psi_{\chi^-})
        \qty(\overline{\Psi}_f\gamma_\mu P_L\Psi_{f'})\ .
\end{align}
Then the renormalized operators at the scale $\mu$ are given by
\begin{align}
    \mathcal{O}_i = Z_{ij}(\mu)\qty[\mathcal{O}_j]_\mu\,\,\,(i, j = V, A)\ .
\end{align}
Explicit calculations of photon loops provide us the renormalization factors,
\begin{align}
    Z_{VV}(\mu) &= 1 -\frac{\alpha}{4\pi}
     \frac{3}{2}\qty(\frac{1}{\bar{\epsilon}} + \log\mu^2) = Z_{AA}(\mu)\ ;\\
     Z_{VA}(\mu) &= \frac{\alpha}{4\pi}
     \frac{3}{2}\overline{Q}\qty(\frac{1}{\bar{\epsilon}} + \log\mu^2) = Z_{AV}(\mu)\ ,
\end{align}
where we have used $Q_\chi = -1$ and $Q_{f}-Q_{f'} = -1$. Note that the QCD corrections
of $O(\alpha^0\alpha_s)$ are absent from 
the renormalization factors even if $f$ and $f'$
are quarks.
Therefore, the anomalous dimension matrix for $(\mathcal{O}_V, \mathcal{O}_A)$
at one-loop level
is given by
\begin{align}
    \gamma_\mathcal{O}(\alpha) = 
    Z(\mu)^{-1}\,\mu\frac{\partial}{\partial\mu}Z(\mu)
    =\frac{\alpha}{4\pi}
    \cdot3\mqty(-1 & \overline{Q} \\ \overline{Q} &-1)\ .
\end{align}
The operator basis to diagonalize the matrix is given by
\begin{align}
    \mathcal{O}_V^L = \qty(\mathcal{O}_V - \mathcal{O}_A)/2\ ;\,\,\,
    \mathcal{O}_Y^R = \qty(\mathcal{O}_V + \mathcal{O}_A)/2\ .
\end{align}
The Wilson coefficients of the effective Lagrangian,
\begin{align}
    \mathcal{L}_\mathrm{FF} 
    = -2\sqrt{2}G_{F1}(\mu)\qty[\mathcal{O}_V^L]_\mu 
    + 4\sqrt{2}G_{F2}(\mu)\qty[\mathcal{O}_Y^R]_\mu\ ,
\end{align}
obey the RG equations
\begin{align}
    \mu\frac{d}{d\mu}G_{F1}(\mu) = \gamma_V^L(\alpha)G_{F1}(\mu)\ ;\,\,\,
    \mu\frac{d}{d\mu}G_{F2}(\mu) = \gamma_Y^R(\alpha)G_{F2}(\mu)\ ,
\end{align}
where
\begin{align}
    \gamma_V^L(\alpha) = -\frac{\alpha}{4\pi}\cdot 3(1+\overline{Q})\ ; \,\,\,
    \gamma_Y^R(\alpha) = -\frac{\alpha}{4\pi}\cdot 3(1-\overline{Q})\ .
\end{align}
By solving the RG equations 
under the condition,
$G_{F1}(m_W) = G_{F1}(m_W) = G_F^0$,
at the leading-order of $\alpha$,
we obtain
\begin{align}
    G_{F1}(\mu_\mathrm{IR}) 
    &= \qty[1+\frac{\alpha}{4\pi}\cdot 
    3\qty(1+\overline{Q})\log\frac{m_W}{\mu_\mathrm{IR}}]\GFtree\ ; \cr
    G_{F2}(\mu_\mathrm{IR}) 
    &= \qty[1+\frac{\alpha}{4\pi}\cdot 
    3\qty(1-\overline{Q})\log\frac{m_W}{\mu_\mathrm{IR}}]\GFtree\ .
\end{align}
The decay amplitude defined by 
the coupling at the renormalization scale $\mu_\mathrm{IR}$,
\begin{align}
    \mathcal{M}_{\overline{\mathrm{MS}},\mathrm{1-loop}}
    =\, &-2\sqrt{2}G_{F1}(\mu_\mathrm{IR}) \bar{u}_{f}(p_3)\gamma^\mu P_L u_{\chi^-}(p_1)\bar{u}_{\chi^0}(p_2)\gamma_\mu P_L v_{f'}(p_4)\cr
    &+4\sqrt{2}G_{F2}(\mu_\mathrm{IR})\bar{u}_f(p_3)P_R u_{\chi^-}(p_1)\bar{u}_{\chi^0}(p_2)P_L v_{f'}(p_4)\ ,
\end{align}
provides the leading-log decay rate,
\begin{align}
    \Gamma_\mathrm{LL}\qty(\chi^-\to\chi^0ff')
    &= \qty{1+\frac{\alpha}{4\pi}\cdot 
    \qty[3\qty(1+\overline{Q})+3\qty(1-\overline{Q})]\log\frac{m_W}{\mu_\mathrm{IR}}}
    \Gamma_\mathrm{tree}
    \qty(\chi^-\to\chi^0ff')\cr
    &= \qty{1+\frac{\alpha}{4\pi}\cdot 
    6\log\frac{m_W}{\mu_\mathrm{IR}}}
    \Gamma_\mathrm{tree}\qty(\chi^-\to\chi^0ff')\ .
\end{align}
Now we observe that the leading log correction to the decay rate is independent of 
$\overline{Q}$.

\section{QED Correction to Single Pion Decay Rate}
\label{app: not MSbar}
\renewcommand{\theequation}{\thesection.\arabic{equation}}
The QED correction to the pion decay rate 
including the real photon emissions
is given in Ref.~\cite{Knecht:1999ag} (see also Refs.~\cite{Marciano:1993sh, Decker:1994ea}\,\footnote{We need to correct $19\to 17$ and $13\to 11$ in 
Eq.\,(7b)
of Ref.\,\cite{Marciano:1993sh}.}),
\begin{align}
\label{eq:GpiNLO}
  \frac{\delta\Gamma_\pi}
   {\Gamma_{\pi}} =\,\,
   & \frac{\alpha}{2\pi}
    \left[
    -\frac{3}{2}\left(\frac{1}{\bar{\epsilon}} + \log \frac{\bar{\mu}^2}{m_{\pi^\pm}^2}\right)
    +6\log \frac{m_\mu}{m_{\pi^\pm}} + \frac{11}{4}-\frac{2}{3}\pi^2 + f_\pi\left(\frac{m_\mu}{m_{\pi^\pm}}\right)
    \right]  \cr
    &
    + 
    e^2 
    \Bigg[\frac{8}{3} K_1 + \frac{20}{9} K_5+4 K_{12}\cr &
    \hspace{1cm}-\hat{X}_6 -\frac{4}{3} (X_1+\hat{X}_1)
    -4 (X_2+ \hat{X}_2)
    +4  X_3
    \Bigg]\ , 
    \end{align}
where\,\footnote{We omit the contributions from the LEC $K_{2,6}$,
which are eventually canceled 
by taking the ratio between
the NLO chargino decay rate 
and the NLO charged pion decay rate.}
    \begin{align}
        f_\pi(r)=\,\,
        &4 \left(\frac{1+r^2}{1-r^2}\log r - 1\right)\log(1-r^2) + 4 \frac{1+r^2}{1-r^2}
    \mathrm{Li}_2(r^2) \cr
& -\frac{r^2(8-5r^2)}{(1-r^2)^2}\log r - \frac{r^2}{1-r^2} \left(\frac{3}{2}+\frac{4}{3}\pi^2\right)\ .
\end{align}
In D\&M analysis, three 
mass scales, $\mu$, $\mu_0$, $\mu_1$
are introduced, 
although final result does not depend on $\mu$'s.
Here, we set $\mu = \mu_0=\mu_1=\bar{\mu}$ for simplicity.
The expression for $K_{12}$ is given in Eq.\,\eqref{eq:K12}, and the expressions for the 
relevant $X$'s are given by,
\begin{align}
    e^2(X_1+\hat{X}_1) =& \frac{3\alpha}{16\pi}
    \left(
    \log\frac{M_V^2}{m_Z^2}
    + 1-\frac{c_V}{2M_V^2}\right)\ ; \\
    e^2(X_2 + \hat{X}_2)=&
    \frac{\alpha}{16\pi}
    \Bigg[\frac{5}{\bar{\epsilon}}+
    \frac{3M_V^2M_A^2}{(M_A^2-M_V^2)^2}
    \log\frac{M_A^2}{M_V^2}
    -\frac{3M_A^2}{M_A^2-M_V^2}
    +\frac{15}{2}
    \Bigg]\cr
    &-\frac{1}{4}e^2(g_{02}^{r(\mathrm{PV})}(\bar{\mu})-g_{03}^{r(\mathrm{PV})}(\bar{\mu}))\ ,
    \\
    e^2 X_3 =&
        \frac{3\alpha}{8\pi}
    \left(\frac{1}{\bar{\epsilon}}+\log\frac{\bar{\mu}^2}{M_V^2} 
    + \frac{M_V^2}{M_A^2-M_V^2}\log\frac{M_A^2}{M_V^2} +
    \frac{5}{6}
    \right)\ ;\\
    e^2 \hat{X}_6 =& - \frac{\alpha}{4\pi}
        \frac{1}{\bar{\epsilon}} -2 e^2g_{00}^{r(\mathrm{PV})}(\bar{\mu})+ \frac{\alpha}{8\pi}\ .
\end{align}
Note that  we misinterpreted the subtraction scheme for the $K$-terms and $X$-terms in Ref.\,\cite{Descotes-Genon:2005wrq} as the minimal subtraction in Ref.\,\cite{Ibe:2022lkl}. 
However, the error of the decay rates 
of the single pion mode 
caused by the changes of the constants
is negligible compared to the other uncertainties associated with the hadron models.

\section{Hadronic Decay of Tau Lepton}
\label{sec: Hadronic Decay of Tau}
\renewcommand{\theequation}{\thesection.\arabic{equation}}
\subsection{Inclusive Multi-Hadron Decay}
The amplitude of the decay, $\tau^-(p_1)\rightarrow\nu_\tau(p_2)+h^-(p_f)$, 
is given by
\begin{align}
    \label{eq: inclusive tau amplitude}
    \mathcal{M}(\tau^-\rightarrow\nu_\tau h^-)
    &= -\frac{G_F V^*_{uD}}{\sqrt{2}}\bar{u}_{\nu_\tau}(p_2)\gamma^\mu(1-\gamma_5)u_{\tau}(p_1)
    \mel{h^-(p_f)}{U_\mu(0)}{0}\ ,
\end{align}
where $p_f$ collectively denotes final state momenta, 
and $U_\mu(x)$ is the vector or axial current of the light quarks,
that is, $U_\mu = V_\mu = \overline{\Psi}_D\gamma^\mu \Psi_u$ or 
$U_\mu = A_\mu = \overline{\Psi}_D\gamma^\mu \gamma_5 \Psi_u\ (D = d, s)$.

The matrix element of the current $U_\mu$ in Eq.~\eqref{eq: inclusive tau amplitude}
is responsible for hadronization.
In the decay rates,
this matrix element is characterized by functions of the invariant mass of
the hadrons in the final state, $q^2 = p_f^2 = (p_1-p_2)^2$, via
the optical theorem.
Those function are called 
the spectral functions $v_J(q^2)$ and $a_J(q^2)$~\cite{Tsai:1971vv},
which are defined by
\begin{align}
&\sum_h \frac{1}{S_h}\int\prod_{f}\frac{d^3p_f}{(2\pi)^32p_f^0}
    \mel{0}{V^{\mu\dagger}(0)}{h^-(p_f)}\mel{h^-(p_f)}{V^\nu(0)}{0} (2\pi)^4 
    \delta^4(q-p_f)\cr    
   &\phantom{abcd} = \frac{1}{\pi}\qty{(q^\mu q^\nu-q^2g^{\mu\nu})v_1(q^2) 
    + q^\mu q^\nu v_0(q^2)}\times \theta(q^0)\ ; \\
&\sum_h \frac{1}{S_h}\int\prod_{f}\frac{d^3p_f}{(2\pi)^32p_f^0}
    \mel{0}{A^{\mu\dagger}(0)}{h^-(p_f)}\mel{h^-(p_f)}{A^\nu(0)}{0} (2\pi)^4 
    \delta^4(q-p_f)\cr    
   &\phantom{abcd} = \frac{1}{\pi}\qty{(q^\mu q^\nu-q^2g^{\mu\nu})a_1(q^2) 
    + q^\mu q^\nu a_0(q^2)}\times \theta(q^0)\ ,
\end{align}
where $S_h$ is the symmetry factor for the hadronic final state (for instance, $S_{\pi^+2\pi^-}=2$).
The normalization of our spectral function is the same as that of Ref.~\cite{ALEPH:1999uux}.
Note that this convention is different from
that of Ref.~\cite{Tsai:1971vv} by a factor $\pi$,
that is, $\left.(v/a)_J\right|_\mathrm{ours} = \pi\times\left.(v/a)_J\right|_\mathrm{Tsai}$.

In terms of these functions, we obtain the decay rate as a integral
over the momentum transfer,
\begin{align}
    \Gamma(\tau^-\rightarrow\nu_\tau+\mathrm{hadrons})
    =&\, \frac{G_F^2\abs{V_{uD}}^2m_\tau^3}{32\pi^3}
    \int dq^2\,\qty(1-\frac{q^2}{m_\tau^2})^2\qty(1+\frac{2q^2}{m_\tau^2})
    v_1(q^2)
    \notag \\
    &\hspace{20pt}
    +\frac{G_F^2\abs{V_{uD}}^2m_\tau^3}{32\pi^3}
    \int dq^2\,\qty(1-\frac{q^2}{m_\tau^2})^2
    v_0(q^2)\ .
\end{align}
The decay rates 
for the axial channels
can be obtained 
by replacing the vector spectral functions with the axial ones.
By including short-distance effects, we get 
\begin{align}
    \label{eq: decay to V J=1}
    \frac{d\Gamma(\tau^-\rightarrow\nu_\tau V^-_{J=1}(q)) }{dq^2}
     &= S_\mathrm{EW}^\tau\times\Gamma_e\frac{6\abs{V_{uD}}^2}{m_\tau^2}
   \qty(1-\frac{q^2}{m_\tau^2})^2\qty(1+\frac{2q^2}{m_\tau^2})v_1(q^2); \\
    \label{eq: decay to A J=1}
   \frac{d\Gamma(\tau^-\rightarrow\nu_\tau A^-_{J=1}(q)) }{dq^2}
     &= S_\mathrm{EW}^\tau\times\Gamma_e\frac{6 \abs{V_{uD}}^2}{m_\tau^2}
   \qty(1-\frac{q^2}{m_\tau^2})^2\qty(1+\frac{2q^2}{m_\tau^2})a_1(q^2);  \\
    \label{eq: decay to V J=0}
   \frac{d\Gamma(\tau^-\rightarrow\nu_\tau V^-_{J=0}(q)) }{dq^2}
     &= S_\mathrm{EW}^\tau\times\Gamma_e\frac{6 \abs{V_{uD}}^2}{m_\tau^2}
   \qty(1-\frac{q^2}{m_\tau^2})^2v_0(q^2); \\
    \label{eq: decay to A J=0}
   \frac{d\Gamma(\tau^-\rightarrow\nu_\tau A^-_{J=0}(q)) }{dq^2}
     &= S_\mathrm{EW}^\tau\times\Gamma_e\frac{6 \abs{V_{uD}}^2}{m_\tau^2}
   \qty(1-\frac{q^2}{m_\tau^2})^2a_0(q^2),
\end{align}
where $\Gamma_e = \left.S^\tau_\mathrm{EW}\right|_\mathrm{lepton} G_F^2m_\tau^5/(192\pi^3)$.
The radiative correction factors $S_\mathrm{EW}^\tau$ and $\left.S_\mathrm{EW}^\tau\right|_\mathrm{lepton}$ are given by
Eqs.~\eqref{eq: Belle SEW} and \eqref{eq: lepton SEW}, respectively.

\subsection{Two Pseudo-Scalar Meson Mode}
Let us restrict our attention to the two pseudo-scalar meson mode. 
The amplitude of the decay, $\tau^-(p_1)\rightarrow\nu_\tau(p_2)+P^-(p_3)+P^0(p_4)$, 
is given by
\begin{align}
    \mathcal{M}(\tau^-\rightarrow\nu_\tau P^-P^0)
    &= -\frac{G_F V^*_{uD}}{\sqrt{2}}\bar{u}_{\nu_\tau}(p_2)\gamma^\mu(1-\gamma_5)u_{\tau}(p_1)
    \mel{P^-(p_3) P^0(p_4)}{V^\mu(q)}{0}\ .
\end{align}
The matrix element of the charged vector current can be written by 
two form factors~\cite{Finkemeier:1996dh},
\begin{align}
    \mel{P^-(p_3) P^0(p_4)}{V^\mu(q)}{0} 
    = C_{P^-P^0}
    \qty[F^{P^-P^0}_V(q^2)\qty(g^{\mu\nu}-\frac{q^\mu q^\nu}{q^2})(p_3-p_4)_\nu
    + \frac{\Delta_{P^-P^0}}{q^2}F^{P^-P^0}_S(q^2)q^\mu]\ ,
\end{align}
where 
$\Delta_{P^-P^0}=m_{P^-}^2-m_{P^0}^2$
and 
$C_{P^-P^0}$ is a normalization constant chosen 
as
\begin{align}
   \abs{C_{\pi^-\pi^0}} = \sqrt{2}\ ; \quad
   \abs{C_{K^-K^0}} = 1\ ; \quad
   \abs{C_{\pi^-\overline{K^0}}} = 1\ ;\quad
   \abs{C_{K^-\pi^0}} = \frac{1}{\sqrt{2}}\ ; \quad
   \abs{C_{K^-\eta}} = \sqrt{\frac{3}{2}}\ .
\end{align}
Except for these group theoretical factors, the matrix elements in the same isospin multiplet
are expressed by the same form factors. Therefore
\begin{align}
    F^{\pi^-\overline{K^0}}_V = F^{K^-\pi^0}_V (=: F^{\pi K}_V)\ ; \quad
    F^{\pi^-\overline{K^0}}_S = F^{K^-\pi^0}_S (=: F^{\pi K}_S)\ .
\end{align}

Then the decay rate,
\begin{align}
\Gamma(\tau^-\rightarrow\nu_\tau P^-P^0) 
    &= \frac{1}{2m_\tau}
    \int \frac{d^3p_2}{(2\pi)^32p_2^0}\frac{d^3p_3}{(2\pi)^32p_3^0}\frac{d^3p_4}{(2\pi)^32p_4^0}
    (2\pi)^4\delta^4(q-p_3-p_4)\notag \\
   &\hspace{60pt}\times\overline{\abs{\mathcal{M}(\tau^-\rightarrow\nu_\tau P^-P^0)}^2} \cr
   &=: \int_{(m_{P^-}+m_{P^0})^2}^{m_\tau^2} dq^2\, \frac{d\Gamma_{P^-P^0}}{dq^2}\ , 
\end{align}
is given by
\begin{align}
   \frac{d\Gamma_{P^-P^0}}{dq^2}
   &=S_\mathrm{EW}^\tau\times 
   \frac{\abs{V_{uD}}^2\Gamma_e}{4m_\tau^2}
   C_{P^-P^0}^2\qty(1-\frac{q^2}{m_\tau^2})^2\notag \\
   &\hspace{10pt}
   \times\qty[\qty(1+\frac{2q^2}{m_\tau^2})\lambda^{3/2}_{P^-P^0}(q^2)\abs{F_V^{P^-P^0}(q^2)}^2
   +3\qty(\frac{\Delta_{P^-P^0}}{q^2})^2\lambda^{1/2}_{P^-P^0}(q^2)\abs{F_S^{P^-P^0}(q^2)}^2],
\end{align}
where $\lambda_{P^-P^0}(q^2):=\lambda(1, m_{P^-}^2/q^2, m_{P^0}^2/q^2)$.
By comparing this with Eqs.~\eqref{eq: decay to V J=1} and \eqref{eq: decay to V J=0}, 
we obtain the relations to the spectral functions:
\begin{align}
\label{eq: two meson v J=1 to form factor}
    v^{P^-P^0}_1(q^2) 
    &= \frac{1}{24}\lambda^{3/2}_{P^-P^0}(q^2)\abs{C_{P^-P^0}F_V^{P^-P^0}(q^2)}^2; \\
\label{eq: two meson v J=0 to form factor}
    v^{P^-P^0}_0(q^2) &= \frac{1}{8}\qty(\frac{\Delta_{P^-P^0}}{q^2})^2
    \lambda^{1/2}_{P^-P^0}(q^2)\abs{C_{P^-P^0}F_S^{P^-P^0}(q^2)}^2.
\end{align}

The mass squared difference, $\Delta_{P^-P^0}$, reflects isospin breaking effects 
and we can ignore it when $P^-$ and $P^0$ have the same strangeness.
Hence we obtain the decay rate for each mode,
\begin{align}
   \frac{d\Gamma_{\pi^-\pi^0}}{dq^2}
   &=
   S_\mathrm{EW}^\tau\times\frac{\abs{V_{ud}}^2\Gamma_e}{2m_\tau^2}
   \qty(1-\frac{q^2}{m_\tau^2})^2
   \qty(1+\frac{2q^2}{m_\tau^2})
   \qty(1-\frac{4m_{\pi^\pm}^2}{q^2})^{3/2}
   \abs{F_V^{\pi^-\pi^0}(q^2)}^2\ ; \\
   \frac{d\Gamma_{K^-K^0}}{dq^2}
   &=
   S_\mathrm{EW}^\tau\times
   \frac{\abs{V_{ud}}^2\Gamma_e}{4m_\tau^2}
   \qty(1-\frac{q^2}{m_\tau^2})^2
   \qty(1+\frac{2q^2}{m_\tau^2})
   \qty(1-\frac{4m_K^2}{q^2})^{3/2}
   \abs{F_V^{K^-K^0}(q^2)}^2\ ; \\
   \frac{d\Gamma_{\pi^-\overline{K^0}}}{dq^2}
   &=
   S_\mathrm{EW}^\tau\times
   \frac{\abs{V_{us}}^2\Gamma_e}{4m_\tau^2}
   \qty(1-\frac{q^2}{m_\tau^2})^2\notag \\
   &\hspace{10pt}
   \times\qty[\qty(1+\frac{2q^2}{m_\tau^2})\lambda^{3/2}_{\pi^-\overline{K^0}}(q^2)\abs{F_V^{\pi K}(q^2)}^2
   +3\qty(\frac{\Delta_{\pi^-\overline{K^0}}}{q^2})^2\lambda_{\pi^-\overline{K^0}}(q^2)\abs{F_S^{\pi K}(q^2)}^2]\ ;
   \\
   \frac{d\Gamma_{K^-\pi^0}}{dq^2}
   &=
   S_\mathrm{EW}^\tau\times
   \frac{\abs{V_{us}}^2\Gamma_e}{8m_\tau^2}
   \qty(1-\frac{q^2}{m_\tau^2})^2\notag \\
   &\hspace{10pt}
   \times\qty[\qty(1+\frac{2q^2}{m_\tau^2})\lambda^{3/2}_{K^-\pi^0}(q^2)\abs{F_V^{\pi K}(q^2)}^2
   +3\qty(\frac{\Delta_{K^-\pi^0}}{q^2})^2\lambda_{K^-\pi^0}(q^2)\abs{F_S^{\pi K}(q^2)}^2]\ ; \\
   \frac{d\Gamma_{K^-\eta}}{dq^2}
   &=
   S_\mathrm{EW}^\tau\times
   \frac{3\abs{V_{us}}^2\Gamma_e}{8m_\tau^2}
   \qty(1-\frac{q^2}{m_\tau^2})^2\notag \\
   &\hspace{10pt}
   \times\qty[\qty(1+\frac{2q^2}{m_\tau^2})\lambda^{3/2}_{K^-\eta}(q^2)\abs{F_V^{K^-\eta}(q^2)}^2
   +3\qty(\frac{\Delta_{K^-\eta}}{q^2})^2\lambda_{K^-\eta}(q^2)\abs{F_S^{K^-\eta}(q^2)}^2]\ .
\end{align}

\bibliographystyle{apsrev4-1}
\bibliography{ref}

\end{document}